\documentclass[11pt,letterpaper]{article}
\usepackage[T1]{fontenc}
\usepackage[utf8]{inputenc}
\usepackage{lmodern}
\usepackage[margin=1in]{geometry}
\usepackage[english]{babel}
\usepackage{amsmath,amssymb,amsthm,mathtools}
\usepackage{enumitem,booktabs,array,multirow,microtype,xurl,xcolor,placeins,framed,float}
\usepackage{tikz}
\usetikzlibrary{arrows.meta}
\usepackage[ruled,linesnumbered,vlined]{algorithm2e}
\usepackage[noadjust]{cite}
\usepackage[colorlinks=true,allcolors=blue!45!black,linktoc=all]{hyperref}

\definecolor{shadecolor}{gray}{0.9}
\definecolor{diagramblue}{HTML}{246C86}
\definecolor{diagramamber}{HTML}{A05B18}
\setlist{itemsep=2pt,topsep=4pt,leftmargin=*}
\allowdisplaybreaks[2]
\numberwithin{equation}{section}
\DeclarePairedDelimiter\floor{\lfloor}{\rfloor}
\DeclarePairedDelimiter\ceil{\lceil}{\rceil}

\newtheorem{theorem}{Theorem}[section]
\newtheorem{lemma}[theorem]{Lemma}
\newtheorem{proposition}[theorem]{Proposition}

\theoremstyle{definition}
\newtheorem{definition}[theorem]{Definition}

\theoremstyle{remark}

\newcommand{\bR}{\mathbb{R}}
\newcommand{\bX}{\mathbf{X}}

\newcommand{\bc}{\mathbf{c}}
\newcommand{\bv}{\mathbf{v}}
\newcommand{\bt}{\mathbf{t}}
\newcommand{\bx}{\mathbf{x}}
\newcommand{\by}{\mathbf{y}}
\newcommand{\bw}{\mathbf{w}}
\newcommand{\cP}{\mathcal{P}}
\newcommand{\cT}{\mathcal{T}}
\newcommand{\Comp}{\mathsf{Comp}}
\newcommand{\Pre}{\mathsf{Pre}}
\newcommand{\Suf}{\mathsf{Suf}}

\DeclareMathOperator*{\argmin}{argmin}
\DeclareMathOperator*{\argmax}{argmax}
\DeclareMathOperator{\Rank}{Rank}

\PrintSemicolon
\SetAlFnt{\small}
\SetAlCapFnt{\small}
\SetAlgoCaptionSeparator{.}

\title{Logarithmic Comparison-Query Complexity for EF1 Allocations}
\author{}
\date{}

\author{%
  \begin{tabular}{@{}c@{\qquad\qquad}c@{}}
    Zehan Lin & Shengxin Liu \\
    University of Macau & Harbin Institute of Technology, Shenzhen \\
    \texttt{yc47490@um.edu.mo} & \texttt{sxliu@hit.edu.cn} \\
    \noalign{\vskip 1.0em}
    Biaoshuai Tao & Shengwei Zhou \\
    Shanghai Jiao Tong University & Nanyang Technological University \\
    \texttt{bstao@sjtu.edu.cn} & \texttt{s.arthur.zhou@gmail.com}
  \end{tabular}%
}

\begin{document}
\maketitle

\begin{abstract}
We study the comparison-query complexity of computing allocations that are \emph{envy-free up to one item} (EF1) for $n$ agents with heterogeneous additive preferences over $m$ indivisible items.
In this model, the algorithm may only query which of two bundles an agent prefers, without ever observing numerical valuations or costs.
Prior work established $O(\log m)$ query bounds for at most three agents, leaving open whether query complexity logarithmic in $m$ is achievable for every fixed $n \ge 4$.
We resolve this open question in the affirmative: for every fixed $n\ge2$, we design algorithms that compute EF1 allocations using $O(\log m)$ comparison queries, matching the known $\Omega(\log m)$ lower bound.
Our results hold for both chores with nonnegative costs and goods with nonnegative valuations.
\end{abstract}


\clearpage
\pdfbookmark[1]{\contentsname}{toc}
\tableofcontents
\clearpage

\pagenumbering{arabic}

\section{Introduction}\label{sec:intro}
Fair division is a fundamental problem that has been extensively studied in mathematics, economics, and computer science~\cite{steinhaus1948problem}, which considers the allocation of a set $M$ of $m$ indivisible items among a set $N$ of $n$ agents with heterogeneous preferences.
When the valuations are positive, the items are viewed as \emph{goods} (e.g., resources), whereas when the valuations are negative, the items are regarded as \emph{chores} (e.g., tasks).
The central objective of the fair division problem is to compute an allocation that ensures fairness for all agents.
%

Various criteria have been proposed to evaluate fairness, among which \emph{envy-freeness} (EF)~\cite{Foley1967} is a fundamental notion that requires every agent to weakly prefer her own bundle to that of any other agent.
Since EF cannot always be achieved with indivisible items (even for two agents and a single item), a standard relaxation is \emph{envy-freeness up to one item} (EF1)~\cite{LiptonEtAl2004,BhaskarEtAl2021}.
For goods, EF1 requires that any envy can be eliminated by the removal of one good from the envied agent's bundle, whereas for chores it requires the removal of one chore from the envious agent's own bundle.
Prior work has shown that EF1 allocations always exist in both settings and can be computed via the envy-cycle procedure for goods~\cite{LiptonEtAl2004} and the top-trading envy-cycle elimination algorithm for chores~\cite{BhaskarEtAl2021}.

However, most existing algorithmic work on fair division assumes cardinal access to preferences, under which the algorithm is given or can query exact numerical valuations or costs.
This assumption imposes a strong requirement in practice, because an agent's evaluation of a bundle often depends on subjective factors like effort, expertise, and time constraints, and numerical reports may also expose sensitive private information.
At the same time, agents can still answer pairwise comparisons reliably, since an agent only needs to decide which of two proposed bundles she prefers without the need to assign a precise numerical value or cost to either.
These considerations motivate the \emph{comparison-query model}, which was introduced for indivisible goods by Bu, Li, Liu, Song, and Tao~\cite{conf/wine/BuLLST24} and subsequently studied for both goods~\cite{journals/corr/YamagataS26} and chores~\cite{LinEtAl2026}.
In this model, the algorithm accesses preferences exclusively through pairwise bundle comparisons, and we measure complexity by the number of queries made to the agents.

Since classic EF1 algorithms such as round-robin and envy-cycle elimination~\cite{LiptonEtAl2004,BhaskarEtAl2021,journals/siamdm/OhPS21} allocate items sequentially, their direct implementations in the comparison-query model require $\Omega(m)$ queries.
This linear dependence on $m$ is particularly undesirable when a few agents divide many items.
For instance, \cite{journals/teco/CaragiannisKMPS19} reported that more than 95\% of Spliddit instances involved at most three agents, and they also documented an inheritance-division instance with roughly $1{,}400$ goods and ten siblings.
Moreover, approximate algorithms for cake division often discretize a heterogeneous resource into small intervals that are treated as indivisible items, where finer discretizations yield smaller approximation errors and increase $m$ while $n$ remains fixed~\cite{conf/aaai/BalkanskiBKP14,conf/nips/BranzeiN22}.
These applications motivate the pursuit of comparison-query bounds that are sublinear, or even logarithmic, in $m$ for every fixed $n$.

To improve upon the $O(m)$ query complexity of classic algorithms and achieve sublinear query bounds is not immediate.\footnote{Note that even for two agents with identical additive preferences, a deterministic algorithm requires $\Theta(m)$ comparison queries in the worst case to verify whether a prescribed allocation is EF1 (see Appendix~\ref{sec:ef1-verification}).}
A na\"ive partition of the items into groups followed by an EF1 algorithm on the groups does not by itself guarantee EF1 for the original items, since the removal of an entire group may eliminate envy whereas the removal of a single original item may not.
With heterogeneous preferences, a given partition may leave several agents with only the same few bundles that are acceptable under EF1.
Any adjustment to these bundles for one agent may then invalidate the EF1 guarantees for others.
The central challenge is therefore to find, with few bundle comparisons, a partition that admits an EF1 assignment for all agents.

For two or three agents with additive preferences, $O(\log m)$-query EF1 algorithms have been developed for goods~\cite{conf/wine/BuLLST24} and chores~\cite{LinEtAl2026}, which match the $\Omega(\log m)$ lower bound in both settings.
When agents have identical additive valuations for goods, Yamagata and Sumita~\cite{journals/corr/YamagataS26} also achieved $O(\log m)$ comparison queries for any fixed number of agents.
For heterogeneous additive preferences, however, whether $O(\log m)$ queries suffice has remained open even for four agents.
This gap leads to our main question.
\begin{center}
\begin{minipage}{0.97\linewidth}
\begin{shaded}
\noindent\textbf{Main Question.}
\textit{For every fixed number of agents, can an EF1 allocation of indivisible items be computed using only $O(\log m)$ comparison queries under heterogeneous additive preferences?}
\end{shaded}
\end{minipage}
\end{center}

\subsection{Our Contributions and Techniques}
In this paper, we answer this question affirmatively for both chores and goods, with a query bound that is polynomial in the number of agents.

\begin{center}
\begin{minipage}{0.97\linewidth}
\begin{shaded}
\noindent\textbf{Main Result.}
(Theorems~\ref{thm:chores} and~\ref{thm:goods})\textbf{.}
\emph{For both chores and goods with nonnegative additive costs or valuations, EF1 allocations can be computed using $O(n^8\log m)$ comparison queries.}
\end{shaded}
\end{minipage}
\end{center}

For every fixed number of agents, both bounds simplify to $O(\log m)$ and match the known $\Omega(\log m)$ lower bound, which holds even for identical binary preferences~\cite{conf/wine/BuLLST24,LinEtAl2026}.
Throughout this paper, we write $O_n(\cdot)$ to hide factors that depend only on $n$.
Since the underlying techniques for both settings are similar, we focus on chores in the main text and defer the algorithm and analysis for goods to Appendix~\ref{sec:goods}.
In the following, we present our main ideas in this paper.

We first review the three-agent algorithm in~\cite{LinEtAl2026}, which starts with a partition $(A,B,C)$ found in $O(\log m)$ queries where every bundle is EF1 for agent~$1$.
The only nontrivial case arises when agents~$2$ and~$3$ share a unique favorite bundle, say $A$.\footnote{Otherwise, agents~$2$ and~$3$ can receive distinct favorite bundles and agent~$1$ takes the third bundle, which immediately gives an EF1 allocation.}
Their algorithm transfers chores from $B$ and $C$ to $A$ while $B$ and $C$ remain EF1 for agent~$1$, which locates a preference transition for agents~$2$ and~$3$ in $O(\log m)$ queries and yields an EF1 allocation via a case analysis on the two adjacent partitions.
For $n > 3$, however, preference conflicts can arise simultaneously across multiple subsets of agents and bundles, so a transition between two agents no longer resolves all of them and this argument does not directly extend to general $n$.

From their algorithm, we nevertheless obtain some useful observations.
When $n$ is large, an initial partition based solely on one agent's EF1 condition is no longer a desirable foundation.
Intuitively, such a partition is advantageous for the divider, but it provides no guarantee for the other agents.
In the worst case, every other agent may find only a single bundle acceptable under EF1 and view all other bundles as excessively costly.
Therefore, we need to establish a more balanced global structure that serves as a favorable initial state for all agents.
Moreover, we observe that heavy chores can cause abrupt jumps in bundle costs under reallocation, which makes gradual adjustments difficult.
Ideally, bundle adjustments should move only light chores so that agent preferences evolve smoothly.

Based on these observations, we first reduce the instance to one without heavy chores.
Intuitively, under a suitable threshold, the number of heavy chores is only $O_n(1)$, so we can temporarily set them aside and later insert them via envy-cycle procedure.\footnote{Recall that our goal is a query bound that is logarithmic in $m$ for every fixed $n$, so an additional $O_n(1)$ overhead does not affect the asymptotic dependence on $m$.}
For the light chores, we then construct $n-1$ bundles that are balanced for all agents and gradually transfer chores from them into an initially empty $n$-th bundle to achieve an EF1 allocation.

\paragraph{Constructing Small Blocks and Balanced Bundles.}
To support these transfers, we maintain a residual pool alongside a set of isolated heavy chores reserved for later reinsertion.
Our goal is to partition the residual pool into $n-1$ bundles such that, for every agent, every bundle still costs more than $1/n$ of her residual pool even after the removal of any single chore.
The main difficulty is that heaviness must be measured relative to the current residual pool, since the heaviness threshold drops as the pool shrinks and a previously small chore may become heavy.
We bound the total number of removals by $O_n(1)$ through an early termination rule combined with binary splitting trees that are dynamically reused across removals.
Specifically, whenever some agent finds $n-1$ isolated chores that are individually at least as costly as the entire residual pool, we can assign the residual pool to her and one such chore to every other agent to obtain an EF1 allocation directly.
Because every cut in the splitting trees is located via binary search and the total number of splits and updates depends only on $n$, this reduction takes $O_n(\log m)$ comparison queries.
When the reduction does not terminate early, the union of all agents' split points partitions the residual chores into $\mathrm{poly}(n)$ contiguous blocks that are small for every agent.
Under a uniform random assignment that places every block independently into one of the $n-1$ bundles, every bundle has expected cost $1/(n-1)$ of every agent's residual pool.
Since every block is small, a probabilistic argument based on a variance bound ensures that some partition keeps all bundle costs close to these expectations simultaneously, and the slack between $1/(n-1)$ and $1/n$ preserves the $1/n$ lower bound after the removal of any single chore.
We then convert this existence argument into a deterministic construction in the comparison model via carefully chosen reference bundles that serve as benchmarks for balance.

\paragraph{Establishing EF1 via Continuous Matching.}
Given the $n-1$ balanced bundles, all residual chores initially lie in them, while the $n$-th bundle is empty.
We then gradually transfer chores from the ends of these bundles into the $n$-th bundle, which leaves a prefix of every original bundle.
As the prefixes shrink, their costs weakly decrease while the cost of the $n$-th bundle weakly increases.
Thus, the set of agents who regard the $n$-th bundle as a favorite bundle shrinks monotonically as chores are transferred.
We aim to identify a transition where this set is nonempty before one more transfer and becomes empty afterward.
The main challenge is to prove that a transition of this form can also support an EF1 allocation, since the other $n-1$ agents must simultaneously receive distinct prefixes while ensuring EF1.
We address this challenge by a continuous interpolation of every agent's prefix preferences, which allows the cuts to move continuously between consecutive item positions.
A topological theorem of Meunier and Su~\cite{MeunierSu2019} then yields a transition where distinct favorite prefixes can be assigned to the other $n-1$ agents regardless of who receives the $n$-th bundle.
Our continuous extension is designed so that, after rounding the cuts back to whole items, every prefix recipient can eliminate envy toward the other prefixes through the removal of the last chore from her own bundle.
Together with the preference transition for the $n$-th bundle, this guarantee ensures that at least one of the two adjacent integral partitions admits an EF1 allocation.

\paragraph{Searching via Root Monotonicity and Item Grouping.}
While the topological argument is nonconstructive, we show that the desired transition can be located within $O(\mathrm{poly}(n)\log m)$ comparison queries.
By the design of our continuous extension, the graph linking agents to their favorite prefixes at any valid transition forms a tree with a distinguished root bundle.
Our key structural result is root monotonicity: for a fixed root bundle, all other cuts can only move in the same direction as the root cut, even when the underlying tree changes.
In particular, the root cut uniquely determines all other cuts.
Although this monotonicity along a single coordinate suggests a binary search over the root cut, a direct implementation would need to guess the tree structure and run nested binary searches for the other $n-2$ bundles, which takes $O_n(\log^2 m)$ queries with an exponential factor in $n$.
Instead, we maintain candidate prefix-length vectors for every choice of the root and rule them out in rounds.
In every round, we group consecutive chores into $O(n)$ blocks in each bundle and compare only prefixes that end at block boundaries.
We choose these boundaries according to the remaining candidate cuts, with finer divisions in regions where many candidates are concentrated.
The key is to ensure that only a small fraction of candidates are ``exceptional'', for which grouping may prevent a direct use of root monotonicity.
We keep these candidates and use root monotonicity to filter the ``non-exceptional'' ones, ruling out many candidates at once while preserving every candidate that could yield the desired transition.
Each round then either locates a valid transition or eliminates a constant fraction of the candidates, which gives an overall query bound of $O(\mathrm{poly}(n)\log m)$.

\subsection{Related Work}
We review prior work on query and communication complexity, with an emphasis on fair division.
For broader background, see~\cite{journals/ai/AmanatidisABFLMVW23,journals/sigecom/AzizLMW22,liu2024mixed,Moulin2003}.

\paragraph{Query Complexity for Indivisible Items.}
Comparison-query algorithms have also been developed for other fairness guarantees.
For additive goods, Bu, Li, Liu, Song, and Tao~\cite{conf/wine/BuLLST24} gave an $O(n^4\log m)$-query algorithm guaranteeing both proportionality up to one item (PROP1) and at least half of each agent's maximin share (MMS).
Yamagata and Sumita~\cite{journals/corr/YamagataS26} improved this bound to $O(n^3\log m)$.
For additive chores, Lin, Liu, Tao, and Zhou~\cite{LinEtAl2026} obtained an $O(n^3\log m)$-query algorithm for PROP1 and, for fixed $n$ and $\varepsilon>0$, an $O(\log m)$-query algorithm for $(13/11+\varepsilon)$-MMS.
Other comparison-based work on goods considered envy-freeness up to one less-preferred good (EFL), including the completion of partial EFL allocations~\cite{AkramiEtAl2026} and simultaneous EFL and constant-factor MMS guarantees under submodular valuations~\cite{FeigeFine2026}.
Li, Manurangsi, Scarlett, and Suksompong~\cite{journals/iandc/LiMSS25} proved an $\Omega(nm+m\log m)$ lower bound for exactly implementing round-robin using comparisons between individual items, and also studied noisy queries.
Noisy-query models were also studied for envy-free allocation~\cite{journals/corr/abs-2602-06361}.
In the value-query model, a query reveals an agent's numerical value for a specified bundle.
Oh, Procaccia, and Suksompong~\cite{journals/siamdm/OhPS21} gave EF1 algorithms for goods with additive valuations using $O(\log m)$ value queries for two or three agents.
They also gave an $O(nm)$-query implementation of envy-cycle elimination for arbitrary $n$.
For envy-freeness up to any good (EFX), value-query lower bounds are $\Omega(m)$ for deterministic algorithms under additive valuations~\cite{journals/siamdm/OhPS21} and exponential in $m$ even for randomized algorithms under submodular valuations~\cite{journals/siamdm/PlautR20}.

\paragraph{Communication Complexity for Indivisible Items.}
Communication complexity measures the number of bits exchanged by agents who know their own valuations and can send arbitrary messages.
Plaut and Roughgarden~\cite{journals/siamcomp/PlautR20} characterized polynomial versus exponential communication for finding envy-free or proportional allocations, or determining that none exist, across several valuation classes and multiplicative relaxations.
For additive goods with $m\ge 2n$, Feige~\cite{conf/sigecom/Feige25} gave protocols simultaneously guaranteeing PROP1 and an $n/(2n-1)$ fraction of the truncated proportional share (TPS), where TPS is at least MMS.
These protocols used $O(n\log m)$ expected bits with randomization and $O(n\log m\log n)$ bits deterministically.
He also proved an $\Omega(n\log(m/n))$ randomized communication lower bound for PROP1, even with identical binary additive valuations.
For EF1, he noted communication upper bounds of $O(m\log m)$ under additive valuations and $O(n\log(m/n))$ under identical additive valuations.
Since a comparison-query algorithm can be simulated using one communicated bit per query, comparison-query upper bounds imply communication upper bounds, while communication lower bounds imply comparison-query lower bounds for the same allocation task.
Communication upper bounds, however, need not yield comparison-query upper bounds, since communication protocols allow messages beyond bundle comparisons.

\paragraph{Query Complexity for Divisible Items.}
For cake cutting, the Robertson--Webb model permits evaluation queries for interval values and cut queries for locating endpoints with prescribed values~\cite{books/daglib/RobertsonW98}.
Proportional division has deterministic total query complexity $\Theta(n\log n)$~\cite{journals/dam/EvenP84,journals/talg/EdmondsP11}, with the same lower bound on the expected number of queries for randomized protocols allocating connected pieces~\cite{ArndtPruhsTran2026}.
Counting the two query types separately, Arndt, Moseley, Im, and Pruhs~\cite{conf/soda/ArndtMIP27} gave a deterministic protocol using $O(n)$ cut queries and $O(n^2)$ evaluation queries.
For exact envy-freeness, Procaccia~\cite{conf/ijcai/Procaccia09} proved an $\Omega(n^2)$ query lower bound.
Bounded protocols allowing disconnected pieces were established for goods~\cite{conf/stoc/AzizM16,conf/focs/AzizM16} and chores~\cite{conf/soda/DehghaniFHY18}.
Recent work on goods obtained single-exponential deterministic query bounds~\cite{YeBai2026} and polynomial expected query bounds using randomization~\cite{MackenzieSuzuki2026}.
Psomas~\cite{Psomas2026} gave a bounded protocol with an elementary, self-contained correctness proof.
With connected pieces, Stromquist~\cite{journals/combinatorics/Stromquist08} ruled out finite Robertson--Webb protocols for exact envy-freeness with three or more agents.
For connected $\varepsilon$-envy-free allocations with additive valuations, Br{\^a}nzei and Nisan~\cite{conf/nips/BranzeiN22} established a tight deterministic query bound of $\Theta(\log(1/\varepsilon))$ for three agents, while Hollender and Rubinstein~\cite{conf/focs/HollenderR23} gave an $O(\log^2(1/\varepsilon))$-query algorithm for four agents.
Other work studied simultaneous preference elicitation~\cite{conf/aaai/BalkanskiBKP14}, bounded-round protocols~\cite{BranzeiPaparasRecker2020}, and communication complexity~\cite{conf/ec/BranzeiN19}.

\paragraph{Query Complexity Beyond Fair Division.}

Comparison queries have also been studied in optimization, learning, and social choice.
\cite{conf/stoc/Cohen-AddaddGGL26} studied linear combinatorial optimization using an oracle that compares the total weights of two feasible solutions without revealing their numerical values.
They established general query bounds and gave algorithms for minimum spanning trees, bipartite matchings, and shortest paths.
Fox and Woodruff~\cite{FoxWoodruff2026} studied symmetric submodular minimization using only comparisons between function values.
Related work studied learning combinatorial functions from pairwise comparisons~\cite{BalcanVitercikWhite2016} and solving $k$-SUM and related problems using comparisons between subset sums~\cite{KaneLovettMoran2019}.
In linear social choice, Ge, Halpern, Kehne, and Vorobeychik~\cite{conf/soda/GeHKV27} studied pairwise preference queries for selecting candidates with high social welfare.
Recent work also studied sorting using comparisons when some order relations are already known~\cite{conf/soda/HaeuplerHIRTT25,conf/focs/Rutschmann26}.
Beyond comparison queries, recent work studied fixed-point computation using function-evaluation queries~\cite{journals/jacm/ChenLY25,conf/focs/ChenLY26}, submodular maximization using value and independence oracles~\cite{conf/focs/BuchbinderF24,conf/stoc/BuchbinderF25}, and graph edge connectivity using cut queries~\cite{conf/soda/0001SW25}.


\section{Preliminaries}\label{sec:model}

We consider the problem of fairly allocating a set $M = \{e_1, \dots, e_m\}$ of $m$ indivisible chores among a set $N = [n]$ of $n$ agents, where $[t] := \{1, \dots, t\}$ for any integer $t \ge 1$. We refer to any subset $X \subseteq M$ as a \emph{bundle}.
Each agent $i\in N$ has a cost function $c_i:2^M\to\bR_{\ge0}$ and prefers bundles of smaller cost. We write $\bc=(c_1,\dots,c_n)$ for the cost profile.
A set function $f: 2^M \to \bR_{\ge 0}$ is \emph{additive} if $f(S) = \sum_{e \in S} f(\{e\})$ for all $S \subseteq M$. We write $f(e)$ for $f(\{e\})$.
Throughout, we assume that all cost functions are additive.
An \emph{allocation} $\bX=(X_1,\dots,X_n)$ consists of pairwise disjoint bundles whose union is $M$, with bundle $X_i$ assigned to agent $i$. Empty bundles are allowed. A \emph{partial allocation} is an $n$-tuple of pairwise disjoint subsets of $M$ that need not cover all items. We denote the residual set of unallocated items by $R=M\setminus\bigcup_{i\in N}X_i$.
For an ordered sequence of items $S = (s_1, \dots, s_r)$ and an integer $t \in \{0, \dots, r\}$, we define its prefix and suffix of lengths $t$ and $r-t$, respectively, by
$$
    \Pre(S, t) := \{s_1, \dots, s_t\}
    \quad\text{and}\quad
    \Suf(S, t) := \{s_{t+1}, \dots, s_r\},
$$
with $\Pre(S, 0) = \Suf(S, r) = \emptyset$. When items are arranged in a fixed linear order, a partition is \emph{contiguous} if every bundle is a contiguous interval of that order.

Next, we introduce several fairness notions.
\begin{definition}[EF]
An allocation $\bX$ is envy-free (EF) if for any agents $i,j\in N$, we have $c_i(X_i)\le c_i(X_j)$.
\end{definition}

\begin{definition}[EF1 for Chores]\label{def:chore-ef1}
An allocation $\bX$ is \emph{envy-free up to one item} (EF1) for chores if for all agents $i,j\in N$, either $X_i=\emptyset$, or there exists an item $e\in X_i$ such that $c_i(X_i\setminus\{e\})\le c_i(X_j)$.
Furthermore, any such $e$ is called a \emph{witness item} for agent $i$ with respect to agent $j$.
\end{definition}
The same definitions apply to partial allocations.

\begin{definition}[Comparison Model]\label{def:comparison}
Algorithms access the cost functions only through comparison queries.
For any agent $i\in N$ and bundles $X,Y\subseteq M$, the query $\Comp_i(X,Y)$ returns the bundle with lower cost, breaking ties in favor of the first argument:
$$
    \Comp_i(X,Y)=
    \begin{cases}
        X, & \text{if }c_i(X)\le c_i(Y),\\
        Y, & \text{if }c_i(X)>c_i(Y).
    \end{cases}
$$
\end{definition}

In this paper, each invocation of $\Comp_i$ counts as one comparison query. 
%
We fix an arbitrary public linear order $M=(e_1,\dots,e_m)$ on the items in advance. 
Unless specified otherwise, the items within each bundle inherit this order. 
%
When $n=1$ or $m\le n$, EF1 allocations require no queries.
For $n=2$, the binary-search cut-and-choose method~\cite{conf/wine/BuLLST24} also gives an EF1 allocation of chores under additive costs using $O(\log m)$ comparisons.
This bound is also known for $n=3$~\cite{LinEtAl2026}.
For the remainder of the main text, we assume that $n\ge3$ and $m>n$.

\section{From Balanced Partitions to EF1 Allocations}\label{sec:shared}
\begin{figure}[!htb]
\centering
\begin{tikzpicture}[
    x=0.88cm,y=0.88cm,font=\small,
    bundleoutline/.style={draw=black!55,line width=0.7pt},
    suffixflow/.style={draw=diagramblue,line width=0.9pt}
]
\foreach \y/\cut/\j in {3/2.2/1,1.8/1.5/2,0/2.8/{n-1}} {
    \fill[black!2] (0,{\y-0.29}) rectangle (4.4,{\y+0.29});
    \fill[diagramblue!18] (\cut,{\y-0.29}) rectangle (4.4,{\y+0.29});
    \draw[bundleoutline] (0,{\y-0.29}) rectangle (4.4,{\y+0.29});
    \draw[suffixflow] (\cut,{\y-0.29}) -- (\cut,{\y+0.29});
    \node[anchor=east] at (-0.2,\y) {$P_{\j}$};
    \draw[suffixflow] (4.4,\y) -- (5.8,\y);
}
\node at (2.2,0.9) {$\vdots$};
\node[anchor=base] at (2.2,-0.85) {Balanced bundles};
\draw[suffixflow] (5.8,0) -- (5.8,3);
\draw[suffixflow,-{Stealth[length=2.2mm]}] (5.8,1.5) -- (7.1,1.5);
\draw[bundleoutline,fill=diagramblue!18]
    (7.1,1.21) rectangle (11.5,1.79);
\node at (9.3,0.9) {$R$};
\node[anchor=base] at (9.3,-0.85) {Remaining bundle};
\end{tikzpicture}
\caption{The blue suffixes are combined into a single bundle $R$, while the unshaded prefixes form the other $n-1$ bundles. Chores are ordered from left to right, and lengths are for illustration only.}
\label{fig:bundle-overview}
\end{figure}
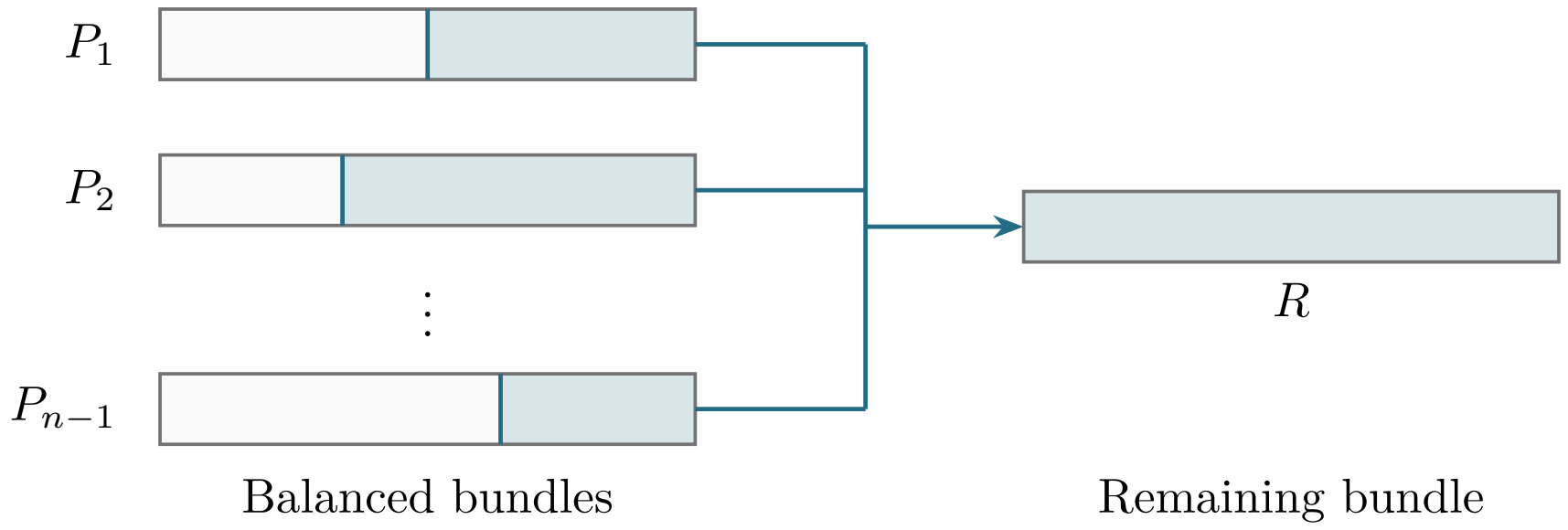

In this section, we reduce the computation of an EF1 allocation under nonnegative additive costs to the task of locating suitable cuts in $n-1$ balanced bundles (see Figure~\ref{fig:bundle-overview}).
In the following, we first show how to construct these balanced bundles after we set aside a bounded number of heavy chores, which can be reinserted via envy-cycle procedure (Section~\ref{subsec:(n-1)-partition}).
We then combine continuous interpolation with the topological matching theorem of Meunier and Su~\cite{MeunierSu2019} and rounding to establish the existence of an EF1 allocation in which $n-1$ agents receive prefixes of distinct balanced bundles and the remaining agent receives the combined suffixes (Section~\ref{subsec:balance}).
The structural properties established in this existence proof then guide our efficient search procedure, which we present in Section~\ref{sec:search}.

\subsection{The Construction of Small Blocks and Balanced Bundles}
\label{subsec:(n-1)-partition}
As the first step, we design an algorithm that constructs $n-1$ bundles $P_1,\ldots,P_{n-1}$ that are sufficiently balanced for every agent after we set aside a small number of chores.
Specifically, we choose $k$ as the smallest power of two that is at least $16n^5$, which satisfies $16n^5 \le k < 32n^5$.

\begin{samepage}
\begin{theorem}\label{thm:(n-1)-partition}
   There exists an algorithm that makes $O(n^8\log m)$ comparison queries and outputs a heavy set $E\subseteq M$ with $\vert{}E\vert{}=O(nk\log k)$ together with one of the outcomes below for $M'=M\setminus E$:
    \begin{itemize}
        \item an EF1 allocation of $M'$, or
        \item a partition $(P_1,\ldots,P_{n-1})$ of $M'$ satisfying both of the following conditions:
        \begin{enumerate}
            \item $c_i(M')>0$ for each agent $i$, and $c_i(e)\leq  c_i(M') / k$ for each agent $i$ and each $e\in M'$;
            \item for each $j\in[n-1]$, we have $|P_j|\geq2$ and $c_i(P_j)>(\frac1n+\frac1k) \cdot c_i(M')$ for each agent $i \in N$.
        \end{enumerate}
    \end{itemize}
\end{theorem}
\end{samepage}
Note that once an EF1 allocation of $M' = M\setminus E$ is obtained, the remaining chores in $E$ can be inserted one by one.
The following lemma, based on the top-trading envy-cycle elimination procedure of Bhaskar, Sricharan, and Vaish~\cite{BhaskarEtAl2021}, bounds the additional query cost by $O(n^2|E|)$, which is $O(n^8\log n)$ by  Theorem~\ref{thm:(n-1)-partition}.
\begin{lemma}[\cite{BhaskarEtAl2021}]\label{lem:chore-insert}
Any EF1 partial allocation $\bX$ can be extended to include an unallocated chore $e$ while preserving EF1 using $O(n^2)$ comparison queries.
\end{lemma}
Therefore, we only need to focus on the second branch of the algorithm described in Theorem~\ref{thm:(n-1)-partition}.
The techniques for proving Theorem~\ref{thm:(n-1)-partition} are largely independent of the technique of our main algorithm.
We therefore defer the proof of Theorem~\ref{thm:(n-1)-partition} to Appendix~\ref{append:(n-1)-partition} and instead highlight the main ideas here.

The algorithm first identifies the \emph{heavy} chores that obstruct a balanced partition into $n-1$ bundles.
Suppose that the chores are arranged along a line.
To divide the line into $k$ parts of roughly equal cost, every agent $i$ places $k-1$ cut points along this line, where every cut point falls on a single boundary chore and splits the segment into a prefix, that singleton chore, and a suffix (see Figure~\ref{fig:interval-tree}).
Across all $k-1$ cuts, this process yields a partition into $2k-1$ contiguous blocks, which consist of $k-1$ singletons at the cut points together with $k$ interior blocks between them.
We can obtain this partition for every agent via $k-1$ binary searches, which requires $O(k\log m)$ comparison queries per agent.
Then, through comparisons between the singletons and the interior blocks, we can effectively detect heavy chores.

One complication is that heaviness is measured relative to the current residual pool $M'$ rather than the original chore set $M$ (as Theorem~\ref{thm:(n-1)-partition} requires $c_i(e)\leq c_i(M') /k$ rather than $c_i(e)\leq c_i(M)/k$).
Therefore, we must repeat the above procedure whenever a heavy chore is detected and removed.
New heavy chores may emerge in $M'$ after a removal in every iteration, which could otherwise lead to an excessive number of removals.
To prevent this issue, we introduce an early termination rule based on a simple observation.
If, at some point, an agent $i$ finds that the $n-1$ most costly chores in $E$ are individually more costly than $M\setminus E$, then we can immediately construct an EF1 allocation in which agent $i$ receives $M\setminus E$, every other agent receives one of these $n-1$ chores in $E$, and the rest of $E$ is allocated via cycle elimination on the envy graph.

The query complexity for this step can be further reduced by some techniques that reuse the ``splitting tree'' for the $(2k-1)$-partition step.
This can avoid redoing the whole splitting procedure after every removal of a heavy chore.
At a cost of this, we need a finer partition but still with $O(k)$ cuts.
The above-mentioned techniques are available in Appendix~\ref{subsec:compression}.

After the above-mentioned procedure, we end up with $M'$ that contains only small enough chores (if the procedure does not terminate with an EF1 allocation).
In addition, each agent has $O(k)$ cut points on the interval.
Combining the cut points of all agents gives a partition of $M'$ into $O(nk)$ blocks where each block has cost at most $c_i(M')/k$ for each agent $i$.
We view each block as a single chore, so the number of chores now is $\text{poly}(n)$, which is independent of $m$ (notice that the item-grouping here is only used for Theorem~\ref{thm:(n-1)-partition}; we will consider original chores in later subsections).
The existence of the required partition $(P_1,\ldots,P_{n-1})$ can be proved by a standard probabilistic argument, and the lower bound $16n^5$ has been chosen large enough to make the probabilistic argument work.

A more refined probabilistic argument based only on \emph{pairwise independence} can reduce the space of all partitions into $n-1$ bundles to a family of size $\mathrm{poly}(n)$ that still contains a valid partition.
Thus, we can simply enumerate all partitions in this reduced space to identify a valid one.
One difficulty is that the validity condition $c_i(P_j) > \bigl(\frac{1}{n}+\frac{1}{k}\bigr) \cdot c_i(M')$ cannot be directly verified via comparison queries.
To resolve this issue, we construct a \emph{reference bundle} for every agent $i$ with cost slightly larger than $\bigl(\frac{1}{n}+\frac{1}{k}\bigr) \cdot c_i(M')$ (and we show that the replacement of the threshold $\bigl(\frac{1}{n}+\frac{1}{k}\bigr)\cdot c_i(M')$ by the cost of the reference bundle for every agent still preserves a valid partition).

Finally, we apply an additional refinement to further reduce the query complexity.
During the exhaustive search, we can avoid direct queries on every candidate partition.
The algorithm maintains normalized \emph{auxiliary costs} and first selects a partition that is sufficiently balanced under those costs.
Whenever a preselected partition fails a reference bundle comparison, we update and project the auxiliary costs so that the squared distance to the normalized true costs drops by a guaranteed margin, which bounds the total number of queried candidates.
The full details of this construction appear in Appendix~\ref{subsec:balanced-construction}.

\subsection{The Existence of EF1 Allocations}\label{subsec:balance}
In this subsection, we fix the $n-1$ balanced bundles $(P_1,\dots,P_{n-1})$ from Theorem~\ref{thm:(n-1)-partition} and show that removing suitable suffixes from them and merging these suffixes into an $n$-th bundle always yields an EF1 allocation.
Specifically, one agent receives the combined suffix bundle while the remaining $n-1$ agents are matched to distinct prefix bundles so that EF1 holds between every pair of agents.

For $j \in [n - 1]$, let $(e_{j,1},\dots,e_{j,|P_j|})$ denote the chores in each bundle $P_j$ in order.
Given a prefix-length vector $\bt=(t_1,\dots,t_{n-1})$ with $t_j\in[|P_j|]$ for each $j\in[n-1]$, we maintain the prefix $\Pre(P_j,t_j)$ of each bundle $P_j$ and combine the remaining suffixes into a new bundle
$$
    R(\bt)=\bigcup_{j=1}^{n-1}\bigl(P_j\setminus\Pre(P_j,t_j)\bigr).
$$

Our main goal in this subsection is to establish the following theorem.

\begin{theorem}\label{thm:existence-prefix-EF1}
    There exists a prefix-length vector $\bt$ such that $\Pre(P_1,t_1),\ldots,\Pre(P_{n-1},t_{n-1}),R(\bt)$ admit an EF1 allocation among $n$ agents.
\end{theorem}
Theorem~\ref{thm:existence-prefix-EF1} guarantees that a desired prefix-length vector exists, though it does not immediately provide a constructive algorithm.
In fact, toward proving Theorem~\ref{thm:existence-prefix-EF1} in this subsection, we establish several structural properties of $\bt$ that are crucial for our later algorithm design (Section~\ref{sec:search}).

Given a prefix-length vector $\bt$ and any index $r\in[n-1]$ with $t_r\ge 2$, we define the neighboring vector obtained by decreasing coordinate $r$ by one as
$$
    \bt^{-r}=(t_1,\dots,t_{r-1},t_r-1,t_{r+1},\dots,t_{n-1}),
$$
which differs from $\bt$ only in that the $r$-th prefix bundle drops its last chore.
We denote the $j$-th coordinate of $\bt^{-r}$ by $t_j^{-r}$, so that $t_r^{-r}=t_r-1$ and $t_j^{-r}=t_j$ for all $j\ne r$.
A Boolean predicate $\Phi$ taking values in $\{0,1\}$ on prefix-length vectors is called monotone if $\Phi(\bt) \le \Phi(\bt')$ whenever $\bt \le \bt'$ coordinatewise.
In the following, we define the transition predicate $\Phi$ by setting $\Phi(\bt)=1$ if and only if some agent weakly prefers the remaining bundle to every prefix bundle:
$$
    \exists i\in N \quad\text{such that}\quad
    c_i(R(\bt))\le\min_{\ell\in[n-1]}\{c_i(\Pre(P_\ell,t_\ell))\}.
$$

Note that as $\bt$ increases coordinatewise, each prefix bundle expands while $R(\bt)$ shrinks. 
By the monotonicity of the costs, this implies that $\Phi$ is monotone.
For each agent, identifying a minimum-cost prefix bundle and comparing it against $R(\bt)$ takes $O(n)$ queries, so evaluating $\Phi(\bt)$ requires $O(n^2)$ queries in total.
Our goal is to find two neighboring vectors $\bt^{-r}$ and $\bt$ for some $r\in[n-1]$ with $\Phi(\bt^{-r})=0$ and $\Phi(\bt)=1$, such that the resulting prefix bundles can be matched to $n-1$ of the agents (leaving the remaining agent with the combined suffix bundle).

We first extend the discrete domain $\prod_{j=1}^{n-1}[|P_j|]$ of $\bt$ to the continuous domain $\prod_{j=1}^{n-1}[1,|P_j|]$ to allow fractional prefix lengths.
We assign ranks to the nonempty prefixes and extend these ranks continuously between consecutive integer prefix lengths.
For each agent, distinct labeled prefixes have distinct ranks at integer lengths, whereas interpolated ranks may coincide at fractional lengths.
The extension is constant on the first half of each unit interval.

For each agent $i\in N$, $j\in[n-1]$, and $t\in[|P_j|]$, we order all prefixes $\Pre(P_j,t)$ lexicographically by their cost $c_i(\Pre(P_j,t))$, their length $t$, and the index $j$, and denote the resulting positive integer rank of $\Pre(P_j,t)$ by $\Rank_i(P_j,t)$.
By construction, these ranks satisfy
$$
    0 < \Rank_i(P_j,1) < \cdots < \Rank_i(P_j,|P_j|),
$$
and $\Rank_i(P_j,t)\neq\Rank_i(P_\ell,s)$ whenever $j\neq\ell$.
We can compare $\Rank_i(P_j,t)$ and $\Rank_i(P_\ell,s)$ using a single call to $\Comp_i$, without explicitly computing either rank value.

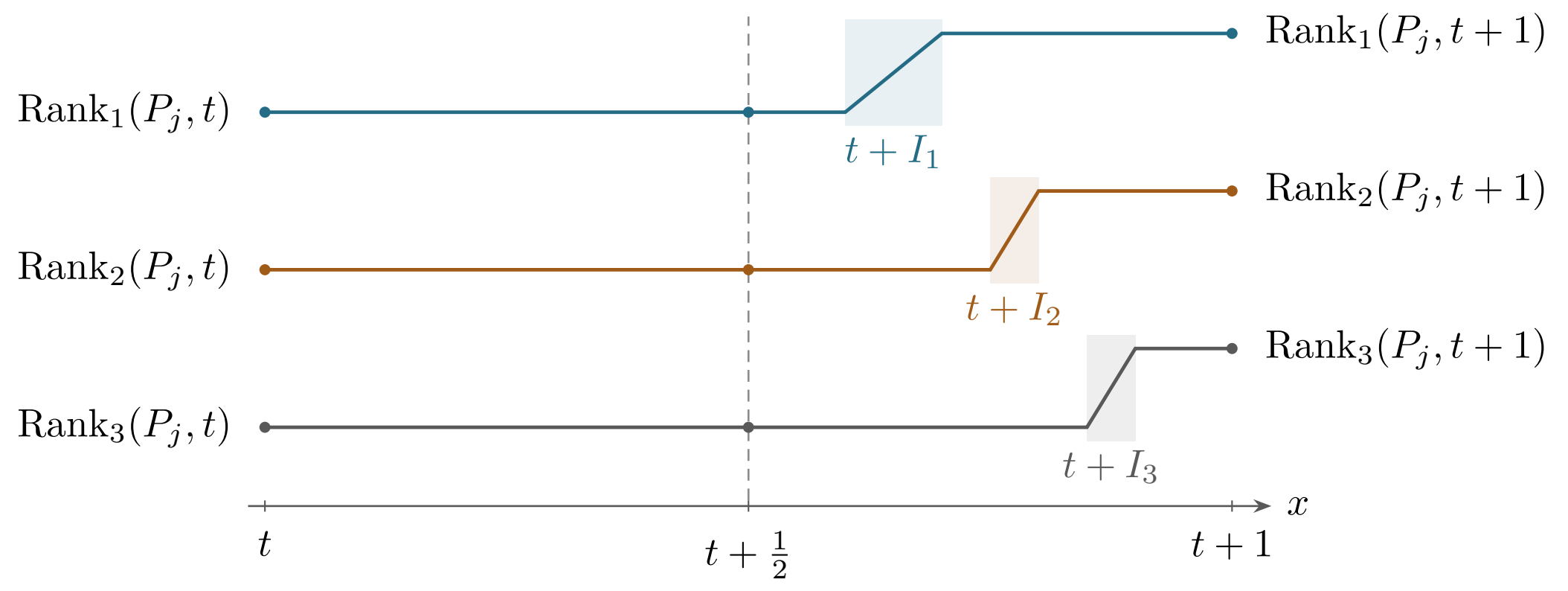
\begin{figure}[!htb]
\centering
\begin{tikzpicture}[x=1cm,y=1cm,font=\small]
\draw[draw=black!45,densely dashed,line width=0.5pt]
    (4.3,-0.65) -- (4.3,3.65);
\foreach \i/\y/\a/\b/\col in {
    1/2.8/5.16/6.02/diagramblue,
    2/1.4/6.45/6.88/diagramamber,
    3/0/7.31/7.74/black!65
} {
    \fill[\col,opacity=0.1] (\a,{\y-0.12}) rectangle (\b,{\y+0.82});
    \draw[draw=\col,line width=0.9pt]
        (0,\y) -- (\a,\y) -- (\b,{\y+0.7}) -- (8.6,{\y+0.7});
    \foreach \x/\height in {0/\y,4.3/\y,8.6/{\y+0.7}} {
        \fill[\col] (\x,\height) circle (1.4pt);
    }
    \node[anchor=east] at (-0.16,\y) {$\Rank_{\i}(P_j,t)$};
    \node[anchor=west] at (8.76,{\y+0.7}) {$\Rank_{\i}(P_j,t+1)$};
    \node[text=\col] at ({(\a+\b)/2},{\y-0.35}) {$t+I_{\i}$};
}
\draw[-{Stealth[length=1.6mm]},draw=black!65,line width=0.5pt]
    (-0.15,-0.7) -- (8.95,-0.7) node[right] {$x$};
\foreach \x/\label in {0/t,4.3/{t+\frac12},8.6/{t+1}} {
    \draw[draw=black!65] (\x,-0.65) -- (\x,-0.75);
    \node[anchor=north] at (\x,-0.79) {$\label$};
}
\end{tikzpicture}
\caption{Rank interpolation for one fixed bundle $P_j$ and three agents. Shaded intervals are disjoint and lie to the right of $t+1/2$, where each agent's rank still equals its value at $t$. Vertical scales are schematic and separate for each agent.}
\label{fig:rank-interpolation}
\end{figure}

We next extend these ranks continuously between consecutive integer prefix lengths.
We fix $n$ pairwise disjoint closed intervals $I_1<\cdots<I_n\subset(1/2,1)$ of positive length.
For each $i$ and $j$, we extend $\Rank_i(P_j,x)$ continuously to real $x\in[1,|P_j|]$.
On each unit interval $[t,t+1]$, the value $\Rank_i(P_j,x)$ equals $\Rank_i(P_j,t)$ to the left of the active interval $t+I_i$, increases affinely to $\Rank_i(P_j,t+1)$ across $t+I_i$, and stays at $\Rank_i(P_j,t+1)$ to its right (see Figure~\ref{fig:rank-interpolation}).

\begin{samepage}
By construction, this interpolation has two key properties.
\begin{itemize}
    \item Within a fixed bundle, at most one agent's rank function varies at a given prefix coordinate, because $I_1,\dots,I_n$ are disjoint.
    \item For every integer
    $t\in\{2,\dots,|P_j|\}$, we have
    $\Rank_i(P_j,t-1/2)=\Rank_i(P_j,t-1)$ for $i\in N$.
\end{itemize}
\end{samepage}

To establish the existence of a valid transition, we use the following matching theorem of Meunier and Su on the standard simplex $\Delta=\{\by\in\bR_{\ge 0}^{n-1}:\sum_{j=1}^{n-1}y_j=1\}$.

\begin{lemma}[{\cite{MeunierSu2019}}]\label{lem:continuous-matching}
Suppose that at each point $\by\in\Delta$, every agent $i\in N$ selects a nonempty subset of $[n-1]$.
These selections have closed graphs, and index $j$ is selected only when $y_j>0$.
Then there exists a point $\by^*\in\Delta$ at which removing any single agent leaves a perfect matching between the remaining $n-1$ agents and $[n-1]$, with each agent matched to an index she selects at $\by^*$.
\end{lemma}

In the following, we fix a partition $(P_1,\dots,P_{n-1})$ satisfying the cost bounds in Theorem~\ref{thm:(n-1)-partition}.
\begin{definition}[Minimum-Rank Graph]\label{def:minimum-rank-graph}
For each real prefix-length vector $\bx$, we define a bipartite graph $G(\bx)$ between the agents and the bundles.
An edge $(i,j)$ belongs to $G(\bx)$ if and only if
$$
    j\in\argmin_{\ell\in[n-1]}
    \{\Rank_i(P_\ell,x_\ell)\}.
$$
\end{definition}
Our goal is to find a real prefix-length vector $\bx$ such that $G(\bx)$ admits a perfect matching after any single agent is removed, so that the recipient of the remaining bundle need not be fixed in advance.
A \emph{nearest-integer rounding} of $\bx$ is obtained by rounding each coordinate to a nearest integer, breaking ties arbitrarily at half-integers.
Furthermore, for a real vector $\bx$, we write $\floor{\bx}$ and $\ceil{\bx}$
for coordinatewise rounding down and up, respectively.

\begin{figure}[!htb]
\centering
\begin{tikzpicture}[
    x=1cm,y=1cm,font=\small,
    agent/.style={circle,draw=black!65,fill=white,minimum size=0.44cm,inner sep=1pt},
    prefixbundle/.style={draw=diagramblue,fill=diagramblue!6,minimum width=0.62cm,minimum height=0.46cm,inner sep=2pt},
    matching/.style={draw=diagramblue,line width=1pt},
    rounding/.style={-{Stealth[length=1.4mm]},draw=black!50,densely dashed,line width=0.5pt}
]
\node[font=\small\bfseries] at (3,5.35) {(a) A point on the transition boundary};
\fill[diagramblue!9]
    (1.7,4.5) -- (1.7,3.35) -- (2.8,3.35) -- (2.8,2.25)
    -- (3.9,2.25) -- (3.9,1.15) -- (5.5,1.15) -- (5.5,4.5) -- cycle;
\draw[draw=black!25,line width=0.45pt]
    (0.45,4.5) -- (5.5,4.5) -- (5.5,0.35);
\draw[draw=diagramblue,line width=0.9pt]
    (1.7,4.5) -- (1.7,3.35) -- (2.8,3.35) -- (2.8,2.25)
    -- (3.9,2.25) -- (3.9,1.15) -- (5.5,1.15);
\draw[-{Stealth[length=1.6mm]},draw=black!65,line width=0.55pt]
    (0.45,0.35) -- (5.95,0.35) node[right] {$x_1$};
\draw[-{Stealth[length=1.6mm]},draw=black!65,line width=0.55pt]
    (0.45,0.35) -- (0.45,4.9) node[left] {$x_2$};
\node[anchor=north east] at (0.48,0.31) {$(1,1)$};
\foreach \a/\label in {2.25/{t_1-1},3.35/{t_1},5.5/{|P_1|}} {
    \draw[draw=black!60] (\a,0.3) -- (\a,0.4);
    \node[anchor=north] at (\a,0.19) {$\label$};
}
\foreach \a/\label in {2.8/{t_2},4.5/{|P_2|}} {
    \draw[draw=black!60] (0.4,\a) -- (0.5,\a);
    \node[anchor=east] at (0.3,\a) {$\label$};
}
\draw[draw=black!20,densely dotted,line width=0.45pt]
    (0.45,2.8) -- (3.35,2.8);
\draw[draw=black!20,densely dotted,line width=0.45pt]
    (2.25,0.35) -- (2.25,2.8);
\draw[draw=black!20,densely dotted,line width=0.45pt]
    (3.35,0.35) -- (3.35,2.8);
\foreach \a/\b in {1.7/3.9,4.6/1.15} {
    \draw[-{Stealth[length=1.4mm]},draw=black!30,line width=0.55pt]
        (5.5,4.5) -- (\a,\b);
}
\draw[-{Stealth[length=1.7mm]},draw=diagramblue,line width=0.9pt]
    (5.5,4.5) -- (2.8,2.6);
\fill[black!70] (5.5,4.5) circle (1.8pt);
\node[text=diagramblue] at (4.5,3.3) {$\Omega$};
\draw[rounding] (2.8,2.6) -- (2.25,2.8);
\draw[rounding] (2.8,2.6) -- (3.35,2.8);
\fill[black!85] (2.8,2.6) circle (1.9pt);
\node[anchor=north east] at (2.68,2.51) {$\bx$};
\draw[draw=diagramamber,fill=white,line width=0.9pt]
    (2.25,2.8) circle (2pt);
\fill[diagramblue] (3.35,2.8) circle (2pt);
\node[text=diagramamber,anchor=east] at (2.09,3.08) {$\Phi=0$};
\node[text=diagramblue,anchor=west] at (3.52,2.65) {$\Phi=1$};

\node[font=\small\bfseries] at (11.15,5.35) {(b) Matchings at the same point};
\foreach \case/\shift in {1/7.4,2/10.1,3/12.8} {
    \begin{scope}[xshift=\shift cm]
    \node[font=\footnotesize] at (0.95,4.45) {Agent $\case$ removed};
    \foreach \i/\a in {1/0,2/0.95,3/1.9} {
        \ifnum\i=\case
            \node[agent,draw=black!30,fill=black!5,text=black!35]
                (c\case a\i) at (\a,3.6) {$\i$};
        \else
            \node[agent] (c\case a\i) at (\a,3.6) {$\i$};
        \fi
    }
    \node[prefixbundle] (c\case p1) at (0.45,1.5) {$P_1$};
    \node[prefixbundle] (c\case p2) at (1.45,1.5) {$P_2$};
    \foreach \i/\j in {1/1,2/1,2/2,3/2} {
        \draw[draw=black!25,line width=0.45pt] (c\case a\i) -- (c\case p\j);
    }
    \ifnum\case=1
        \draw[matching] (c\case a2) -- (c\case p1);
        \draw[matching] (c\case a3) -- (c\case p2);
    \fi
    \ifnum\case=2
        \draw[matching] (c\case a1) -- (c\case p1);
        \draw[matching] (c\case a3) -- (c\case p2);
    \fi
    \ifnum\case=3
        \draw[matching] (c\case a1) -- (c\case p1);
        \draw[matching] (c\case a2) -- (c\case p2);
    \fi
    \end{scope}
}
\end{tikzpicture}
\caption{The illustration of Lemma~\ref{lem:continuous-transition} for $n=3$. At the same point $\bx$, nearest-integer roundings give both values of $\Phi$ in (a), and $G$ admits a perfect matching after any agent removal in (b). Shading marks $\Omega$, and blue edges mark the matchings.}
\label{fig:continuous-transition}
\end{figure}
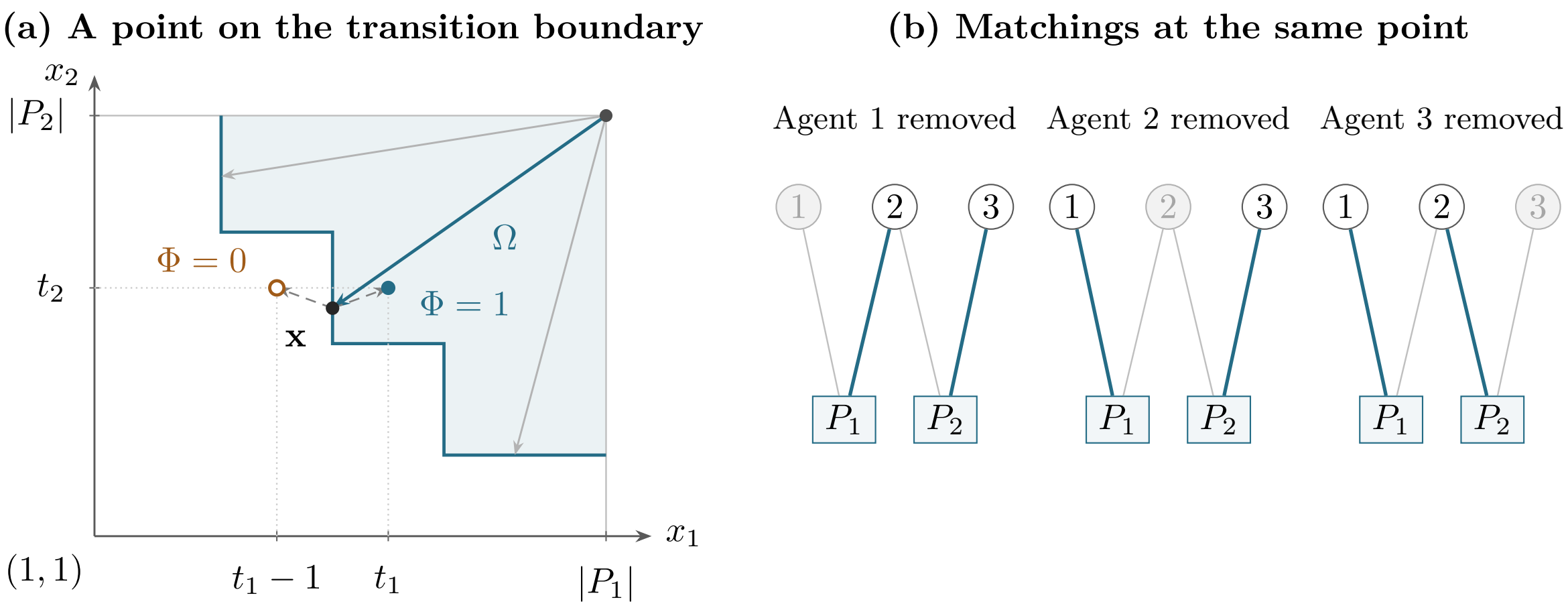

Given the definition above, we first establish the boundary conditions for the predicate $\Phi$\footnote{Recall that $\Phi$ is the transition predicate defined by $\Phi(\bt)=1$ if and only if some agent weakly prefers the remaining bundle to every prefix bundle.}.
\begin{samepage}
\begin{lemma}\label{lem:chore-boundaries}
The predicate $\Phi$ satisfies the following boundary properties:
\begin{enumerate}[label=(\arabic*)]
    \item For any integer vector $\bt$, if $t_j=1$ for some $j\in[n-1]$, then $\Phi(\bt)=0$.
    \item For any real vector $\bx$, if $x_j=|P_j|$ and $(i,j)$ is an edge of $G(\bx)$, then $\Phi(\floor{\bx})=1$.
\end{enumerate}
\end{lemma}
\end{samepage}
\begin{proof}
By Theorem~\ref{thm:(n-1)-partition}, for every agent $i \in N$, each individual chore in $M'$ has cost at most $c_i(M')/k$, whereas each bundle $P_j$ has cost strictly exceeding $(1/n+1/k) \cdot c_i(M')$.

For (1), we consider any prefix-length vector $\bt$ with $t_j=1$ for some $j\in[n-1]$.
Since $R(\bt)$ contains $P_j\setminus \Pre(P_j,1)$ and $\Pre(P_j,1)$ consists of a single chore, for any $i \in N$, we have
$$
    c_i(R(\bt)) \ge c_i(P_j) - c_i(\Pre(P_j,1)) > \frac{c_i(M')}{n} > \frac{c_i(M')}{k} \ge c_i(\Pre(P_j,1)),
$$
which immediately yields that $\Phi(\bt)=0$.

For (2), we consider a vector $\bx$ with $1\le x_\ell\le |P_\ell|$ for all $\ell\in[n-1]$.
We assume that $x_j=|P_j|$ and $(i,j)$ is an edge of $G(\bx)$, for some $i\in N$ and $j\in[n-1]$.
By the definition of $G(\bx)$, we have
$$
    \Rank_i(P_j,|P_j|)\le\Rank_i(P_\ell,x_\ell)
    \qquad\text{for every }\ell\in[n-1].
$$
Then for every $\ell\in[n-1]$, we have $\Rank_i(P_j,|P_j|)\le \Rank_i(P_\ell,x_\ell)\le \Rank_i(P_\ell,\ceil{x_\ell})$, which implies $c_i(P_j)\le c_i(\Pre(P_\ell,\ceil{x_\ell}))$.
Since $\Pre(P_\ell,\floor{x_\ell})$ contains at most one fewer chore than $\Pre(P_\ell,\ceil{x_\ell})$ and any single chore costs at most $c_i(M')/k$, it follows that
$$
    c_i(\Pre(P_\ell,\floor{x_\ell}))\ge c_i(P_j)-\frac{c_i(M')}{k}>\frac{c_i(M')}{n}.
$$
Finally, because $P_1,\dots,P_{n-1}$ partition $M'$, we obtain $\Phi(\floor{\bx})=1$ as
\begin{equation*}
    c_i(R(\floor{\bx}))=c_i(M')-\sum_{\ell=1}^{n-1}c_i(\Pre(P_\ell,\floor{x_\ell}))
    <\frac{c_i(M')}{n}<\min_{\ell\in[n-1]} c_i(\Pre(P_\ell,\floor{x_\ell})). \qedhere
\end{equation*}
\end{proof}
With these boundary properties, we can now apply
Lemma~\ref{lem:continuous-matching} to establish the existence of a point on the transition boundary with the required matching property.
\begin{lemma}\label{lem:continuous-transition}
There exists a vector $\bx$ whose nearest-integer roundings yield both values of $\Phi$.
Moreover, $G(\bx)$ admits a perfect matching upon removing any single agent.
\end{lemma}
\begin{proof}
We first construct a continuous boundary separating the regions where $\Phi=0$ and $\Phi=1$.
Specifically, each prefix-length vector $\bt$ with $\Phi(\bt)=1$ defines the closed box
$$
    t_j-\frac12\le x_j\le |P_j|
    \qquad\text{for every }j\in[n-1],
$$
and we denote by $\Omega$ the union of all such boxes.
By Lemma~\ref{lem:chore-boundaries}(1), every point $\bx\in\Omega$ satisfies $3/2\le x_j\le |P_j|$ for all $j\in[n-1]$.
Because $\Phi(|P_1|,\dots,|P_{n-1}|)=1$, the set $\Omega$ is a nonempty finite union of boxes sharing the upper corner $(|P_1|,\dots,|P_{n-1}|)$.
For each direction $\by\in\Delta$, we define $\bx(\by)$ as the last point in $\Omega$ along the ray $|P_j|-\lambda y_j$ as $\lambda\ge 0$ increases.
Since the ray exits the box corresponding to $\bt$ at $\lambda=1/\max_j\{y_j/(|P_j|-t_j+1/2)\}$ and $\Omega$ is a finite union of such boxes, the map $\by\mapsto\bx(\by)$ is continuous on $\Delta$.
Since $\bx(\by)\in\Omega$, monotonicity of $\Phi$ ensures that at least one nearest-integer rounding of $\bx(\by)$ satisfies $\Phi=1$.
On the other hand, if every nearest-integer rounding satisfied $\Phi=1$, the box of the smallest nearest-integer rounding would contain $\bx(\by)$ in its interior along the ray, contradicting that $\bx(\by)$ is the last point in $\Omega$.
Thus, $\bx(\by)$ has nearest-integer roundings with both $\Phi=0$ and $\Phi=1$ (see Figure~\ref{fig:continuous-transition}(a)).

At each $\by\in\Delta$, agent $i$ selects the set of bundles $\argmin_{j\in[n-1]}\{\Rank_i(P_j,x_j(\by))\}$.
By continuity of the interpolated ranks and $\bx(\by)$, these selections are nonempty and have closed graphs.
Moreover, if $y_j=0$, then $x_j(\by)=|P_j|$, so bundle $j$ cannot be selected by any agent $i$ (otherwise Lemma~\ref{lem:chore-boundaries}(2) would give $\Phi(\floor{\bx(\by)})=1$, forcing every nearest-integer rounding of $\bx(\by)$ to satisfy $\Phi=1$).
Lemma~\ref{lem:continuous-matching} therefore yields a point $\bx=\bx(\by^*)$ whose minimum-rank graph $G(\bx)$ admits a perfect matching after any single agent is removed (see Figure~\ref{fig:continuous-transition}(b)).
\end{proof}

While Lemma~\ref{lem:continuous-transition} yields a real prefix-length vector $\bx$, we still need to show that rounding $\bx$ appropriately produces a partition that admits an EF1 allocation. To formalize this rounding step, we first introduce the notion of a robust vector.\begin{samepage}
\begin{definition}[Robust Vector]\label{def:robustness}
A real prefix-length vector $\bx$ is called \emph{robust} if both of the following conditions hold.
\begin{itemize}
\item The graph $G(\bx)$ admits a perfect matching after any single agent is removed.
\item At least one coordinate of $\bx$ is a half-integer.
\end{itemize}
\end{definition}
\end{samepage}
Note that Lemma~\ref{lem:continuous-transition} directly guarantees the existence of such a robust vector $\bx$.\begin{definition}[Consistent Pair]
A pair $(\bt,r)$, where $\bt$ is a prefix-length vector and $r\in[n-1]$, is \emph{consistent} if there exists a robust vector $\bx$ satisfying the following two conditions:
\begin{itemize}
\item The coordinate $x_r$ is a half-integer.
\item Both $\bt$ and $\bt^{-r}$ are nearest-integer roundings of $\bx$.
\end{itemize}
The pair $(\bt,r)$ is \emph{at the boundary} if $\Phi(\bt^{-r})=0$ and $\Phi(\bt)=1$, and we refer to $r$ as the \emph{root} of $\bt$.
\end{definition}
Note that Lemma~\ref{lem:continuous-transition} guarantees that the nearest-integer roundings of $\bx$ attain both values of $\Phi$.
With every non-half-integer coordinate rounded to its unique nearest integer, monotonicity ensures that rounding all half-integer coordinates downward yields $\Phi=0$, whereas rounding them all upward yields $\Phi=1$.
As we flip these half-integer coordinates from downward to upward one by one, some consecutive step must produce a consistent pair $(\bt,r)$ at the boundary.

Finally, we prove that any consistent pair $(\bt,r)$ at the boundary yields an EF1 allocation.

\begin{lemma}\label{lem:allocation-from-transition}
For any consistent pair $(\bt,r)$ at the boundary, at least one of the two partitions
$$\bigl(\Pre(P_1,t_1^{-r}),\dots,\Pre(P_{n-1},t_{n-1}^{-r}),R(\bt^{-r})\bigr) \quad\text{and}\quad \bigl(\Pre(P_1,t_1),\dots,\Pre(P_{n-1},t_{n-1}),R(\bt)\bigr)$$
admits a bijective EF1 assignment to the agents for the instance $M'$.
\end{lemma}
\begin{proof}
Let $\bx$ be a robust vector whose nearest-integer roundings include both $\bt$ and $\bt^{-r}$.
By definition, we have $\Phi(\bt^{-r})=0$ and $\Phi(\bt)=1$, where the latter implies $t_j\ge2$ for all $j\in[n-1]$ by Lemma~\ref{lem:chore-boundaries}(1).
Let $e$ denote the last chore in $\Pre(P_r,t_r)$.

We first establish a cost bound for every edge $(i,j)$ in $G(\bx)$.
Since $\bt$ is a nearest-integer rounding of $\bx$, we have $t_j-1\le x_j$.
For any bundle $j'\in[n-1]$, rounding $x_{j'}$ upward cannot decrease its rank, whereas rounding $x_{j'}$ downward leaves its rank unchanged under our interpolation because $0\le x_{j'}-t_{j'}\le 1/2$.
Combined with the edge condition in $G(\bx)$, these properties yield
$$
    \Rank_i(P_j,t_j-1)\le\Rank_i(P_j,x_j)\le\Rank_i(P_{j'},x_{j'})\le\Rank_i(P_{j'},t_{j'}).
$$
Moreover, since $x_r=t_r-1/2$ and $I_i\subseteq(1/2,1)$, our interpolation gives $\Rank_i(P_r,x_r)=\Rank_i(P_r,t_r-1)$.
For $j'\ne r$, we have $t_{j'}^{-r}=t_{j'}$, so the preceding rank bound is unchanged.
For $j'=r$, the preceding rank chain and the equality above give $\Rank_i(P_j,t_j-1)\le\Rank_i(P_r,x_r)=\Rank_i(P_r,t_r^{-r})$.
Hence, $\Rank_i(P_j,t_j-1)\le\Rank_i(P_{j'},t_{j'}^{-r})$ for every $j'\in[n-1]$.
Since ranks preserve the cost ordering of integer prefixes, we obtain
$$
    c_i(\Pre(P_j,t_j-1))\le\min_{\ell\in[n-1]}\{c_i(\Pre(P_\ell,t_\ell^{-r}))\}.
$$
Under either partition, if agent $i$ receives the $j$-th prefix bundle, her assigned prefix without its last chore is contained in $\Pre(P_j,t_j-1)$.
Thus, any assignment of the prefix bundles along edges of $G(\bx)$ guarantees EF1 among their recipients.
We now complete the allocation by distinguishing two cases.
\begin{itemize}
    \item {\textbf{Case 1: For all agents $i \in N$, it holds that $\min_{\ell\in[n-1]}\{c_i(\Pre(P_\ell,t_\ell^{-r}))\}<c_i(R(\bt))$.}}
    In this case, we choose the partition induced by $\bt$.
    Since $\Phi(\bt)=1$, there exists an agent $i^*\in N$ with $c_{i^*}(R(\bt))\le\min_{\ell\in[n-1]}\{c_{i^*}(\Pre(P_\ell,t_\ell))\}$.
    We assign $R(\bt)$ to $i^*$, who envies no prefix bundle, and assign the prefix bundles to the remaining agents according to a perfect matching in $G(\bx)$ after removing $i^*$, which exists because $\bx$ is robust.
    Combined with the condition of this case, the prefix cost bound established above ensures that every prefix recipient also satisfies EF1 toward $R(\bt)$ after removing her last chore.
    Thus, the resulting allocation is EF1.

    \item {\textbf{Case 2: There exists some agent $i^* \in N$, such that $c_{i^*}(R(\bt))\le\min_{\ell\in[n-1]}\{c_{i^*}(\Pre(P_\ell,t_\ell^{-r}))\}$.}}
    In this case, we choose the partition induced by $\bt^{-r}$ and assign $R(\bt^{-r})$ to $i^*$.
    Because $R(\bt^{-r})=R(\bt)\cup\{e\}$, removing $e$ from $R(\bt^{-r})$ eliminates her envy toward every prefix bundle.
    As in Case 1, the robustness of $\bx$ yields a perfect matching in $G(\bx)$ after removing $i^*$, according to which we assign the prefix bundles to the remaining agents.
    Since $\Phi(\bt^{-r})=0$, every agent $i\in N$ satisfies $\min_{\ell\in[n-1]}\{c_i(\Pre(P_\ell,t_\ell^{-r}))\}<c_i(R(\bt^{-r}))$.
    Together with the prefix cost bound above, this ensures that every prefix recipient also satisfies EF1 toward $R(\bt^{-r})$ after removing her last chore.
    Thus, the resulting allocation is EF1.\qedhere
\end{itemize}
\end{proof}

We therefore only need to find a consistent pair $(\bt,r)$ at the boundary using $\operatorname{poly}(n) \cdot \log m$ comparison queries, which we address in Section~\ref{sec:search}.

\section{The Computation of EF1 Allocations}\label{sec:search}

By the results of Section~\ref{sec:shared}, there exists a consistent pair $(\bt,r)$ at the boundary that induces an EF1 allocation of $M'=M\setminus E$, where $E$ is the set of heavy chores isolated in Section~\ref{subsec:(n-1)-partition}.
In this section, we show how to find a valid boundary pair $(\bt,r)$ in $O(\mathrm{poly}(n)\log m)$ comparison queries and reinsert the chores in $E$ to produce an EF1 allocation of $M$.

Crucially, our search relies on the structural property that the minimum-rank graph is a tree at every robust vector (Lemma~\ref{lem:robust-x-properties}).
Applying this tree structure, we establish \emph{root monotonicity}: for a fixed root, its prefix length uniquely determines a consistent pair (if one exists), and increasing this length never decreases any other prefix length, even when the underlying tree changes (Lemma~\ref{lem:root-monotonicity}).

\paragraph{Algorithm Overview.}
For each root $r$, our goal is to find a consistent pair $(\bt,r)$ at the boundary.
While root monotonicity naturally suggests a binary search over $t_r$, determining the remaining prefix lengths via nested binary searches would incur $O_n(\log^2 m)$ queries.
To bypass these inner searches, we instead maintain a set of candidate prefix-length vectors and prune them in rounds.
In each round, we partition consecutive chores into $O(n)$ blocks per bundle, choosing the block boundaries based on the remaining candidates (Lemma~\ref{lem:candidate-weighted-grouping}).
Treating each block as a single meta-chore yields a coarse system.
After suitably extending the prefix ranges, we can elicit all agents' preference orders over the coarse prefixes using $\mathrm{poly}(n)$ queries, which then allows us to compute a consistent coarse pair at the median candidate root length with zero additional queries.
Although grouping can change consistency, root monotonicity allows us to compare the coarse pair with every non-exceptional candidate that is consistent in the original system (Lemma~\ref{lem:exceptional}).
Consequently, our block construction guarantees that each round either identifies a valid boundary pair or eliminates a constant fraction of the candidates without discarding any valid boundary pair.
This yields an $O(\mathrm{poly}(n)\log m)$ query bound (Theorem~\ref{thm:coarse-prefix-search}).

\subsection{Root Monotonicity and a Preliminary $O_n(\log^2m)$ Algorithm}\label{subsec:root-monotonicity}

In this subsection, we establish root monotonicity for consistent pairs and use it to obtain an algorithm for finding a pair at the boundary.
We first characterize the minimum-rank graphs (Definition~\ref{def:minimum-rank-graph}) of the robust vectors that witness consistency.
Recall that a robust vector $\bx$ has a half-integer coordinate $x_r$ and that $G(\bx)$ admits a perfect matching after any one agent is removed.
Given $\bx$, an edge $(i,j)$ in $G(\bx)$ is said to be \emph{active} if $x_j-\lfloor x_j\rfloor\in I_i$. Otherwise, it is \emph{inactive}.

\begin{lemma}\label{lem:robust-x-properties}
Let $\bx$ be a robust vector, and suppose that $x_r$ is a half-integer. We have the following:
\begin{enumerate}
\item The graph $G(\bx)$ is a tree.
\item Every agent is incident to exactly one inactive edge.
\item Every bundle in $[n-1]\setminus\{r\}$ is incident to exactly one active edge.
\end{enumerate}
\end{lemma}
\begin{proof}
We first establish that $G(\bx)$ is connected.
Since robustness precludes isolated bundle vertices, every connected component of $G(\bx)$ contains at least one agent.
Consider a connected component with $a$ agents and $b$ bundles, and remove one agent from it.
By robustness (Definition~\ref{def:robustness}), the remaining $n-1$ agents can be matched bijectively to all $n-1$ bundles in $G(\bx)$.
Because the $b$ bundles in this component have no neighbors outside it, they must be matched to $b$ distinct agents among the remaining $a-1$ agents within the same component.
Thus, $a-b\geq 1$ for every connected component.
Summing $a-b$ over all components yields $n-(n-1)=1$, which implies that $G(\bx)$ consists of a single connected component.

Next, we bound the number of inactive and active edges.
Fix an agent $i$. If $(i,j)$ is an inactive edge, then $\Rank_i(P_j,x_j)$ equals the rank of a prefix of $P_j$.
If $i$ were incident to two inactive edges $(i,j)$ and $(i,k)$, both would attain $i$'s minimum rank and thus have equal rank.
This is impossible because prefixes from distinct bundles have distinct ranks under our prefix order.
Consequently, every agent is incident to at most one inactive edge, meaning there are at most $n$ inactive edges in total.
Similarly, because the active intervals are pairwise disjoint, every bundle is incident to at most one active edge.
Moreover, since the root coordinate $x_r$ is a half-integer and every active interval lies within $(1/2,1)$, all edges incident to bundle $r$ are inactive.
Hence, there are at most $n-2$ active edges.
Combining these bounds, $G(\bx)$ has at most $n+(n-2) = 2n-2$ edges.
On the other hand, since $G(\bx)$ is connected and has $n+(n-1)=2n-1$ vertices, it must have at least $2n-2$ edges.
Therefore, $G(\bx)$ has exactly $2n-2$ edges and is a tree, establishing~(1).

Furthermore, all upper bounds above must hold with equality: $G(\bx)$ contains exactly $n$ inactive edges and $n-2$ active edges.
By the degree bounds established earlier, every agent is incident to exactly one inactive edge, and every bundle in $[n-1]\setminus\{r\}$ is incident to exactly one active edge, proving~(2) and~(3).
\end{proof}

By Lemma~\ref{lem:robust-x-properties}(3), every nonroot coordinate has fractional part in $(1/2,1)$, so $x_r$ is the unique half-integer coordinate of $\bx$.
We next show that, for two robust vectors with the same root, increasing the root coordinate cannot decrease any other coordinate, even when their minimum-rank graphs differ.
In particular, a fixed root coordinate determines at most one robust vector.

\begin{lemma}\label{lem:root-monotonicity-fractional}
    Fix $r\in[n-1]$, and let $\mathbf{x}$ and $\mathbf{x}'$ be robust vectors whose $r$-th coordinates are half-integers.
    If $x_r\leq x_r'$, then $x_j\leq x_j'$ for every $j\in[n-1]$.
\end{lemma}

For a high-level idea of the proof, suppose for contradiction that \(A=\{j\in[n-1]:x_j>x_j'\}\) is nonempty.
We have \(r\notin A\).
We aim to derive a contradiction to the robustness of $\bx'$.
Consider the agents with a neighbor in \(A\) in \(G(\bx)\).
After switching to \(\bx'\), the ranks of bundles in \(A\) weakly decrease for every agent, while those of bundles outside \(A\) weakly increase.
If every agent originally adjacent to \(A\) has all her neighbors in \(A\) in \(G(\bx')\), then \(\bx'\) cannot be robust.
Indeed, \(A\) has at least \(|A|+1\) neighbors in \(G(\bx)\), since \(A\) must still be matched after one of its neighbors is removed.
Removing a neighbor of \(r\) in \(G(\bx')\) would then leave all these agents to be matched to only \(|A|\) bundles.
Therefore, some agent \(i\) has a neighbor \(j\in A\) in \(G(\bx)\) and a neighbor \(k\notin A\) in \(G(\bx')\).
This can only happen if the two corresponding bundles attain the same minimum rank for \(i\) at both vectors.
Since \(x_j>x_j'\), agent \(i\)'s edge to \(j\) must be inactive in \(G(\bx)\).
Since each agent has exactly one inactive edge (Lemma~\ref{lem:robust-x-properties}(2)), the edge \((i,k)\) is active in \(G(\bx)\), and we must have \(x_k=x_k'\) for the rank of \(k\) to remain unchanged.
We say that agent \(i\) extends \(A\) to \(A\cup\{k\}\).
By Lemma~\ref{lem:robust-x-properties}(3), \((i,k)\) is the unique active edge incident to \(k\), so \(i\) is the only agent through whom \(k\) can be added.
In each extension, an agent whose minimum rank is unchanged and whose unique inactive neighbor is in the current set adds her active neighbors outside the set whose coordinates are unchanged.
Since \(r\) is inactive for every agent, these extensions cannot reach \(r\).
In the full proof, we show that once the extensions stop, every agent adjacent to the resulting set in \(G(\bx)\) has all her neighbors in that set in \(G(\bx')\), giving the same contradiction.

We defer the full proof of Lemma~\ref{lem:root-monotonicity-fractional} to Appendix~\ref{app:root-monotonicity}.
The following lemma gives a crucial observation on the monotonicity of $\mathbf{t}$ with respect to the root prefix length $t_r$ for consistent pairs, which follows from Lemma~\ref{lem:root-monotonicity-fractional} by rounding upward.

\begin{lemma}\label{lem:root-monotonicity}
    For any consistent pairs $(\mathbf{t},r)$ and $(\mathbf{t}',r)$, if $t_r\geq t_r'$, then $t_j\geq t_j'$ for all $j\in[n-1]$.
    In particular, $r$ and $t_r$ uniquely determine $\mathbf{t}$.
\end{lemma}
\begin{proof}
By consistency, there exist robust vectors $\mathbf{x}$ and $\mathbf{x}'$ witnessing $(\mathbf{t},r)$ and $(\mathbf{t}',r)$, respectively.
Their root coordinates satisfy $x_r=t_r-\frac12\ge t_r'-\frac12=x_r'$.
Applying Lemma~\ref{lem:root-monotonicity-fractional} to $\mathbf{x}'$ and $\mathbf{x}$ yields $x_j'\le x_j$ for every $j\in[n-1]$.
By Lemma~\ref{lem:robust-x-properties}(3), the fractional part of each nonroot coordinate lies in some $I_i\subset(1/2,1)$, so each such coordinate rounds upward.
Since $t_r$ and $t_r'$ are the upper roundings at the root, we have $\mathbf{t}=\lceil\mathbf{x}\rceil$ and $\mathbf{t}'=\lceil\mathbf{x}'\rceil$.
Taking coordinatewise ceilings gives $t_j'=\lceil x_j'\rceil\le \lceil x_j\rceil=t_j$ for every $j\in[n-1]$.
Finally, if $t_r=t_r'$, applying the conclusion in both directions gives $\mathbf{t}=\mathbf{t}'$.
Hence, a fixed root and its prefix length determine at most one consistent pair.
\end{proof}

\begin{figure}[t]
\centering
\begin{tikzpicture}[x=1.60cm,y=1.08cm,font=\small,
    axis/.style={-{Stealth[length=1.7mm]},draw=black!65,line width=0.55pt},
    gridline/.style={draw=black!18,line width=0.4pt},
    possible/.style={draw=diagramblue,fill=white,line width=0.9pt},
    pics/excluded/.style={code={
        \draw[diagramamber,line width=0.9pt,line cap=round]
            (-2.7pt,-2.7pt)--(2.7pt,2.7pt)
            (-2.7pt,2.7pt)--(2.7pt,-2.7pt);
    }}]
    \fill[diagramblue!8] (3.79,1.76) rectangle (4.21,6.22);
    \foreach \k in {2,...,6} {
        \draw[gridline] (\k,2)--(\k,6);
        \draw[gridline] (2,\k)--(6,\k);
        \draw[black!60] (\k,1.71)--(\k,1.81);
        \draw[black!60] (1.72,\k)--(1.79,\k);
        \node[anchor=north] at (\k,1.69) {$\k$};
        \node[anchor=east] at (1.70,\k) {$\k$};
    }
    \draw[axis] (1.76,1.76)--(6.48,1.76) node[right] {$t_1$};
    \draw[axis] (1.76,1.76)--(1.76,6.42) node[left] {$t_2$};

    \foreach \u in {2,...,6} {
        \foreach \v in {2,...,6} {
            \pgfmathtruncatemacro{\allowed}{(\u<4 && \v<=4) ||
                (\u>4 && \v>=4) || (\u==4 && \v==4)}
            \ifnum\allowed=1
                \draw[possible] (\u,\v) circle[radius=3pt];
            \else
                \pic at (\u,\v) {excluded};
            \fi
        }
    }
    \fill[diagramblue] (4,4) circle[radius=4pt];
    \node[anchor=south west,inner sep=2pt,text=diagramblue]
        at (4.05,4.06) {$(4,4)$};

    \node[anchor=west,font=\small\bfseries,inner sep=0pt]
        at (6.82,5.12) {Comparison with $(4,4)$};
    \fill[diagramblue] (6.91,4.52) circle[radius=4pt];
    \draw[possible] (6.91,3.92) circle[radius=3pt];
    \pic at (6.91,3.32) {excluded};
    \node[anchor=west] at (7.12,4.52) {Known consistent vector};
    \node[anchor=west] at (7.12,3.92) {Still possible};
    \node[anchor=west] at (7.12,3.32) {Ruled out};
\end{tikzpicture}
\caption{Root monotonicity for $n=3$ and root $r=1$. Suppose $(4,4)$ is consistent for this root.
For any other consistent vector with the same root, $t_1<4$ implies $t_2\le4$, and $t_1>4$ implies $t_2\ge4$.
In the shaded column, $t_1=4$ forces $t_2=4$.
Open circles indicate vectors not ruled out by comparison with $(4,4)$; they need not be consistent.}
\label{fig:root-order}
\end{figure}
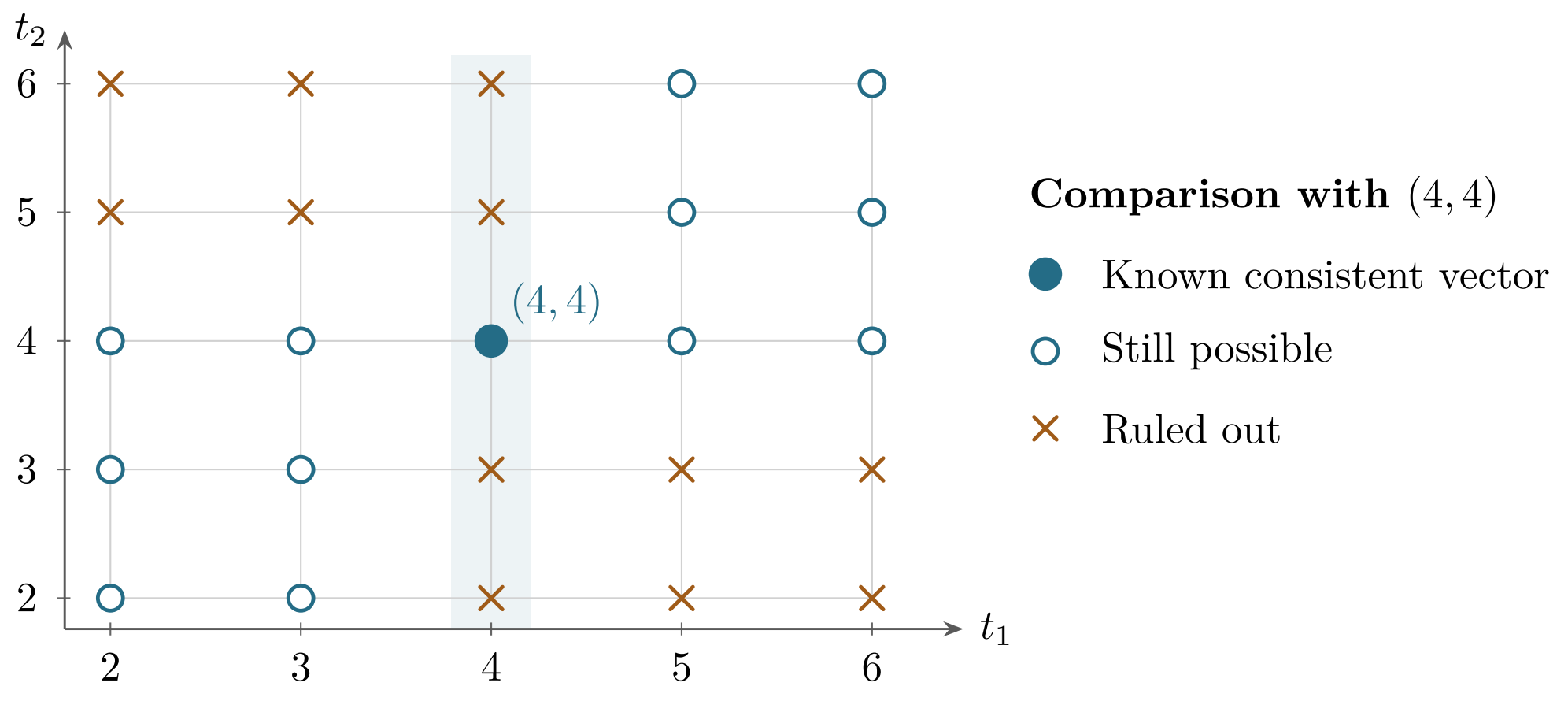

Figure~\ref{fig:root-order} illustrates Lemma~\ref{lem:root-monotonicity} for $n=3$ and root $r=1$.
Once $(4,4)$ is known to be consistent for this root, every other vector with $t_1=4$ is ruled out for the same root. This root monotonicity property then suggests a natural binary search over the root prefix length.
We first enumerate the possible roots $r\in[n-1]$ and, for each root, consider a binary search over $t_r\in\{2,\ldots,|P_r|\}$.
For a fixed root, each trial length $t_r$ determines at most one consistent vector $\bt$.
Whenever these vectors exist, their coordinates, and hence their values of $\Phi$, are nondecreasing in $t_r$.
Our goal is to find a root length whose corresponding pair $(\bt,r)$ is consistent and at the boundary, with $\Phi(\bt^{-r})=0$ and $\Phi(\bt)=1$.
The search range starts at $2$ because Lemma~\ref{lem:chore-boundaries}(1) rules out boundary pairs with any coordinate equal to $1$.
Section~\ref{subsec:balance} guarantees that such a pair exists for some root.
However, obtaining an $O(\mathrm{poly}(n)\log m)$-query algorithm requires addressing two issues:
\begin{enumerate}
    \item A trial root length $t_r$ may have no corresponding consistent vector $\bt$.
    In this case, we cannot evaluate $\Phi$ at such a vector to determine whether $t_r$ is too small or too large.
    \item For each trial root length $t_r$, a na\"ive implementation enumerates the trees $G$ with active and inactive edges satisfying Lemma~\ref{lem:robust-x-properties}.
For each tree, it determines the remaining prefix lengths by binary search, proceeding from the root toward the leaves.
Here, we call the two prefixes defining an interpolation interval its \emph{endpoint prefixes}.
For an active edge $(i,j)$, the endpoint prefix at agent $i$'s inactive edge is already determined.
The search finds the consecutive prefixes of $P_j$ between which that endpoint prefix lies strictly in $i$'s prefix order, rejecting the tree if no such pair exists.
After determining these lengths, it verifies the required comparisons between endpoint prefixes and checks that $G$ admits a perfect matching after any agent is removed.
The comparison checks use $O(n^2)$ additional queries per tree, while the matching checks require no queries.
    Testing these trees separately introduces an exponential factor in $n$ into the query bound.
    Moreover, the outer binary search over $t_r$, together with the inner searches for the other $n-2$ prefix lengths, yields an $O_n(\log^2 m)$ query bound.
\end{enumerate}

To handle the first issue, we extend the prefix ranges so that every trial root length admits a consistent vector.
For example, consider $n=3$ with bundles $P_1,P_2$ and root $r=1$.
Suppose that, for a trial length $T$, every agent finds the entire bundle $P_2$ less costly than $\Pre(P_1,T-1)$.
Then, for any vector $\bx$ within the original prefix ranges satisfying $x_1=T-\frac12$, bundle $P_1$ is isolated in $G(\bx)$, so $\bx$ cannot be robust by Lemma~\ref{lem:robust-x-properties}.
Appending a sufficiently costly \emph{artificial item} to $P_2$ allows its prefix ranks to exceed the fixed root rank.
To handle the lower end of the prefix ranges as well, we also include labeled empty prefixes, which rank below every nonempty prefix.

We extend the prefix ranges of the balanced bundles $P_1,\dots,P_{n-1}$ as follows.
For each $j\in[n-1]$, we append an artificial item $g_j$ with $c_i(g_j)=c_i(M)$ for every agent $i\in N$, and denote the resulting bundle by $P_j^+$.
We include the labeled empty prefixes $\Pre(P_j^+,0)$ in the ordering by cost, prefix length, and bundle index.
We then extend the rank functions continuously to $[0,|P_j^+|]$ using the same active intervals $I_1,\dots,I_n$.

\begin{lemma}\label{lem:fixed-root-extension}
For every root $r\in[n-1]$ and every integer $T\in\{2,\dots,|P_r|\}$, the extended system admits a robust vector $\bx$ satisfying
$$
    x_r=T-\frac12
    \qquad\text{and}\qquad
    0<x_j<|P_j^+|
    \quad\text{for all }j\in[n-1].
$$
\end{lemma}
We defer the full proof of Lemma~\ref{lem:fixed-root-extension} to Appendix~\ref{app:fixed-root-extension}.
By Lemma~\ref{lem:fixed-root-extension}, every trial root length $t_r=T$ admits a consistent pair $(\bt,r)$ in the extended system.
Every robust vector witnessing consistency in the original system remains robust in the extended system, since adding the empty prefixes increases all original interpolated ranks by the same constant $n-1$.
The proofs of Lemmas~\ref{lem:robust-x-properties} and~\ref{lem:root-monotonicity} also apply to the extended system.
Hence, if the pair in the extended system uses an artificial item, Lemma~\ref{lem:root-monotonicity} directs the search toward smaller root lengths.
Otherwise, root monotonicity and the monotonicity of $\Phi$ allow us to determine the search direction or identify a boundary pair by evaluating $\Phi(\bt)$ and $\Phi(\bt^{-r})$.
If $\Phi(\bt^{-r})=0$ and $\Phi(\bt)=1$, Lemma~\ref{lem:chore-boundaries}(1) gives $t_j\ge2$ for every $j\in[n-1]$.
The robust vector witnessing consistency therefore lies within the original prefix ranges, yielding a consistent pair at the boundary in the original system.
This resolves the first issue.
Comparisons involving the \emph{full extended bundles} $P_j^+=\Pre(P_j^+,|P_j|+1)$ can be simulated using the original bundles.
Each $P_j^+$ costs more than every prefix of the original bundles, and comparing $P_j^+$ with $P_\ell^+$ is equivalent to comparing $P_j$ with $P_\ell$, since both contain an artificial item of cost $c_i(M)$.
Together with the tree enumeration and binary searches outlined above, this yields an $O_n(\log^2 m)$-query search with an exponential dependence on $n$.

To handle the second issue, we maintain candidate prefix-length vectors for each root and progressively eliminate those that cannot form a consistent pair at the boundary.
In each round, we partition each extended bundle $P_j^+$ into $O(n)$ blocks of consecutive items according to the surviving candidates and treat each block as a single item.
For a trial root length $T$, we make the $T$-th chore of $P_r$ a singleton block.
The resulting coarse system has $O(n^2)$ prefixes across the $n-1$ bundles.
For each agent, we obtain a full ranking of these prefixes using $O(n^2\log n)$ comparison queries, giving $O(n^3\log n)$ queries in total.
These rankings contain all information needed to find a consistent pair in the coarse system whose root prefix ends at the $T$-th original chore, without further comparison queries.
However, consistency in the coarse system need not imply consistency in the original system.
In Sections~\ref{subsec:items-grouping} and~\ref{subsec:grouping-construction}, we show how to use these coarse pairs to eliminate candidates while preserving every consistent pair at the boundary in the original system.

\subsection{Candidate Elimination by Item Grouping}\label{subsec:items-grouping}

In this subsection, we use item grouping to avoid the inner binary searches mentioned in Section~\ref{subsec:root-monotonicity}.
For each root $r\in[n-1]$, we maintain candidate prefix-length vectors and progressively eliminate those that cannot form a consistent pair at the boundary, while keeping the balanced bundles $(P_1,\dots,P_{n-1})$ fixed.
In each round, we partition each bundle into $O(n)$ blocks of consecutive chores according to the surviving candidates.
We do not seek a consistent pair at the boundary in the coarse system and then try to obtain a consistent boundary pair in the original system by further subdivision.\footnote{We initially explored this approach but were unable to make it work.}
Instead, we find a consistent pair at a median root prefix length in the coarse system and use root monotonicity to filter the candidates in the original system, as in binary search.


An item grouping for a bundle $P_j$ is defined by a set of integer \emph{cut-points} in $\{0,1,\dots,|P_j|+1\}$.
Here, we use the extended bundle $P_j^+$ from Lemma~\ref{lem:fixed-root-extension}, obtained by appending an artificial item $g_j$ with $c_i(g_j)=c_i(M)$ for all $i\in N$.
Cut-point $0$ represents the empty prefix, and cut-point $|P_j|+1$ represents the full extended bundle $P_j^+$.
After sorting the cut-points, each pair of adjacent cut-points $a_j,b_j$ defines the \emph{block} $\Pre(P_j^+,b_j)\setminus\Pre(P_j^+,a_j)$.
After choosing the cut-points for every bundle, we view each block as a single item in the resulting \emph{system}.
The \emph{original system} includes neither empty prefixes nor artificial items, and its cut-points for each $P_j$ are $\{1,\dots,|P_j|\}$.
A prefix length $t_j$ is \emph{in the system} if it is a cut-point of bundle $j$.
Similarly, a prefix-length vector $\bt$ is in the system if every coordinate $t_j$ is in the system.
For each agent, the selected prefixes in every system inherit the same strict order by cost, prefix length, and bundle index.
The length used for tie-breaking is the prefix length in $P_j^+$ rather than the number of blocks.
Within each system, we assign positive integer ranks according to this order of all selected prefixes across all bundles.
Thus, prefixes shared by two systems have the same relative order, although their rank values may differ.
We extend these ranks continuously using the same interpolation rule and the same intervals $I_1,\dots,I_n$ in every system.
Continuous coordinates and nearest-integer rounding are measured in block indices, so each pair of consecutive cut-points corresponds to a unit interpolation interval.
The vectors $\bt$ and $\bw$ record prefix lengths in the extended bundles $P_j^+$ rather than block indices.
We use Definition~\ref{def:robustness} in each system, with its own minimum-rank graph and with coordinates measured in block indices.
A pair $(\bt,r)$ is consistent in a system if there exists a robust vector in block coordinates whose $r$-th coordinate is a half-integer and whose nearest-integer roundings correspond to $\bt$ and the vector obtained by replacing $t_r$ with the immediately preceding cut-point of bundle $r$.
When $t_r-1$ is the cut-point immediately preceding $t_r$, the latter vector is $\bt^{-r}$.
For adjacent cut-points $a_j,b_j$, the endpoint prefixes of the corresponding interpolation interval are $\Pre(P_j^+,a_j)$ and $\Pre(P_j^+,b_j)$.
The proofs of Lemmas~\ref{lem:robust-x-properties} and~\ref{lem:root-monotonicity} therefore apply to each system.

To apply root monotonicity to consistent pairs from different systems, we need a \emph{common system} in which both pairs remain consistent.
The following lemma gives a condition under which changing the grouping preserves consistency.
\begin{lemma}\label{lem:agreeing-in-two-systems}
Let $\bt$ be a common prefix-length vector of two systems. Suppose that for every bundle $j\in[n-1]$, one of the following holds:
\begin{itemize}
    \item $t_j$ is the first cut-point of bundle $j$ in both systems, or
    \item the cut-point of bundle $j$ immediately preceding $t_j$ is identical in both systems.
\end{itemize}
Then, for any $r\in[n-1]$, $(\bt,r)$ is consistent in one system if and only if it is in the other.
\end{lemma}
The insight behind this lemma is simple.
For a consistent pair $(\bt,r)$, consider the box with corners ``$\bt-\mathbf1$'' and $\bt$, where subtracting one means moving one \emph{block} backward in each bundle, rather than removing one original item.
Consistency asks whether the corresponding interpolation intervals contain a robust vector whose $r$-th coordinate is a half-integer.
The assumption ensures that these intervals have the same endpoint prefixes in both systems.
Although their rank values may differ, the relative order of these endpoint prefixes is unchanged.
The proof uses this common order to adjust the fractional coordinates while preserving the minimum-rank graph.
We defer the full proof of Lemma~\ref{lem:agreeing-in-two-systems} to Appendix~\ref{sec:agreeing-in-two-systems}

For the remainder of this subsection, we fix a root $r\in[n-1]$ and consider one round of the search.
The algorithm considers the roots in turn until a consistent pair at the boundary is found.
We maintain a candidate set $\mathcal S\subseteq\prod_{j=1}^{n-1}\{2,\dots,|P_j|\}$, initialized to the full product.
Lemma~\ref{lem:chore-boundaries}(1) ensures that every boundary pair in the original system has its prefix-length vector in this initial set.

When $\mathcal S$ is nonempty, we choose a root prefix length $T\in\{2,\dots,|P_r|\}$ that is ``at the median'': at least half of the candidates have their $r$-th coordinate at least $T$, and at least half have it at most $T$.
In each extended bundle $P_j^+$, we keep $0,1,|P_j|$, and $|P_j|+1$ as cut-points.
For each nonroot bundle, we introduce additional cut-points so that the resulting coarse system has $O(n)$ blocks per bundle.
For the root bundle, we add only $T-1$ and $T$, making the $T$-th original chore a singleton block and leaving $O(1)$ blocks.
Section~\ref{subsec:grouping-construction} describes the exact choice of cut-points.
Here, we establish the properties that allow the coarse pair to filter candidates in the original system.

Given $T$ and this coarse system, we find a consistent pair $(\bw,r)$ with $w_r=T$.
The empty prefixes rank below the root prefix at $T-1$, while the full extended bundles rank above it.
The proof of Lemma~\ref{lem:fixed-root-extension} therefore applies to the coarse system and guarantees the existence of such a pair.
To find it, each agent ranks the $O(n^2)$ selected nonempty prefixes of the original bundles, using $O(n^2\log n)$ comparison queries.
The empty prefixes are ordered by bundle index, and the full extended bundles follow all prefixes of the original bundles and have the same relative order as the full original bundles.
They can therefore be placed in the rankings without additional queries.
These rankings provide all information needed to find a consistent pair in this system without further comparison queries.
Thus, finding $(\bw,r)$ uses $O(n^3\log n)$ queries in total.

Consider the box cornered by $\bw$ and ``$\bw-\mathbf1$'', where subtracting one means moving to the preceding cut-point in each bundle:
$$
    \mathcal B_{\bw}
    =[a_1,b_1]\times\cdots\times[a_r,b_r]
    \times\cdots\times[a_{n-1},b_{n-1}].
$$
Here, $b_j=w_j$ and $a_j$ is the cut-point immediately preceding $w_j$ for every $j\in[n-1]$.
In particular, $a_r=T-1$ and $b_r=T$.

\begin{definition}
Let $\bt\in\prod_{j=1}^{n-1}\{2,\dots,|P_j|\}$ be a prefix-length vector in the original system.
\begin{itemize}
    \item $\bt$ is \emph{exceptional} with respect to $\bw$ if inserting the $2(n-1)$ cut-points $t_1-1,t_1,\dots,t_{n-1}-1,t_{n-1}$ into the coarse system subdivides $[a_j,b_j]$ for some $j\in[n-1]$.
    \item $\bt$ is \emph{non-exceptional} with respect to $\bw$ if none of these intervals is subdivided.
\end{itemize}
\end{definition}

\begin{figure}[ht]
\centering
\begin{tikzpicture}[x=1cm,y=1cm,font=\small,
    strip/.style={draw=black!55,line width=0.7pt},
    cell/.style={draw=diagramblue,fill=diagramblue!12,line width=0.9pt},
    coarsejoin/.style={draw=diagramblue,line width=0.65pt},
    candidate/.style={draw=diagramamber,line width=0.9pt},
    predecessor/.style={draw=diagramamber,densely dashed,line width=0.7pt}
]
\begin{scope}[shift={(0.000,0.000)}]
\node[anchor=base,align=center,text width=7.60cm,text height=9pt,text depth=2pt,font=\small\bfseries,inner sep=0pt] at (3.80,3.82) {(a) Non-exceptional vectors};
\draw[strip,fill=black!2] (1.630,0.000) rectangle (2.170,3.240);
\draw[cell] (1.630,1.260) rectangle (2.170,1.440);
\draw[strip,fill=black!2] (3.280,0.000) rectangle (3.820,3.240);
\draw[cell] (3.280,1.260) rectangle (3.820,2.160);
\draw[strip,fill=black!2] (4.930,0.000) rectangle (5.470,3.240);
\draw[cell] (4.930,1.080) rectangle (5.470,1.980);
\draw[strip,fill=black!2] (6.580,0.000) rectangle (7.120,3.240);
\draw[cell] (6.580,1.260) rectangle (7.120,2.160);
\draw[coarsejoin] (2.170,1.260) -- (3.280,1.260);
\draw[coarsejoin] (3.820,1.260) -- (4.930,1.080);
\draw[coarsejoin] (5.470,1.080) -- (6.580,1.260);
\draw[coarsejoin] (2.170,1.440) -- (3.280,2.160);
\draw[coarsejoin] (3.820,2.160) -- (4.930,1.980);
\draw[coarsejoin] (5.470,1.980) -- (6.580,2.160);
\draw[predecessor] (1.630,0.540) -- (2.170,0.540) -- (3.280,0.360) -- (3.820,0.360) -- (4.930,0.720) -- (5.470,0.720) -- (6.580,0.540) -- (7.120,0.540);
\draw[candidate] (1.630,0.720) -- (2.170,0.720) -- (3.280,0.540) -- (3.820,0.540) -- (4.930,0.900) -- (5.470,0.900) -- (6.580,0.720) -- (7.120,0.720);
\draw[candidate,line width=0.6pt] (1.470,0.800) -- (1.630,0.720);
\node[anchor=east,text=diagramamber,font=\footnotesize,inner sep=1pt] at (1.430,0.800) {$\mathbf t^{(1)}$};
\draw[predecessor,line width=0.6pt] (1.470,0.420) -- (1.630,0.540);
\node[anchor=east,text=diagramamber,font=\footnotesize,inner sep=1pt] at (1.430,0.420) {$\mathbf t^{(1)}-\mathbf 1$};
\draw[predecessor] (1.630,2.340) -- (2.170,2.340) -- (3.280,2.700) -- (3.820,2.700) -- (4.930,2.340) -- (5.470,2.340) -- (6.580,2.520) -- (7.120,2.520);
\draw[candidate] (1.630,2.520) -- (2.170,2.520) -- (3.280,2.880) -- (3.820,2.880) -- (4.930,2.520) -- (5.470,2.520) -- (6.580,2.700) -- (7.120,2.700);
\draw[candidate,line width=0.6pt] (1.470,2.600) -- (1.630,2.520);
\node[anchor=east,text=diagramamber,font=\footnotesize,inner sep=1pt] at (1.430,2.600) {$\mathbf t^{(2)}$};
\draw[predecessor,line width=0.6pt] (1.470,2.220) -- (1.630,2.340);
\node[anchor=east,text=diagramamber,font=\footnotesize,inner sep=1pt] at (1.430,2.220) {$\mathbf t^{(2)}-\mathbf 1$};
\draw[coarsejoin] (1.470,1.540) -- (1.630,1.440);
\node[anchor=east,text=diagramblue,font=\footnotesize,inner sep=1pt] at (1.430,1.540) {$\mathbf w$};
\draw[coarsejoin] (1.470,1.160) -- (1.630,1.260);
\node[anchor=east,text=diagramblue,font=\footnotesize,inner sep=1pt] at (1.430,1.160) {$\mathbf a$};
\node[anchor=base,font=\small,inner sep=0pt] at (1.900,-0.310) {$P_1$};
\node[anchor=base,font=\small,inner sep=0pt] at (3.550,-0.310) {$P_2$};
\node[anchor=base,font=\small,inner sep=0pt] at (5.200,-0.310) {$P_3$};
\node[anchor=base,font=\small,inner sep=0pt] at (6.850,-0.310) {$P_4$};
\node[anchor=base,align=center,text width=7.60cm,font=\small,inner sep=0pt] at (3.80,-0.79) {$\mathbf t^{(1)}\le\mathbf a,\qquad\mathbf t^{(2)}-\mathbf 1\ge\mathbf w$};
\end{scope}
\begin{scope}[shift={(8.150,0.000)}]
\node[anchor=base,align=center,text width=7.60cm,text height=9pt,text depth=2pt,font=\small\bfseries,inner sep=0pt] at (3.80,3.82) {(b) An exceptional vector};
\draw[strip,fill=black!2] (1.630,0.000) rectangle (2.170,3.240);
\draw[cell] (1.630,1.260) rectangle (2.170,1.440);
\draw[strip,fill=black!2] (3.280,0.000) rectangle (3.820,3.240);
\draw[cell] (3.280,1.080) rectangle (3.820,2.160);
\draw[strip,fill=black!2] (4.930,0.000) rectangle (5.470,3.240);
\draw[cell] (4.930,1.080) rectangle (5.470,2.160);
\draw[strip,fill=black!2] (6.580,0.000) rectangle (7.120,3.240);
\draw[cell] (6.580,1.260) rectangle (7.120,2.160);
\draw[coarsejoin] (2.170,1.260) -- (3.280,1.080);
\draw[coarsejoin] (3.820,1.080) -- (4.930,1.080);
\draw[coarsejoin] (5.470,1.080) -- (6.580,1.260);
\draw[coarsejoin] (2.170,1.440) -- (3.280,2.160);
\draw[coarsejoin] (3.820,2.160) -- (4.930,2.160);
\draw[coarsejoin] (5.470,2.160) -- (6.580,2.160);
\draw[predecessor] (1.630,0.540) -- (2.170,0.540) -- (3.280,1.440) -- (3.820,1.440) -- (4.930,2.700) -- (5.470,2.700) -- (6.580,2.520) -- (7.120,2.520);
\draw[candidate] (1.630,0.720) -- (2.170,0.720) -- (3.280,1.620) -- (3.820,1.620) -- (4.930,2.880) -- (5.470,2.880) -- (6.580,2.700) -- (7.120,2.700);
\draw[candidate,line width=0.6pt] (1.470,0.800) -- (1.630,0.720);
\node[anchor=east,text=diagramamber,font=\footnotesize,inner sep=1pt] at (1.430,0.800) {$\mathbf t$};
\draw[predecessor,line width=0.6pt] (1.470,0.420) -- (1.630,0.540);
\node[anchor=east,text=diagramamber,font=\footnotesize,inner sep=1pt] at (1.430,0.420) {$\mathbf t-\mathbf 1$};
\draw[coarsejoin] (1.470,1.540) -- (1.630,1.440);
\node[anchor=east,text=diagramblue,font=\footnotesize,inner sep=1pt] at (1.430,1.540) {$\mathbf w$};
\draw[coarsejoin] (1.470,1.160) -- (1.630,1.260);
\node[anchor=east,text=diagramblue,font=\footnotesize,inner sep=1pt] at (1.430,1.160) {$\mathbf a$};
\node[anchor=base,font=\small,inner sep=0pt] at (1.900,-0.310) {$P_1$};
\node[anchor=base,font=\small,inner sep=0pt] at (3.550,-0.310) {$P_2$};
\node[anchor=base,font=\small,inner sep=0pt] at (5.200,-0.310) {$P_3$};
\node[anchor=base,font=\small,inner sep=0pt] at (6.850,-0.310) {$P_4$};
\node[anchor=base,align=center,text width=7.60cm,font=\small,inner sep=0pt] at (3.80,-0.79) {$a_2<t_2-1<t_2<w_2$};
\end{scope}
\begin{scope}[shift={(0.000,-5.550)}]
\node[anchor=base,align=center,text width=7.60cm,text height=9pt,text depth=2pt,font=\small\bfseries,inner sep=0pt] at (3.80,3.82) {(c) A non-exceptional vector};
\draw[strip,fill=black!2] (1.630,0.000) rectangle (2.170,3.240);
\draw[cell] (1.630,1.260) rectangle (2.170,1.440);
\draw[strip,fill=black!2] (3.280,0.000) rectangle (3.820,3.240);
\draw[cell] (3.280,1.080) rectangle (3.820,2.160);
\draw[strip,fill=black!2] (4.930,0.000) rectangle (5.470,3.240);
\draw[cell] (4.930,1.800) rectangle (5.470,1.980);
\draw[strip,fill=black!2] (6.580,0.000) rectangle (7.120,3.240);
\draw[cell] (6.580,1.260) rectangle (7.120,2.160);
\draw[coarsejoin] (2.170,1.260) -- (3.280,1.080);
\draw[coarsejoin] (3.820,1.080) -- (4.930,1.800);
\draw[coarsejoin] (5.470,1.800) -- (6.580,1.260);
\draw[coarsejoin] (2.170,1.440) -- (3.280,2.160);
\draw[coarsejoin] (3.820,2.160) -- (4.930,1.980);
\draw[coarsejoin] (5.470,1.980) -- (6.580,2.160);
\draw[predecessor] (1.630,0.540) -- (2.170,0.540) -- (3.280,0.540) -- (3.820,0.540) -- (4.930,1.800) -- (5.470,1.800) -- (6.580,0.540) -- (7.120,0.540);
\draw[candidate] (1.630,0.720) -- (2.170,0.720) -- (3.280,0.720) -- (3.820,0.720) -- (4.930,1.980) -- (5.470,1.980) -- (6.580,0.720) -- (7.120,0.720);
\draw[candidate,line width=0.6pt] (1.470,0.800) -- (1.630,0.720);
\node[anchor=east,text=diagramamber,font=\footnotesize,inner sep=1pt] at (1.430,0.800) {$\mathbf t$};
\draw[predecessor,line width=0.6pt] (1.470,0.420) -- (1.630,0.540);
\node[anchor=east,text=diagramamber,font=\footnotesize,inner sep=1pt] at (1.430,0.420) {$\mathbf t-\mathbf 1$};
\draw[coarsejoin] (1.470,1.540) -- (1.630,1.440);
\node[anchor=east,text=diagramblue,font=\footnotesize,inner sep=1pt] at (1.430,1.540) {$\mathbf w$};
\draw[coarsejoin] (1.470,1.160) -- (1.630,1.260);
\node[anchor=east,text=diagramblue,font=\footnotesize,inner sep=1pt] at (1.430,1.160) {$\mathbf a$};
\node[anchor=base,font=\small,inner sep=0pt] at (1.900,-0.310) {$P_1$};
\node[anchor=base,font=\small,inner sep=0pt] at (3.550,-0.310) {$P_2$};
\node[anchor=base,font=\small,inner sep=0pt] at (5.200,-0.310) {$P_3$};
\node[anchor=base,font=\small,inner sep=0pt] at (6.850,-0.310) {$P_4$};
\node[anchor=base,align=center,text width=7.60cm,font=\small,inner sep=0pt] at (3.80,-0.79) {$t_3-1=a_3,\qquad t_3=w_3$};
\end{scope}
\begin{scope}[shift={(8.150,-5.550)}]
\node[anchor=base,align=center,text width=7.60cm,text height=9pt,text depth=2pt,font=\small\bfseries,inner sep=0pt] at (3.80,3.82) {(d) Impossible for consistent pairs};
\draw[strip,fill=black!2] (1.630,0.000) rectangle (2.170,3.240);
\draw[cell] (1.630,1.260) rectangle (2.170,1.440);
\draw[strip,fill=black!2] (3.280,0.000) rectangle (3.820,3.240);
\draw[cell] (3.280,1.080) rectangle (3.820,2.160);
\draw[strip,fill=black!2] (4.930,0.000) rectangle (5.470,3.240);
\draw[cell] (4.930,1.260) rectangle (5.470,1.980);
\draw[strip,fill=black!2] (6.580,0.000) rectangle (7.120,3.240);
\draw[cell] (6.580,1.260) rectangle (7.120,1.980);
\draw[coarsejoin] (2.170,1.260) -- (3.280,1.080);
\draw[coarsejoin] (3.820,1.080) -- (4.930,1.260);
\draw[coarsejoin] (5.470,1.260) -- (6.580,1.260);
\draw[coarsejoin] (2.170,1.440) -- (3.280,2.160);
\draw[coarsejoin] (3.820,2.160) -- (4.930,1.980);
\draw[coarsejoin] (5.470,1.980) -- (6.580,1.980);
\draw[predecessor] (1.630,0.540) -- (2.170,0.540) -- (3.280,0.540) -- (3.820,0.540) -- (4.930,2.520) -- (5.470,2.520) -- (6.580,2.520) -- (7.120,2.520);
\draw[candidate] (1.630,0.720) -- (2.170,0.720) -- (3.280,0.720) -- (3.820,0.720) -- (4.930,2.700) -- (5.470,2.700) -- (6.580,2.700) -- (7.120,2.700);
\draw[candidate,line width=0.6pt] (1.470,0.800) -- (1.630,0.720);
\node[anchor=east,text=diagramamber,font=\footnotesize,inner sep=1pt] at (1.430,0.800) {$\mathbf t$};
\draw[predecessor,line width=0.6pt] (1.470,0.420) -- (1.630,0.540);
\node[anchor=east,text=diagramamber,font=\footnotesize,inner sep=1pt] at (1.430,0.420) {$\mathbf t-\mathbf 1$};
\draw[coarsejoin] (1.470,1.540) -- (1.630,1.440);
\node[anchor=east,text=diagramblue,font=\footnotesize,inner sep=1pt] at (1.430,1.540) {$\mathbf w$};
\draw[coarsejoin] (1.470,1.160) -- (1.630,1.260);
\node[anchor=east,text=diagramblue,font=\footnotesize,inner sep=1pt] at (1.430,1.160) {$\mathbf a$};
\node[anchor=base,font=\small,inner sep=0pt] at (1.900,-0.310) {$P_1$};
\node[anchor=base,font=\small,inner sep=0pt] at (3.550,-0.310) {$P_2$};
\node[anchor=base,font=\small,inner sep=0pt] at (5.200,-0.310) {$P_3$};
\node[anchor=base,font=\small,inner sep=0pt] at (6.850,-0.310) {$P_4$};
\node[anchor=base,align=center,text width=7.60cm,font=\small,inner sep=0pt] at (3.80,-0.79) {$t_r<w_r,\qquad t_3>w_3$};
\end{scope}
\end{tikzpicture}
\caption{Exceptional and non-exceptional vectors for $n=5$ and $r=1$, with prefix lengths increasing upward.
Blue intervals mark $[a_j,w_j]$, while solid and dashed amber lines mark $t_j$ and $t_j-1$, respectively.
The vectors in (a) and~(c) are non-exceptional, whereas the inserted cut-points in (b) subdivide $[a_2,w_2]$.
The vector in (d) is also non-exceptional, but the two pairs cannot both be consistent in their respective systems; see Lemma~\ref{lem:exceptional}(1).}
\label{fig:exceptional}
\end{figure}
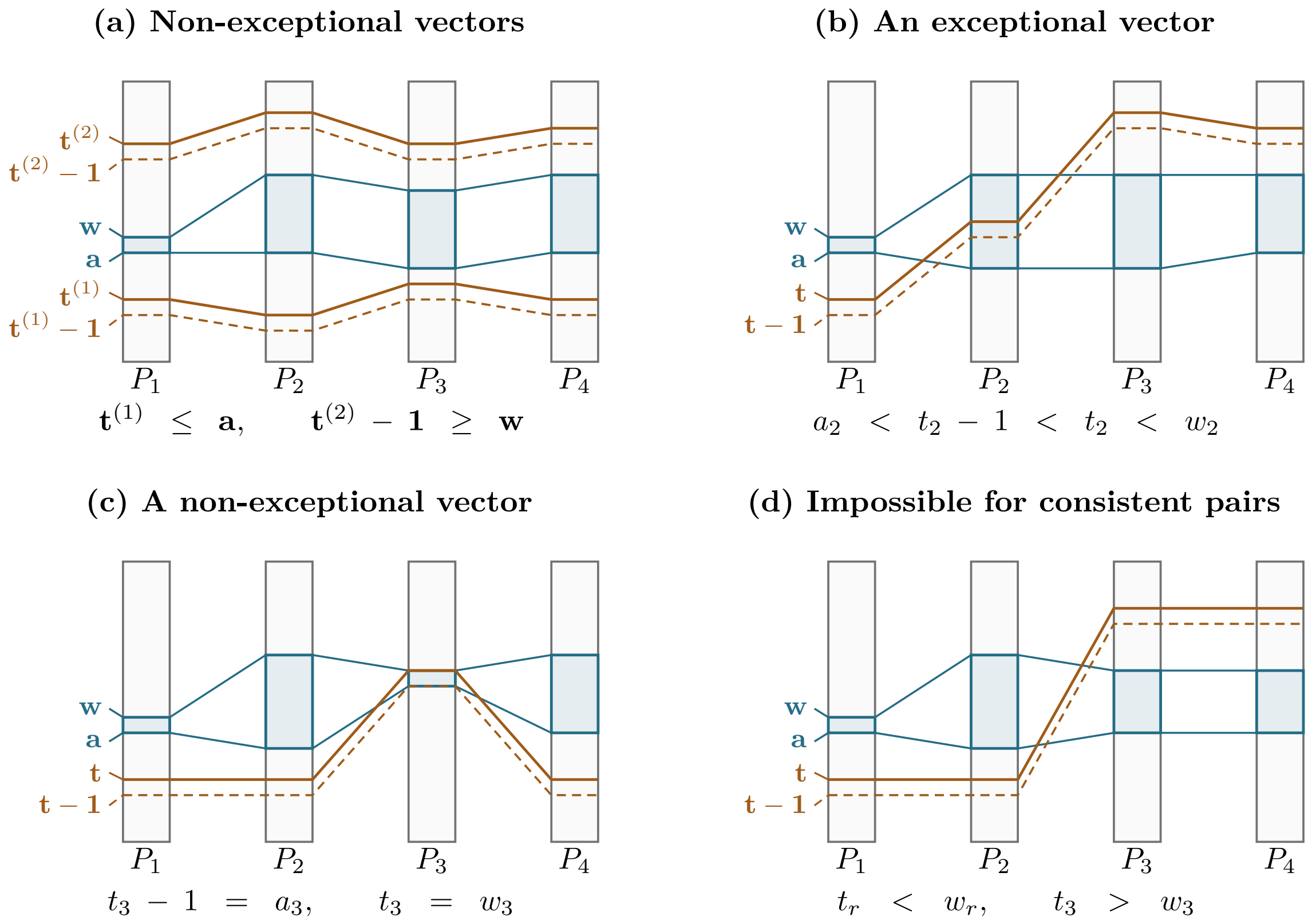

Figure~\ref{fig:exceptional}(a) and~(b) illustrate non-exceptional and exceptional vectors, respectively.
An inserted cut-point subdivides a block only if it lies strictly inside the corresponding interval $[a_j,b_j]$.
In particular, coinciding with the endpoints, as in Figure~\ref{fig:exceptional}(c), does not make a vector exceptional.

Now consider a candidate $\bt$ and add the $2(n-1)$ cut-points $t_1-1,t_1,\dots,t_{n-1}-1,t_{n-1}$ to the coarse system.
Both $\bw$ and $\bt$ now belong to the same system.
If $\bt$ is non-exceptional, the immediate predecessors of $\bw$ remain unchanged, so Lemma~\ref{lem:agreeing-in-two-systems} preserves the consistency of $(\bw,r)$.
If $(\bt,r)$ is consistent in the original system, the same lemma implies that it is also consistent in the new system: each $t_j$ has the same immediate predecessor $t_j-1$ in both systems.
Under these two conditions, we can apply Lemma~\ref{lem:root-monotonicity} to the two consistent pairs in this common system.
In particular, this rules out the configuration in Figure~\ref{fig:exceptional}(d): since $t_r<w_r$, root monotonicity would imply $\bt\le\bw$, contradicting $t_3>w_3$.
For $(\bw,r)$ and $\mathcal B_{\bw}$ as defined above, we obtain the following lemma.

\begin{lemma}\label{lem:exceptional}
The following hold for every pair $(\bt,r)$
consistent in the original system:
\begin{enumerate}
    \item[(1)] If $\bt$ is non-exceptional, then
    $$
        t_r\le T \Longrightarrow \bt\le\bw,
        \qquad
        t_r\ge T \Longrightarrow \bt\ge\bw.
    $$
    \item[(2)] If $t_r=T$ and $b_j-a_j>1$ for some $j\ne r$,
    then $\bt$ is exceptional.
\end{enumerate}
\end{lemma}
\begin{proof}
Part~(1) follows from the argument preceding the lemma.
For part~(2), suppose that $\bt$ is non-exceptional.
Since $t_r=w_r$, part~(1) gives $\bt=\bw$.
If $b_j-a_j>1$ for some $j\ne r$, then $a_j<t_j-1=b_j-1<b_j$.
Thus, inserting the cut-point $t_j-1$ subdivides $[a_j,b_j]$, contradicting the assumption that $\bt$ is non-exceptional.
\end{proof}
Lemma~\ref{lem:exceptional} provides a filter for the search.
Consider the selected $\bw$ with $w_r=T$, where $T$ is a median of the root coordinates of the current candidate set $\mathcal S$.
We first check whether $w_j=|P_j|+1$ for some $j\in[n-1]$.
If so, any non-exceptional candidate $\bt$ with $t_r\ge T$ cannot be consistent in the original system: otherwise, Lemma~\ref{lem:exceptional}(1) would imply $t_j\ge w_j=|P_j|+1$, contradicting $t_j\le|P_j|$.
We therefore remove all such candidates and proceed to the next iteration.
Otherwise, consistency gives $1\le w_j\le|P_j|$ for every $j\in[n-1]$, and $T\ge2$ ensures that $\bw^{-r}$ also lies within the original prefix ranges.
We evaluate $\Phi$ on the original bundles and consider the following cases.
\begin{itemize}
    \item Case 1: $\Phi(\bw)=0$.
    By Lemma~\ref{lem:exceptional}(1), each non-exceptional candidate $\bt$ with $t_r\le T$ such that $(\bt,r)$ is consistent in the original system satisfies $\bt\le\bw$, and hence $\Phi(\bt)\le\Phi(\bw)=0$.
    Thus, no such candidate can form a consistent pair at the boundary, and we can remove all non-exceptional candidates with $t_r\le T$.

    \item Case 2: $\Phi(\bw^{-r})=1$.
    As in Case~1, all non-exceptional candidates $\bt$ with $t_r\ge T$ such that $(\bt,r)$ is consistent in the original system satisfy $\bt\ge\bw$, so $\bt^{-r}\ge\bw^{-r}$.
    Therefore, $\Phi(\bt^{-r})\ge\Phi(\bw^{-r})=1$.
    Again, no such candidate can form a consistent pair at the boundary, and we can remove all non-exceptional candidates with $t_r\ge T$.

    \item Case 3(a): $\Phi(\bw^{-r})=0$ and $\Phi(\bw)=1$; $b_j-a_j=1$ for every $j\in[n-1]$.
    In this case, we have already found a consistent pair at the boundary in the original system, namely, $(\bw,r)$.
    Indeed, $\Phi(\bw)=1$ implies $w_j\ge2$ for every $j$ by Lemma~\ref{lem:chore-boundaries}(1).
    Thus, $a_j=w_j-1\ge1$ is the immediate predecessor of $w_j$ in both systems, and consistency is guaranteed by Lemma~\ref{lem:agreeing-in-two-systems}.

    \item Case 3(b): $\Phi(\bw^{-r})=0$ and $\Phi(\bw)=1$; $b_j-a_j>1$ for some $j\in[n-1]$.
    In this case, we can safely remove all non-exceptional candidates.
    To see this, for every non-exceptional candidate $\bt$ with $t_r<T$ such that $(\bt,r)$ is consistent in the original system, Lemma~\ref{lem:exceptional}(1) implies $\bt\le\bw$.
    Since $t_r\le T-1$, we have $\bt\le\bw^{-r}$.
    Thus, $\Phi(\bt)\le\Phi(\bw^{-r})=0$, so $(\bt,r)$ is not at the boundary.
    Similarly, for every non-exceptional candidate $\bt$ with $t_r>T$ such that $(\bt,r)$ is consistent in the original system, we have $\bt\ge\bw$ and $t_r-1\ge T$.
    Hence, $\bt^{-r}\ge\bw$ and $\Phi(\bt^{-r})\ge\Phi(\bw)=1$, so again $(\bt,r)$ is not at the boundary.
    Finally, if $t_r=T$, Lemma~\ref{lem:exceptional}(2) implies that $\bt$ must be exceptional whenever $(\bt,r)$ is consistent in the original system.
    Therefore, all candidates forming consistent pairs at the boundary must be exceptional.
\end{itemize}
To bound the number of candidates eliminated in this iteration, consider $\mathcal S$ before the update.
As mentioned, we select $T$ at the median, so $|\{\bt\in\mathcal S:t_r\ge T\}|\ge|\mathcal S|/2$ and $|\{\bt\in\mathcal S:t_r\le T\}|\ge|\mathcal S|/2$.
Let $\alpha$ be the fraction of candidates in $\mathcal S$ that are exceptional with respect to $\bw$.
Each of Cases~1 and~2, as well as the case involving an artificial item, eliminates at least a $(1/2-\alpha)$ fraction of $\mathcal S$ if $\alpha\le1/2$.
Case~3(a) returns a solution directly.
Case~3(b) eliminates a $(1-\alpha)$ fraction of $\mathcal S$.
The objective now becomes clear: \emph{after selecting $T$, we need to choose cut-points for the remaining $n-2$ bundles so that $\alpha$ is small.}
If $\alpha$ is bounded above by a constant strictly less than $1/2$ in every iteration, each iteration that does not return a solution eliminates a constant fraction of $\mathcal S$.

Initially, $|\mathcal S|=\prod_{j=1}^{n-1}(|P_j|-1)\le m^{n-1}$.
With item groupings satisfying the above bound on $\alpha$, at most $O(n\log m)$ iterations are needed for each root before finding a solution or exhausting the candidates.
Every deletion preserves all consistent pairs at the boundary with that root.
Since Section~\ref{subsec:balance} guarantees that such a pair exists for some root, the search succeeds for at least one root.
Each iteration uses $O(n^3\log n)$ comparison queries to find $\bw$ and $O(n^2)$ additional queries to evaluate $\Phi$ when needed.
Enumerating the $n-1$ possible roots therefore gives an overall query complexity of $O(n^5\log n\log m)$.

The next subsection constructs the required item grouping for each fixed root $r$ and median $T$, ensuring that the fraction of exceptional candidates is bounded by a sufficiently small constant.

\subsection{Construction of Item Groupings}\label{subsec:grouping-construction}

In this subsection, we construct the item groupings required by Section~\ref{subsec:items-grouping} and present the complete search procedure in Algorithm~\ref{alg:coarse-prefix-search}.
To keep the fraction of exceptional candidates small, we choose cut-points according to \emph{the distribution of the surviving candidates' prefix lengths in each bundle}.
The blocks need not contain comparable numbers of chores: a block containing many chores is harmless when few candidates subdivide it.
If many candidates share the same prefix length $q$ in a bundle, retaining both $q-1$ and $q$ isolates the $q$-th chore as a singleton block and prevents these candidates from subdividing any block in that bundle.
For each fixed root $r$ and median $T$, the construction yields $O(n)$ blocks per bundle.
For every box $\mathcal B_{\bw}$ of the resulting coarse system whose root interval is $[T-1,T]$, fewer than $|\mathcal S|/4$ candidates are exceptional with respect to $\bw$.

Fix a root $r\in[n-1]$, a nonempty candidate set
$\mathcal S\subseteq\prod_{j=1}^{n-1}\{2,\dots,|P_j|\}$,
and a median $T\in\{2,\dots,|P_r|\}$ of $\{t_r : \bt \in \mathcal S\}$ (counted with multiplicity).
The following lemma constructs a coarse system with few exceptional candidates.

\begin{lemma}\label{lem:candidate-weighted-grouping}
There exists a coarse system on $P_1^+,\dots,P_{n-1}^+$ with $O(n)$ blocks per bundle such that
\begin{itemize}
    \item Every bundle $j$ retains the cut-points $0, 1, |P_j|, |P_j|+1$, and the root bundle $r$ additionally retains $T-1$ and $T$.
    \item For every box $\mathcal B_{\bw}$ with $w_r=T$, fewer than $|\mathcal S|/4$ candidates in $\mathcal S$ are exceptional for $\bw$.
\end{itemize}
\end{lemma}
\begin{proof}
Set $h=\ceil{|\mathcal S|/(4(n-1))}$.
For the root bundle $P_r^+$, choose the cut-point set $\{0,1,T-1,T,|P_r|,|P_r|+1\}$.
For each nonroot bundle $j$, list the $j$-th coordinates of the vectors in $\mathcal S$, retaining repetitions, in nondecreasing order as $t_{j,1}\le\cdots\le t_{j,|\mathcal S|}$.
For every $\ell\in\{1,\dots,\floor{|\mathcal S|/h}\}$, keep both $t_{j,\ell h}-1$ and $t_{j,\ell h}$.
Thus, the cut-point set for bundle $j$ is
$$
    \{0,1,|P_j|,|P_j|+1\}
    \cup
    \bigcup_{\ell=1}^{\floor{|\mathcal S|/h}}
    \{t_{j,\ell h}-1,t_{j,\ell h}\},
$$
with duplicate values included only once.
The root bundle has at most six cut-points, and $T-1$ and $T$ are adjacent because they are consecutive integers.
Since $h\ge|\mathcal S|/(4(n-1))$, every nonroot bundle has at most $8(n-1)+4$ cut-points.
Hence, every bundle has $O(n)$ blocks, and the construction requires no comparison queries.
Figure~\ref{fig:candidate-grouping} illustrates the construction for a nonroot bundle.

\begin{figure}[t]
\centering
\begin{tikzpicture}[x=1.2cm,y=1cm,font=\small,
    cut/.style={draw=diagramblue,line width=0.8pt},
    bracket/.style={draw=diagramblue,line width=0.65pt}
]
\node at (3,1.65) {Selected values: $7$ and $12$};

\fill[diagramblue!7] (0,0) rectangle (6,0.5);
\fill[diagramblue!20] (0,0) rectangle (1,0.5);
\fill[diagramblue!20] (5,0) rectangle (6,0.5);
\draw[black!55,line width=0.7pt]
    (0,0) rectangle (6,0.5);

\foreach \x/\t in {0/6,1/7,5/11,6/12} {
    \draw[cut] (\x,-0.07)--(\x,0.57);
    \node[anchor=north,inner sep=2pt]
        at (\x,-0.10) {$\t$};
}

\foreach \y in {0.83,1.02,1.21} {
    \fill[diagramamber] (1,\y) circle[radius=1.8pt];
}
\fill[diagramamber] (4,0.83) circle[radius=1.8pt];
\node[anchor=south,text=diagramamber,inner sep=2pt]
    at (4,0.94) {$10$};
\fill[diagramamber] (6,0.83) circle[radius=1.8pt];

\draw[bracket]
    (0,-0.50)--(0,-0.62)--(1,-0.62)--(1,-0.50);
\draw[bracket]
    (1,-0.50)--(1,-0.62)--(5,-0.62)--(5,-0.50);
\draw[bracket]
    (5,-0.50)--(5,-0.62)--(6,-0.62)--(6,-0.50);

\node[anchor=north,font=\footnotesize] at (0.5,-0.73)
    {One item};
\node[anchor=north,align=center,font=\footnotesize] at (3,-0.73)
    {One candidate in $(7,11]$\\$1=h-1$};
\node[anchor=north,font=\footnotesize] at (5.5,-0.73)
    {One item};
\end{tikzpicture}
\caption{A local example with $h=2$.
Selected values $7$ and $12$ contribute cut-points $6,7,11,12$.
Amber dots represent candidate coordinates.
Repeated coordinates at $7$ do not subdivide $[6,7]$, while the longer block $[7,11]$ has only one candidate coordinate in $(7,11]$.}
\label{fig:candidate-grouping}
\end{figure}
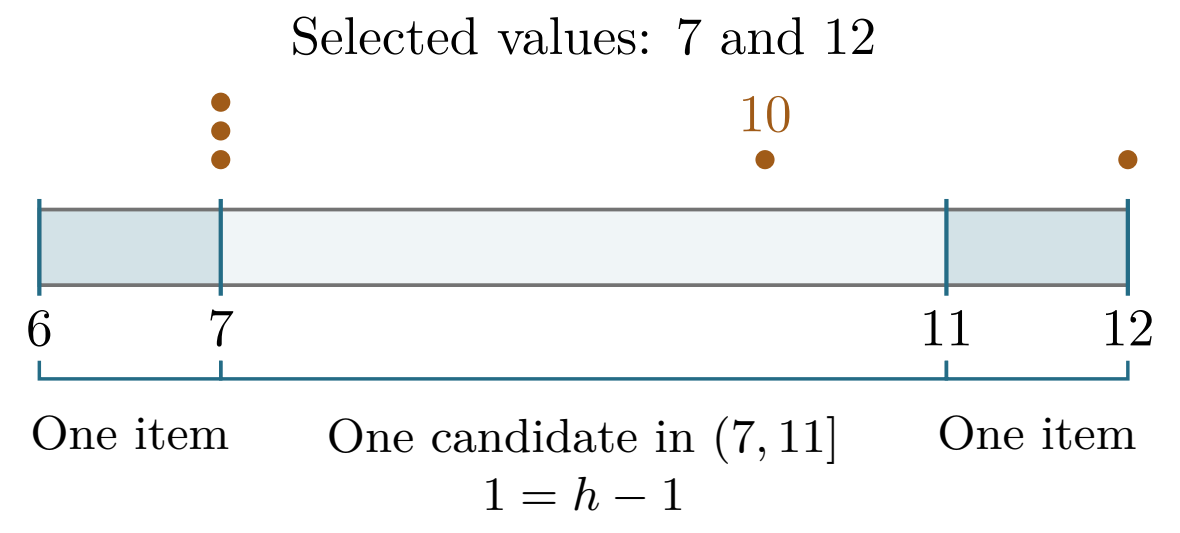

Now fix a cut-point vector $\bw$ as in the statement, and let $\mathcal E\subseteq\mathcal S$ be the candidates that are exceptional with respect to $\bw$.
If $w_j-a_j=1$, no integer cut-point can subdivide $[a_j,w_j]$.
If $w_j-a_j>1$, inserting $t_j-1$ and $t_j$ subdivides this interval if and only if $a_j<t_j\le w_j$.
Indeed, if $a_j<t_j<w_j$, then $t_j$ lies in the interior; if $t_j=w_j$, then $t_j-1=w_j-1$ lies in the interior.
Conversely, if either inserted cut-point lies in the interior, then $a_j<t_j\le w_j$.

We claim that, for every nonroot bundle $j$ with $w_j-a_j>1$, at most $h-1$ candidates satisfy $a_j<t_j\le w_j$.
Suppose otherwise.
In the sorted list of $j$-th coordinates, these candidates occupy at least $h$ consecutive positions, so this interval of positions contains an index $\ell h$.
Consequently, some selected value $q=t_{j,\ell h}$ satisfies $a_j<q\le w_j$.
If $q<w_j$, the selected cut-point $q$ lies strictly between $a_j$ and $w_j$.
If $q=w_j$, the selected cut-point $q-1=w_j-1$ lies strictly between them because $w_j-a_j>1$.
Both cases contradict the adjacency of $a_j$ and $w_j$, proving the claim.

The root interval $[T-1,T]$ has length one.
A union bound over the nonroot bundles thus gives
$$
\begin{aligned}
    |\mathcal E|
    &\le
    \sum_{\substack{j\ne r\\w_j-a_j>1}}
    |\{\bt\in\mathcal S:a_j<t_j\le w_j\}| \le (n-2)(h-1)
    <\frac{n-2}{4(n-1)}|\mathcal S|
    <\frac{|\mathcal S|}{4},
\end{aligned}
$$
where the first strict inequality follows from $h-1<|\mathcal S|/(4(n-1))$.
\end{proof}

The construction also handles small candidate sets without a separate stopping rule.
If $|\mathcal S|\le4(n-1)$, then $h=1$, so the construction includes both $t_j-1$ and $t_j$ as cut-points for every candidate $\bt\in\mathcal S$ and every nonroot bundle $j$.
Thus, no candidate is exceptional with respect to any cut-point vector $\bw$ satisfying the conditions in the lemma.

\paragraph{Computing a Consistent Coarse Pair.}
We now show how to find a consistent pair $(\bw,r)$ with $w_r=T$ in the coarse system.
We first learn the agents' prefix rankings, which provide all information needed to test consistency.
The coarse system has $O(n^2)$ prefixes across all bundles.
For each agent, we sort the selected nonempty prefixes of the original bundles according to the prefix order defined above, using $O(n^2\log n)$ comparison queries.
The empty prefixes precede all nonempty prefixes and are ordered by bundle index.
The full extended bundles follow all prefixes of the original bundles and have the same relative order as the full original bundles: adding $c_i(M)$ to each cost and one to each length preserves this order.
Since all full original bundles are included in the system, the empty prefixes and full extended bundles can be placed in the rankings without additional queries.
Thus, learning all agents' prefix rankings uses $O(n^3\log n)$ comparison queries.

Using these rankings, we next search for a robust vector witnessing the consistency of $(\bw,r)$.
We enumerate cut-point vectors $\bw$ with $w_r=T$ and immediate predecessors $a_1,\dots,a_{n-1}$, together with a designated agent $i$ for each nonroot bundle $j$, proposing $(i,j)$ as its active edge.
Each choice specifies the interpolation intervals and the proposed active edges.
We fix the root coordinate at the midpoint of its interpolation interval.
For each agent $i$, this choice determines an endpoint prefix at every bundle for which she is not the designated agent, as in the proof of Lemma~\ref{lem:agreeing-in-two-systems}.
In particular, the root contributes $\Pre(P_r^+,T-1)$ for every agent.
Let $Q_i$ be agent $i$'s minimum-rank prefix among these endpoint prefixes.
To realize the proposed active edges, we choose each nonroot coordinate so that its designated agent $i$'s interpolated rank equals her rank of $Q_i$.

We therefore check whether $Q_i$ lies strictly between $\Pre(P_j^+,a_j)$ and $\Pre(P_j^+,w_j)$ in agent $i$'s prefix order whenever $i$ is the designated agent of bundle $j$.
If these inequalities hold, we choose each nonroot coordinate within its interpolation interval, with fractional part in the interior of its designated agent's interval $I_i$, so that her interpolated rank equals her rank of $Q_i$.
Such a coordinate exists because the interpolation is strictly increasing across $I_i$.
The resulting minimum-rank graph joins each agent $i$ to the bundle corresponding to $Q_i$ and to every nonroot bundle for which she is the designated agent.
Since the root coordinate is already fixed at a half-integer, it remains to check whether this graph admits a perfect matching after any single agent is removed.
A choice passing these tests yields a robust vector whose coordinatewise ceiling corresponds to $\bw$ and whose root coordinate is a half-integer, and hence a consistent pair $(\bw,r)$.

It remains to show that some enumerated choice passes these tests.
The proof of Lemma~\ref{lem:fixed-root-extension} guarantees a robust vector in the coarse system at the prescribed interpolation interval of the root bundle.
By Lemma~\ref{lem:robust-x-properties}, every nonroot bundle has exactly one active edge in its minimum-rank graph.
Thus, the enumeration includes its interpolation intervals and active edges, and this choice passes all the tests.
The computation therefore finds the required consistent pair using only the learned prefix rankings, without further comparison queries.

The pair $(\bw,r)$ need not be at the boundary.
We use it to eliminate candidates according to the rules in Section~\ref{subsec:items-grouping}.
Algorithm~\ref{alg:coarse-prefix-search} gives the complete search procedure.

\begin{algorithm}[!htbp]
\caption{Candidate-Weighted Coarse Search}
\label{alg:coarse-prefix-search}
\KwIn{Balanced ordered bundles $(P_1,\dots,P_{n-1})$ and transition predicate $\Phi$}
\KwOut{A consistent pair $(\bt,r)$ at the boundary in the original system}
\For{$r\in[n-1]$}{
    $\mathcal S\gets\prod_{j=1}^{n-1}\{2,\dots,|P_j|\}$\;
    \While{$\mathcal S\ne\emptyset$}{
        Choose a median $T\in\{2,\dots,|P_r|\}$ of the $r$-th coordinates in $\mathcal S$, counting repetitions\;
        Construct the coarse system using Lemma~\ref{lem:candidate-weighted-grouping}\;
        Learn each agent's prefix order in this system\;
        Find a consistent pair $(\bw,r)$ with $w_r=T$, and let $a_j$ be the cut-point immediately preceding $w_j$ for every $j\in[n-1]$\;
        $\mathcal E\gets\{\bt\in\mathcal S:\exists j\ne r,\ w_j-a_j>1
            \text{ and }a_j<t_j\le w_j\}$\;
        \uIf{$w_j=|P_j|+1$ for some $j\in[n-1]$}{
            $\mathcal S\gets\mathcal E\cup\{\bt\in\mathcal S:t_r<T\}$\;
        }
        \uElseIf{$\Phi(\bw)=0$}{
            $\mathcal S\gets\mathcal E\cup\{\bt\in\mathcal S:t_r>T\}$\;
        }
        \uElseIf{$\Phi(\bw^{-r})=1$}{
            $\mathcal S\gets\mathcal E\cup\{\bt\in\mathcal S:t_r<T\}$\;
        }
        \uElseIf{$w_j-a_j=1$ for every $j\in[n-1]$}{
            \Return $(\bw,r)$\;
        }
        \Else{
            $\mathcal S\gets\mathcal E$\;
        }
    }
}
\end{algorithm}

\begin{theorem}\label{thm:coarse-prefix-search}
For the balanced bundles $(P_1,\dots,P_{n-1})$, Algorithm~\ref{alg:coarse-prefix-search} finds a consistent pair $(\bt,r)$ at the boundary in the original system using $O(n^5\log n\log m)$ comparison queries.
An EF1 allocation of $M'$ can then be computed using $O(n^2)$ additional comparison queries.
\end{theorem}
\begin{proof}
We first verify that, for each fixed root $r$, the candidate set $\mathcal S$ contains every vector $\bt$ for which $(\bt,r)$ is consistent and at the boundary in the original system.
This holds initially by Lemma~\ref{lem:chore-boundaries}(1).
In each iteration, the set $\mathcal E$ computed by the algorithm consists precisely of the exceptional candidates, as shown in the proof of Lemma~\ref{lem:candidate-weighted-grouping}.
Every update keeps all candidates in $\mathcal E$.
For non-exceptional candidates, the artificial-item case and Cases~1, 2, and~3(b) in Section~\ref{subsec:items-grouping} show that every discarded vector fails either consistency or the boundary condition.
Thus, every update preserves all consistent pairs at the boundary with root $r$.

The algorithm returns only in Case~3(a), when no artificial item is used, $\Phi(\bw^{-r})=0$, $\Phi(\bw)=1$, and $w_j-a_j=1$ for every $j$.
By Lemma~\ref{lem:chore-boundaries}(1), we have $w_j\ge2$, so $a_j=w_j-1\ge1$ is the immediate predecessor of $w_j$ in both systems.
Lemma~\ref{lem:agreeing-in-two-systems} therefore guarantees that $(\bw,r)$ is consistent in the original system.
Together with the two values of $\Phi$, this proves that every returned pair is valid.

By Lemma~\ref{lem:candidate-weighted-grouping}, we have $|\mathcal E|<|\mathcal S|/4$ in every iteration.
The artificial-item case and Cases~1 and~2 each discard at least $|\mathcal S|/2-|\mathcal E|>|\mathcal S|/4$ candidates.
Case~3(b) keeps only the candidates in $\mathcal E$.
Hence, every iteration that does not return a pair reduces the number of candidates to less than three quarters of its previous value.
Section~\ref{subsec:balance} guarantees that a consistent pair at the boundary exists for some root.
Its vector is preserved throughout the search for that root, so this search must return a pair rather than exhaust its candidates.
Thus, the algorithm finds a valid pair.

We next bound the number of comparisons.
For each root, the initial candidate set has size at most $m^{n-1}$, so the constant-factor decrease gives $O(n\log m)$ iterations.
Constructing the grouping requires no comparison queries.
Learning all agents' prefix rankings uses $O(n^3\log n)$ queries per iteration, and the at most two evaluations of $\Phi$ use $O(n^2)$ queries.
Once the rankings are known, finding a consistent coarse pair, checking the matchings, and updating $\mathcal S$ require no further comparison queries.
Summing over the $n-1$ roots gives $O(n^5\log n\log m)$ queries.

Finally, Lemma~\ref{lem:allocation-from-transition} guarantees that one of the two partitions associated with the returned pair admits an EF1 assignment.
The bipartite matching construction in Section~\ref{subsec:chores} finds such an assignment using $O(n^2)$ additional comparison queries.
\end{proof}


\FloatBarrier
\subsection{The Assignment of Bundles and Reinsertion of Chores}\label{subsec:chores}
We now combine the preceding procedures to compute an EF1 allocation of all chores.
We first apply the algorithm in Theorem~\ref{thm:(n-1)-partition}.
If an EF1 allocation of $M'$ is found (the first branch in Theorem~\ref{thm:(n-1)-partition}), we extend it to a complete EF1 allocation by applying the top-trading envy-cycle elimination procedure (Lemma~\ref{lem:chore-insert}).
In the second branch, we have the balanced partition $(P_1,\ldots,P_{n-1})$ of $M'$ containing only small items, and Algorithm~\ref{alg:coarse-prefix-search} finds a consistent pair $(\bt,r)$ at the boundary.

Given this pair, we compute the EF1 assignment guaranteed by Lemma~\ref{lem:allocation-from-transition} using $O(n^2)$ comparison queries.
Specifically, for each of the two partitions in Lemma~\ref{lem:allocation-from-transition}, we denote its bundles by $B_1,\ldots,B_n$, with $B_n$ being the remaining bundle.
For each $j\in[n-1]$, we let $\hat B_j$ be $B_j$ with its last chore removed, leaving it empty if $B_j=\emptyset$, and set $\hat B_n=R(\bt)$.
In the partition induced by $\bt^{-r}$, this removes $e$, the last chore in $\Pre(P_r,t_r)$, from $R(\bt^{-r})$.
In the partition induced by $\bt$, it leaves the remaining bundle unchanged.
We construct a bipartite graph $H$ between agents and bundles, with
    $$(i,j)\in E(H)\quad\Longleftrightarrow\quad c_i(\hat B_j)\le\min_{k\in[n]}\{c_i(B_k)\}.$$
Every perfect matching in $H$ gives an EF1 allocation.
Moreover, the proof of Lemma~\ref{lem:allocation-from-transition} guarantees such a matching for at least one of the two partitions: the prefix recipients satisfy EF1 after removing their last chores, while the remainder recipient either uses $e$ as her witness under $\bt^{-r}$ or is envy-free under $\bt$.
For each agent, finding a minimum-cost full bundle and comparing all $\hat B_j$ against it uses $O(n)$ queries.
Thus, both graphs can be constructed using $O(n^2)$ queries in total, after which a perfect matching yields the allocation without further queries.

Finally, we insert each chore in $E$ using Lemma~\ref{lem:chore-insert}, preserving EF1 throughout.

\begin{algorithm}[!htb]
\caption{Computation of EF1 Allocations for Additive Chores}
\label{alg:chores}
\KwIn{Instance $\mathcal{I}=(N,M,\bc)$ with nonnegative additive cost functions}
\KwOut{An EF1 allocation $\bX=(X_1,\dots,X_n)$}
Run Algorithm~\ref{alg:heavy-items} on $\mathcal I$\;
\eIf{it returns $(E,\cP)$}{
    $M'\gets M\setminus E$\;
    Construct $(P_1,\dots,P_{n-1})$ from $\cP$ using Algorithm~\ref{alg:balanced-partition}\;
    $(\bt,r)\gets$ the output of Algorithm~\ref{alg:coarse-prefix-search} on $(P_1,\dots,P_{n-1})$ with $\Phi$\;
    Compute an EF1 allocation $\bX$ of $M'$ from $(\bt,r)$ using the bipartite matching construction above\;
}{
    $\bX\gets$ the EF1 allocation returned with $E$\;
}
\For{each chore $e\in E$}{
    Insert $e$ into $\bX$ using Lemma~\ref{lem:chore-insert}\;
}
\Return $\bX$\;
\end{algorithm}

The following theorem gives the query bound for the complete procedure.
\begin{theorem}\label{thm:chores}
For instances with nonnegative additive cost functions, Algorithm~\ref{alg:chores} computes EF1 allocations using $O(n^8\log m)$ comparison queries.
\end{theorem}
\begin{proof}
We first show that Algorithm~\ref{alg:chores} returns an EF1 allocation.
By Lemma~\ref{lem:heavy-items}, Algorithm~\ref{alg:heavy-items} returns either an EF1 allocation of $M'=M\setminus E$ or a small-block partition of $M'$.
In the latter case, Algorithm~\ref{alg:balanced-partition} constructs $n-1$ balanced bundles by Lemma~\ref{lem:construct-balanced-partition}.
Theorem~\ref{thm:coarse-prefix-search} guarantees that the search finds a consistent pair $(\bt,r)$ at the boundary.
The bipartite matching construction above then yields an EF1 allocation of $M'$.
In either case, we complete the allocation by inserting the chores in $E$ one at a time, preserving EF1 at each step by Lemma~\ref{lem:chore-insert}.

We next bound the number of comparisons.
By Theorem~\ref{thm:(n-1)-partition}, the reduction and balanced-partition construction together use $O(n^8\log m)$ comparison queries and set aside $|E|=O(nk\log k)$ chores.
If an EF1 allocation has not been found in the preprocessing phase, Theorem~\ref{thm:coarse-prefix-search} gives an EF1 allocation of $M'$ using $O(n^5\log n\log m)$ further queries.
Reinsertion uses $O(n^2|E|)=O(n^8\log n)$ queries by Lemma~\ref{lem:chore-insert}, since $k=\Theta(n^5)$.
As $m>n\ge3$, the total is 
$
    O(n^8\log m).
$
\end{proof}


Our framework extends to the goods setting with the same comparison-query bound.
\begin{theorem}\label{thm:goods}
For indivisible goods with nonnegative additive valuation functions, EF1 allocations can be computed in $O(n^8\log m)$ comparison queries.
\end{theorem}
We defer the algorithm and the proof of Theorem~\ref{thm:goods} to Appendix~\ref{sec:goods}.

\section{Conclusion and Open Problems}\label{sec:conclusion}
In this paper, we studied the computation of EF1 allocations of indivisible items under the comparison-query model.
For indivisible chores with nonnegative additive cost functions, we introduced a deterministic algorithm that computes an EF1 allocation in $O(n^8\log m)$ comparison queries.
We further extended this approach to indivisible goods with nonnegative additive valuation functions, where our algorithm achieves the same query bound.
For every fixed number of agents, our result matches the known $\Omega(\log m)$ lower bound and resolves the open question of whether logarithmic comparison-query complexity is attainable beyond three agents~\cite{conf/wine/BuLLST24,LinEtAl2026}.

A natural direction for future work is to determine the comparison-query complexity of EF1 under general monotone preferences.
Even without additivity, EF1 allocations always exist for monotone preferences and can be computed in $O(n^2m)$ comparison queries via envy-cycle elimination~\cite{LiptonEtAl2004,BhaskarEtAl2021}.
Our logarithmic-query algorithms, however, rely on additivity.
Whether logarithmic query complexity in $m$ remains achievable for every fixed $n$ under general monotone preferences, or whether a superlogarithmic lower bound separates them from additive preferences, remains an open question.


\section*{Declaration of AI Usage}
This project proceeded in two phases.

In the first phase, we developed the framework in Section~\ref{sec:shared} through extensive interactions with GPT-5.6 Pro, GPT-6 Pro, and GPT Astra.
Our initial goal was to achieve a query complexity sublinear in $m$, which improves upon the standard envy-cycle procedure.
During this period, we developed a sequence of algorithms that progressively improved the query complexity from $O_n(m/\log m)$ to $O_n(\sqrt{m})$, then to $O_n(\log^2 m)$, and finally to $O_n(\log m)$.
However, all of these algorithms incurred an exponential dependence on $n$.
The strongest result in this phase achieved $O_n(\log m)$ queries via an exhaustive search over the structure of $G(\bx)$ (see Definition~\ref{def:minimum-rank-graph}).

At the beginning of the second phase, we began to interact with GPT-6 Pro for algorithms with query complexity polynomial in $n$ (while maintaining logarithmic dependency on $m$).
GPT-6 Pro gave a search algorithm that finds $\bx$ with query complexity $O(n^{23}\log m)$.
Unfortunately, we could not understand the algorithm and therefore could not verify the correctness.
On the other hand, we discovered the uniqueness lemma, Lemma~\ref{lem:root-monotonicity-fractional}, on our own.
We later used the idea of coarser-to-finer partitions to obtain the algorithmic framework in Section~\ref{sec:search}.
A search algorithm with query complexity $O(n^5\log n\log m)$ was then obtained, which is the one presented in Section~\ref{sec:search}.
We fed it to GPT-6 Pro, and GPT-6 Pro helped us fill in details of the algorithms and correct some bugs in some corner cases.
GPT-6 Pro further confirmed that our algorithm is substantially different from the one it proposed (the one with complexity $O(n^{23}\log m)$).

\appendix

\section{Verifying EF1 Allocations}\label{sec:ef1-verification}

We consider the problem of determining whether a prescribed allocation is EF1, given comparison access to the agents' preferences.

\begin{proposition}\label{prop:ef1-verification}
For two agents with identical nonnegative additive preferences, determining whether a prescribed allocation is EF1 has deterministic worst-case comparison-query complexity $\Theta(m)$.
This holds both for goods and for chores.
\end{proposition}
\begin{proof}
For the upper bound, we check the EF1 condition for each ordered pair of distinct agents $i,j$.
For goods, we test $v_i(X_i)\ge v_i(X_j\setminus\{e\})$ for each $e\in X_j$ until a witness is found or all items have been checked.
For chores, we similarly test $c_i(X_i\setminus\{e\})\le c_i(X_j)$ for each $e\in X_i$.
Each test uses one comparison query, and the empty-bundle cases are handled directly according to the definition of EF1.
Checking both ordered pairs determines whether the allocation is EF1.
Since there are two agents, this requires at most $|X_1|+|X_2|=m$ comparisons.
For the lower bound, let $m=k+1$ with $k\ge2$, and consider the allocation $(A,B)$, where $A=\{a\}$ and $B=\{b_1,\dots,b_k\}$.
Both agents have the same additive function $f$, interpreted as a valuation for goods and a cost function for chores.
Set $\eta=1/(2k)$, $f(a)=k-1-\eta$, and $f(b_j)=1$ for every $j\in[k]$.
Since $A$ is a singleton, the only nontrivial EF1 condition in either setting is that $f(B\setminus\{b_j\})\le f(A)$ for some $j$.
However, $f(B\setminus\{b_j\})=k-1>f(A)$ for every $j$, so $(A,B)$ is not EF1.

Suppose a correct deterministic algorithm terminates after $q<k-1$ queries on this instance, comparing bundles $(U_r,V_r)$ for $r=1,\dots,q$.
We construct another instance that gives the same answers to these queries but makes $(A,B)$ EF1.
Keep $f(a)$ unchanged and change each $f(b_j)$ to $f(b_j)+\delta_j$.
For each queried pair $(U_r,V_r)$, require $\sum_{j:\,b_j\in U_r}\delta_j-\sum_{j:\,b_j\in V_r}\delta_j=0$, so that the difference $f(U_r)-f(V_r)$ remains unchanged.
Also require $\sum_{j=1}^k\delta_j=0$, so that the value of $B$ remains unchanged.
These are at most $q+1<k$ homogeneous linear equations in $k$ variables, so there is a nonzero solution.
Since its entries sum to zero, at least one entry is positive.
Scale the solution so that $\max_j\delta_j=\eta$.
Then $\delta_j=-\sum_{\ell\ne j}\delta_\ell\ge-(k-1)\eta>-1$ for every $j$.

Let both agents have the additive function $\hat f$ defined by $\hat f(a)=f(a)$ and $\hat f(b_j)=1+\delta_j$.
All item values remain positive.
By construction, the difference between the values of the two bundles in each query remains unchanged, so each query returns the same answer, including ties.
Since the algorithm is deterministic, it asks the same sequence of queries and returns the same answer on both instances.
However, $\hat f(B)=k$, and choosing $j^*$ with $\delta_{j^*}=\eta$ gives $\hat f(B\setminus\{b_{j^*}\})=k-1-\eta=\hat f(A)$.
Thus, $(A,B)$ is EF1 under $\hat f$ in both settings, contradicting the correctness of the algorithm.
Therefore, at least $k-1=m-2$ queries are necessary, giving the $\Omega(m)$ lower bound.
\end{proof}

\section{Algorithm for Balanced Partition with Small Items}
\label{append:(n-1)-partition}
In this section, we will prove Theorem~\ref{thm:(n-1)-partition}.
\subsection{The Reduction to Small Blocks}\label{subsec:compression}
In this subsection, we show how to reduce the instance to one in which the remaining chores are partitioned into blocks of small cost for every agent, while setting aside only $\mathrm{poly}(n)$ heavy chores.
Specifically, we choose $k$ as the smallest power of two that is at least $16n^5$, which means $16n^5\le k<32n^5$.
In the following, we first provide the overview of the reduction.

\begin{figure}[htbp]
\centering
\begin{tikzpicture}[
    x=0.95cm,y=1cm,font=\small,
    block/.style={draw=black!60,fill=black!3,line width=0.9pt},
    containing/.style={draw=diagramblue,fill=diagramblue!10,line width=1.3pt},
    common/.style={draw=diagramblue,fill=diagramblue!6,line width=1pt}
]
\foreach \x/\y in {2/-2,3/0,4/0,6/-1,7/-1} {
    \draw[black!30,densely dotted,line width=0.7pt] (\x,{\y-0.24}) -- (\x,-3.05);
}
\foreach \t in {1,...,10} {
    \node at ({\t-0.5},0.7) {$e_{\t}$};
}
\node at (-1.1,0.7) {$R$};
\foreach \y/\label in {0/\cP_1,-1/\cP_2,-2/\cP_3,-3.3/\cP} {
    \node at (-1.1,\y) {$\label$};
}
\foreach \a/\b in {0/3,3/4,4/10} {
    \draw[block] (\a,-0.24) rectangle (\b,0.24);
}
\draw[containing] (4,-0.24) rectangle (10,0.24);
\foreach \a/\b in {0/6,6/7,7/10} {
    \draw[block] (\a,-1.24) rectangle (\b,-0.76);
}
\draw[containing] (0,-1.24) rectangle (6,-0.76);
\foreach \a/\b in {0/2,2/3,3/10} {
    \draw[block] (\a,-2.24) rectangle (\b,-1.76);
}
\draw[containing] (3,-2.24) rectangle (10,-1.76);
\foreach \a/\b in {0/2,2/3,3/4,4/6,6/7,7/10} {
    \draw[common] (\a,-3.54) rectangle (\b,-3.06);
}
\draw[draw=diagramblue,fill=diagramblue!25,line width=1.4pt]
    (4,-3.54) rectangle (6,-3.06);
\node at (5,-3.3) {$B$};
\end{tikzpicture}
\caption{The illustration of merging cut positions. 
The blocks in the bottom row form the common partition $\cP$, obtained by merging the distinct cut positions of $\cP_1,\cP_2,\cP_3$.
Dotted lines mark these cut positions.
Every block in $\cP$ is contained in one block of each $\cP_i$ and therefore inherits every agent's cost bound.
The darker block $B=\{e_5,e_6\}$ and its containing blocks are highlighted.}
\label{fig:common-partition}
\end{figure}
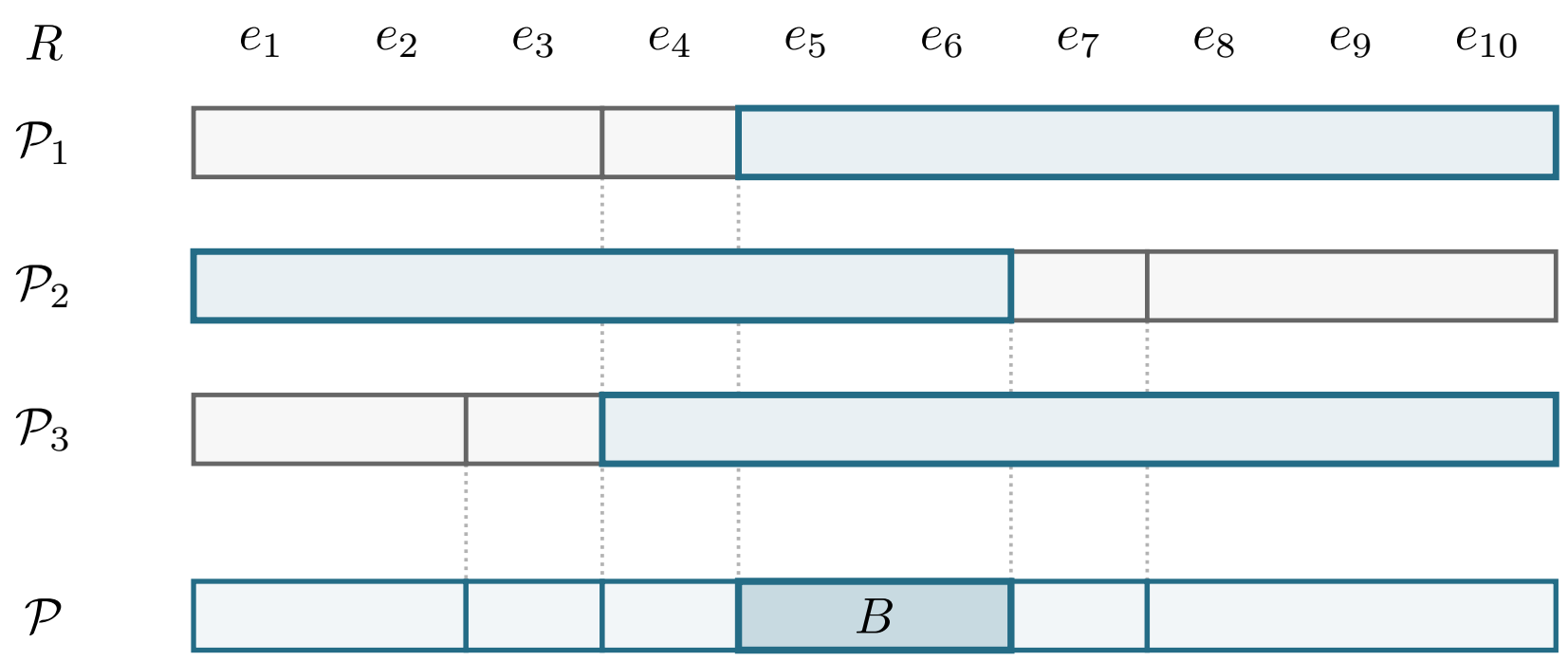

\paragraph{Algorithm Overview.}
Given a chore set $R\subseteq M$, Algorithm~\ref{alg:detector} constructs a binary splitting tree of depth at most $\log_2(2k)$ for each agent.
The algorithm either identifies an agent $i$ and a heavy chore $e$ with $c_i(e)>c_i(R)/(4k)$, or merges the cut positions across all trees to obtain a \emph{small-block partition} of $R$ into $O(nk)$ blocks (see Figure~\ref{fig:common-partition}), each of cost at most $c_i(R)/(2k)$ for every agent $i\in N$ (Lemma~\ref{lem:detector}).
Algorithm~\ref{alg:heavy-items} maintains a remaining pool $R$ (initialized to $M$) and an initially empty set $E$ of set-aside chores.
The primary role of Algorithm~\ref{alg:heavy-items} is to iteratively identify and set aside heavy chores via the splitting procedure of Algorithm~\ref{alg:detector}.
Because the heavy-chore thresholds $c_i(R)/(4k)$ decrease as $R$ shrinks, repeated removal alone cannot bound $|E|$ independently of $m$.
We therefore incorporate early-termination checks that directly construct an EF1 allocation of $M\setminus E$. 
For instance, when some agent views $n-1$ chores in $E$ each at least as much as the entire pool $R$, she receives $R$ while the other agents each receive one of these $n-1$ chores (which are then removed from $E$).
Together with the heavy-chore guarantee, these checks bound the total number of removals by $O(nk\log k)$ (Lemma~\ref{lem:heavy-items}).
\medskip

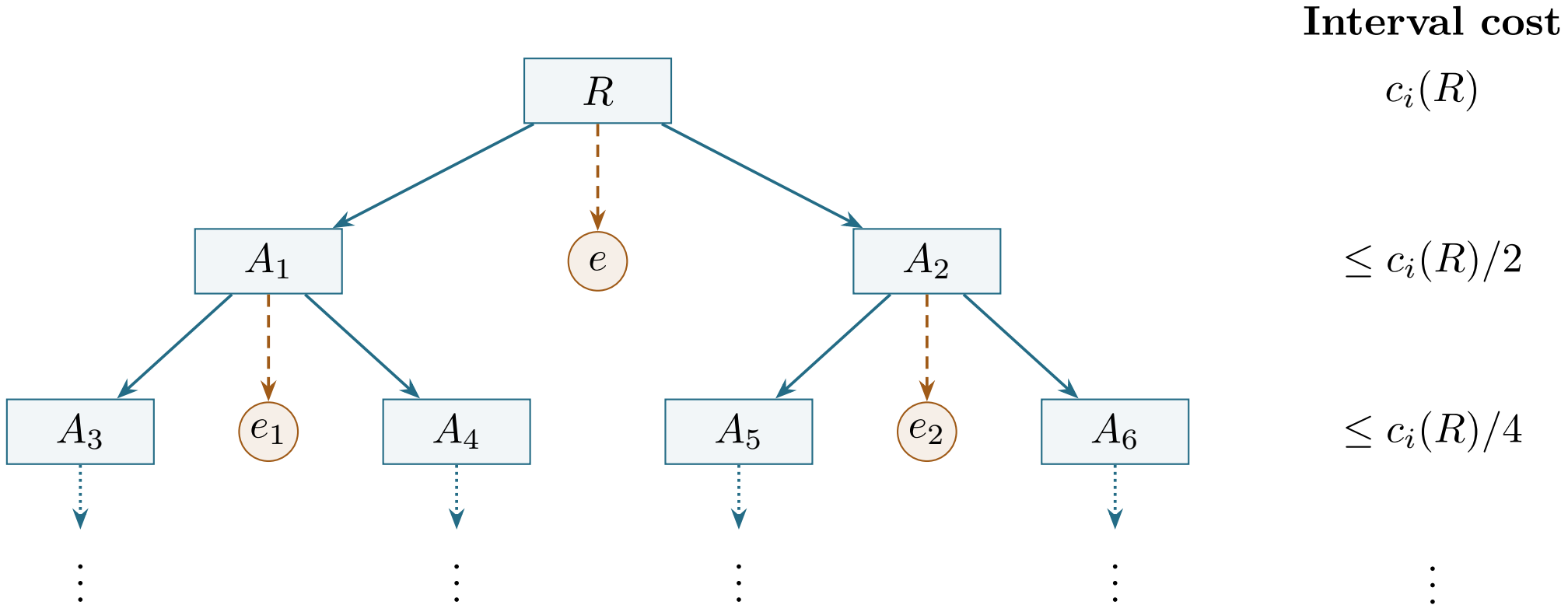
\begin{figure}[!ht]
\centering
\begin{tikzpicture}[
    x=1cm,y=1cm,font=\small,
    interval/.style={draw=diagramblue,fill=diagramblue!6,minimum width=1.25cm,minimum height=0.55cm,inner sep=2pt},
    item/.style={circle,draw=diagramamber,fill=diagramamber!10,minimum size=0.5cm,inner sep=1pt},
    child/.style={-{Stealth[length=1.8mm]},draw=diagramblue,line width=0.7pt},
    isolated/.style={-{Stealth[length=1.8mm]},draw=diagramamber,densely dashed,line width=0.7pt}
]
\node[interval] (root) at (5.5,0) {$R$};
\node[interval] (left) at (2.7,-1.45) {$A_1$};
\node[item] (middle) at (5.5,-1.45) {$e$};
\node[interval] (right) at (8.3,-1.45) {$A_2$};
\draw[child] (root) -- (left);
\draw[child] (root) -- (right);
\draw[isolated] (root) -- (middle);
\node[interval] (ll) at (1.1,-2.9) {$A_3$};
\node[item] (li) at (2.7,-2.9) {$e_1$};
\node[interval] (lr) at (4.3,-2.9) {$A_4$};
\node[interval] (rl) at (6.7,-2.9) {$A_5$};
\node[item] (ri) at (8.3,-2.9) {$e_2$};
\node[interval] (rr) at (9.9,-2.9) {$A_6$};
\foreach \u/\v in {left/ll,left/lr,right/rl,right/rr} {
    \draw[child] (\u) -- (\v);
}
\draw[isolated] (left) -- (li);
\draw[isolated] (right) -- (ri);
\foreach \v in {ll,lr,rl,rr} {
    \draw[child,densely dotted] (\v.south) -- ++(0,-0.55);
    \node at ([yshift=-0.9cm]\v.south) {$\vdots$};
}
\node[font=\small\bfseries] at (12.6,0.6) {Interval cost};
\node at (12.6,0) {$c_i(R)$};
\node at (12.6,-1.45) {$\le c_i(R)/2$};
\node at (12.6,-2.9) {$\le c_i(R)/4$};
\node at (12.6,-4.1) {$\vdots$};
\end{tikzpicture}
\caption{Recursive splitting for one agent $i$. The nodes $A_1,\dots,A_6$ denote subintervals. Each split isolates one item and leaves the intervals before and after it. Dashed edges lead to singleton blocks, which need not satisfy the displayed interval-cost bound. Only interval nodes continue to depth $\log_2(2k)$, where their cost is at most $c_i(R)/(2k)$. Empty intervals are omitted.}
\label{fig:interval-tree}
\end{figure}

We now describe the reduction in detail and establish its guarantees.
The first step is the recursive partitioning procedure in Algorithm~\ref{alg:detector}, which assumes $c_i(R)>0$ for all $i\in N$ and uses the public item order restricted to $R$.
For each agent $i\in N$, the procedure constructs a binary splitting tree over $R$ of depth at most $\log_2(2k)$ (see Figure~\ref{fig:interval-tree}), initialized with the root interval $R$ at depth $0$.
For each nonempty interval $A$ at depth $\ell<\log_2(2k)$, the algorithm performs a binary search to locate a cut item $e\in A$ such that both the prefix of $A$ before $e$ and the suffix of $A$ after $e$ have cost at most $c_i(A)/2$ (Lemma~\ref{lem:halving}).
The cut item $e$ is isolated as a singleton block, while the two remaining subintervals become the children of $A$ at depth $\ell+1$ and are split recursively.

\FloatBarrier

We refer to the nonempty intervals at depth $\log_2(2k)$ as \emph{terminal intervals}, each of which has cost at most $c_i(R)/(2k)$ by construction.
Consequently, in the resulting partition $\cP_i$ of $R$ formed by the terminal intervals and the isolated singletons, only the singletons can have cost exceeding $c_i(R)/(2k)$.
Since the tree produces at most $2k$ terminal intervals and $2k-1$ singletons, we have $|\cP_i|\le 4k-1$.
We call a chore $e\in R$ \emph{heavy} for agent $i\in N$ if $c_i(e)>c_i(R)/(4k)$, where $R$ denotes the current remaining pool\footnote{Note that heaviness is defined relative to the current remaining pool $R$ rather than the initial chore set $M$.}.

The algorithm then finds a maximum-cost block $B_i^*\in\argmax_{B\in\cP_i}\{c_i(B)\}$ for agent $i$.
If $B_i^*=\{e\}$ is a singleton, the bound $|\cP_i|\le 4k-1$ gives the guarantee $c_i(e)>c_i(R)/(4k)$.
Thus, $e$ is heavy relative to the input pool $R$, and the algorithm immediately returns $(i,e)$.
Otherwise, $B_i^*$ is a terminal interval, so every block in $\cP_i$ has cost at most $c_i(B_i^*)\le c_i(R)/(2k)$.
If no heavy chore is returned for any agent, Algorithm~\ref{alg:detector} merges the cut positions of $\cP_1,\dots,\cP_n$ to obtain a common partition $\cP$ of $R$ (see Figure~\ref{fig:common-partition}).

\begin{algorithm}[!htb]
\caption{Recursive Partitioning}
\label{alg:detector}
\KwIn{Instance $\mathcal{I}=(N,R,\bc)$ with $c_i(R)>0$ for all $i\in N$, and a power of two $k\ge 2$}
\KwOut{A block partition $\cP$ of $R$, or a pair $(i,e)\in N\times R$}

\For{$i\in N$}{
    Initialize a binary tree with root interval $R$ at depth $0$\;
    \For{$\ell=0,\dots,\log_2(2k)-1$}{
        \For{each nonempty interval $A$ at depth $\ell$}{
            Binary search for the smallest $t\in[|A|]$ with $\Comp_i(\Suf(A,t),\Pre(A,t))=\Suf(A,t)$\;
            Isolate the $t$-th item $e$ of $A$ and create children $\Pre(A,t-1)$ and $\Suf(A,t)$\;
        }
    }
    $\cP_i\gets$ the nonempty last-level intervals and all isolated singletons\;
    $B_i^*\gets\argmax_{B\in\cP_i}\{c_i(B)\}$\;
    
    \If{$B_i^*=\{e\}$}{ 
        \Return $(i,e)$ \tcp{$B_i^*$ is a singleton}
    }
}
$\cP\gets$ the common partition formed by merging the cut positions of $\cP_1,\dots,\cP_n$\;
\Return $\cP$\;
\end{algorithm}

We first show that each binary-search step in Algorithm~\ref{alg:detector} ensures that the cost of every child interval is at most half that of its parent.
\begin{samepage}
\begin{lemma}\label{lem:halving}
For any agent $i\in N$ and any nonempty ordered bundle $A\subseteq M$, we can use binary search to find an item $e\in A$ at position $t\in[|A|]$ in $O(\log m)$ comparison queries such that
$$
    \max\{c_i(\Pre(A,t-1)),\,c_i(\Suf(A,t))\}\le\frac{c_i(A)}{2}.
$$
\end{lemma}
\end{samepage}
\begin{proof}
Note that as the index $t\in[|A|]$ increases, $c_i(\Pre(A,t))$ is nondecreasing and $c_i(\Suf(A,t))$ is nonincreasing.
Moreover, at $t=|A|$, we have $\Suf(A,|A|)=\emptyset$, so $c_i(\Suf(A,|A|))=0\le c_i(\Pre(A,|A|))$.
Therefore, the predicate $c_i(\Suf(A,t))\le c_i(\Pre(A,t))$ is monotone in $t$ and holds at $t=|A|$, which allows binary search to locate the smallest index $t\in[|A|]$ satisfying $c_i(\Suf(A,t))\le c_i(\Pre(A,t))$ using $O(\log |A|)=O(\log m)$ comparison queries.

Let $t$ be the smallest index returned by the binary search and $e$ be the $t$-th item of $A$.
Since $\Pre(A,t)$ and $\Suf(A,t)$ partition $A$, we have $c_i(\Pre(A,t))+c_i(\Suf(A,t))=c_i(A)$, so $c_i(\Suf(A,t))\le c_i(\Pre(A,t))$ directly implies $c_i(\Suf(A,t))\le c_i(A)/2$.
Similarly, by the minimality of $t$ (or because $\Pre(A,0)=\emptyset$ if $t=1$), we have $c_i(\Pre(A,t-1))\le c_i(\Suf(A,t-1))$. 
Since $\Pre(A,t-1)$ and $\Suf(A,t-1)$ also partition $A$, this yields $c_i(\Pre(A,t-1))\le c_i(A)/2$.
\end{proof}

Lemma~\ref{lem:halving} allows us to establish the correctness and query complexity of Algorithm~\ref{alg:detector}.
\begin{samepage}
\begin{lemma}\label{lem:detector}
Algorithm~\ref{alg:detector} uses $O(n^6\log m)$ comparison queries and returns one of the following:
\begin{itemize}
    \item an item $e\in R$ and an agent $i\in N$ such that $c_i(e)>c_i(R)/(4k)$, or
    \item a partition $\cP$ of $R$ into $O(nk)$ blocks such that $c_i(B)\le c_i(R)/(2k)$ for all $i\in N$ and $B\in\cP$.
\end{itemize}
\end{lemma}
\end{samepage}
\begin{proof}
For a fixed agent $i\in N$, Lemma~\ref{lem:halving} implies inductively that every interval at depth $\ell\le\log_2(2k)$ has cost at most $c_i(R)/2^\ell$.
In particular, every nonempty terminal interval at depth $\log_2(2k)$ has cost at most $c_i(R)/(2k)$.
Note that the binary tree has at most $2k-1$ internal nodes, each isolating one singleton, and at most $2k$ leaves. Thus, the nonempty terminal intervals and isolated singletons form a partition $\cP_i$ of $R$ into at most $4k-1$ blocks.

Let $B_i^*\in\argmax_{B\in\cP_i} \{c_i(B)\}$ be a maximum-cost block in $\cP_i$. 
We consider two cases depending on the structure of $B_i^*$.
If $B_i^*=\{e\}$ is a singleton, then since $\cP_i$ partitions $R$ into $|\cP_i|\le 4k-1$ blocks and $c_i(R)>0$, we have
$$
    c_i(e)=c_i(B_i^*)\ge\frac{c_i(R)}{|\cP_i|}\ge\frac{c_i(R)}{4k-1}>\frac{c_i(R)}{4k}.
$$
Otherwise, $B_i^*$ is a terminal interval at depth $\log_2(2k)$.
Therefore, for every $B\in\cP_i$, we have
$$c_i(B)\le c_i(B_i^*)\le \frac{c_i(R)}{2k}.$$

Suppose that the algorithm finishes the loop over all agents without returning a heavy item.
Then, for every agent $i\in N$, the partition $\cP_i$ divides the ordered set $R$ into at most $4k-1$ contiguous blocks, each of cost at most $c_i(R)/(2k)$.
Since each $\cP_i$ has at most $4k-2$ cut positions along $R$, merging the cut positions across all $n$ agents yields a common partition $\cP$ with at most $n(4k-2)+1=O(nk)$ blocks (see Figure~\ref{fig:common-partition}).
Moreover, every block $B\in\cP$ is contained in a block of
$\cP_i$ for each agent $i\in N$.
Hence, $c_i(B)\le c_i(R)/(2k)$ for all $i\in N$ and $B\in\cP$.

Finally, we analyze the query complexity.
For each agent $i\in N$, at most $2^\ell$ intervals are split at depth $\ell$, giving at most $2k-1$ splits in total.
Furthermore, each split requires $O(\log m)$ comparison queries by Lemma~\ref{lem:halving}, while finding $B_i^*$ takes $O(k)$ additional comparisons.
Since merging the cut positions into $\cP$ requires no further queries, the total number of comparison queries across all $n$ agents is $O(nk\log m) = O(n^6\log m)$.
\end{proof}

To construct blocks that are small for every agent, we repeatedly remove heavy chores while reusing the splitting trees across removals.
Algorithm~\ref{alg:heavy-items} carries out this process, maintaining the remaining pool $R$, initially $M$, and a set $E$ of set-aside heavy chores, initially empty.

\paragraph{Reusing Splitting Trees.}
Note that after each removal, the thresholds $c_i(R)/k$ may decrease, so previously small blocks may no longer satisfy these bounds.
A natural approach is to rebuild the splitting trees after each removal and search again for a heavy chore, which requires $O(nk\log k)$ rounds of reconstruction.
To avoid this overhead, we organize the process into phases and denote by $S$ the remaining pool at the start of each phase.
Although our target is to keep each block's cost within $c_i(R)/k$, the splitting depth in Algorithm~\ref{alg:detector} ensures that every terminal interval of agent $i$ initially costs at most $c_i(S)/(2k)$.
This extra factor of two provides slack to maintain the target bound $c_i(R)/k$ as $R$ shrinks.
Since deletions never increase block costs, this guarantee remains valid as long as $c_i(R)>c_i(S)/2$, while the isolated singletons continue to be checked for heavy chores.
A new phase starts only when some agent's remaining cost drops to at most $c_i(S) / 2$.
Together with the early-termination checks, this bounds the total number of phases by $O(n(n+\log k))$, with each agent's splitting tree built at most once per phase (Lemma~\ref{lem:heavy-items}).

\medskip

\begin{figure}[htbp]
\centering
\begin{tikzpicture}[
    x=1cm,y=1cm,font=\small,
    interval/.style={draw=diagramblue,fill=diagramblue!6,minimum width=1.1cm,minimum height=0.55cm,inner sep=2pt},
    updated/.style={interval,draw=diagramamber,fill=diagramamber!10,line width=0.9pt},
    item/.style={circle,draw=black!55,fill=black!4,minimum size=0.5cm,inner sep=1pt},
    branch/.style={draw=diagramblue,line width=0.7pt},
    isolated/.style={draw=black!45,densely dashed,line width=0.7pt},
    update/.style={-{Stealth[length=1.8mm]},draw=diagramamber,line width=1.1pt}
]
\node[updated] (root) at (10,-0.25) {$R$};
\node[updated] (left) at (7.6,-1.55) {$A_1$};
\node[interval] (right) at (12.4,-1.55) {$A_2$};
\node[item] (middle) at (10,-1.55) {$e_1$};
\node[updated,minimum width=1.45cm] (affected) at (6.25,-3) {$B\setminus\{e\}$};
\node[interval] (lr) at (8.95,-3) {$A_3$};
\node[interval] (rl) at (11.05,-3) {$A_4$};
\node[interval] (rr) at (13.75,-3) {$A_5$};
\node[item] (li) at (7.6,-3) {$e_2$};
\node[item] (ri) at (12.4,-3) {$e_3$};
\foreach \u/\v in {root/right,left/lr,right/rl,right/rr} {
    \draw[branch] (\u) -- (\v);
}
\foreach \u/\v in {root/middle,left/li,right/ri} {
    \draw[isolated] (\u) -- (\v);
}
\draw[update] (affected) -- (left);
\draw[update] (left) -- (root);
\end{tikzpicture}
\caption{Updating a constructed splitting tree when the removed chore $e$ lies in a terminal block $B$. Amber arrows indicate the bottom-up update order. At each ancestor, comparisons select the highest-cost block among its surviving singleton and the records of its nonempty child subtrees, while records outside this path remain unchanged. Gray circles denote isolated singletons.}
\label{fig:tree-update}
\end{figure}
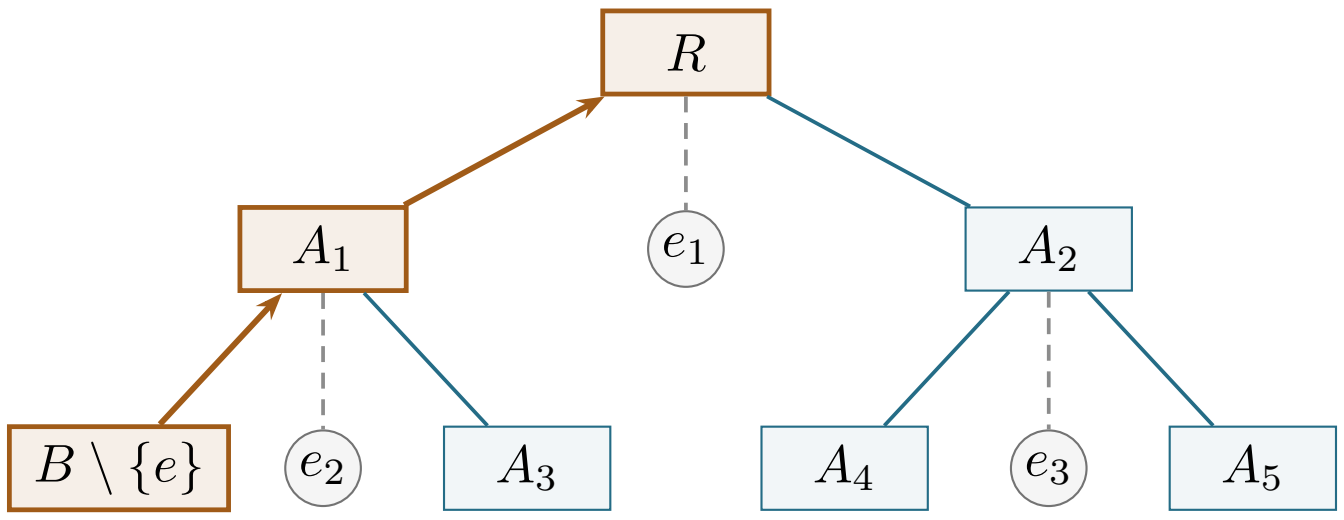

Within each phase, we build each agent $i$'s splitting tree $\cT_i$ on the current pool $R$ only when first needed, following the splitting rule of Algorithm~\ref{alg:detector}.
The nonempty surviving blocks of $\cT_i$ form a partition $\cP_i$ of $R$ with $|\cP_i|\le 4k-1$.
To support dynamic updates, each node in $\cT_i$ tracks the highest-cost surviving block in its subtree.
Specifically, a terminal node stores its surviving interval, whereas an internal split node compares its surviving singleton against the records of its nonempty child subtrees, omitting any empty blocks and subtrees.
Whenever a chore $e$ is removed during the phase, we simply delete it from its containing block in each constructed tree and update the subtree records along the path from the affected node to the root (see Figure~\ref{fig:tree-update}).

As long as $c_i(R)>c_i(S)/2$ for every $i\in N$, each terminal interval $B$ satisfies
$$
    c_i(B\cap R)\le\frac{c_i(S)}{2k}<\frac{c_i(R)}{k}.
$$
Since $R\subseteq S$, we can check the phase condition using a single comparison query per agent:
$$
    c_i(R)\le\frac{c_i(S)}{2} \quad\Longleftrightarrow\quad c_i(R)\le c_i(S\setminus R) \quad\Longleftrightarrow\quad \Comp_i(R,S\setminus R)=R.
$$
Whenever this condition holds for some agent, the algorithm resets $S$ to $R$, discards the existing trees, and starts a new phase.
Otherwise, for each agent $i\in N$, the algorithm builds $\cT_i$ if it is not yet constructed and retrieves the highest-cost surviving block $B_i^*$ recorded at the root of $\cT_i$.

If $B_i^*=\{e\}$ is a singleton, averaging over the at most $4k-1$ blocks in $\cP_i$ yields
$$
    c_i(e)\ge\frac{c_i(R)}{4k-1}>\frac{c_i(R)}{4k},
$$
so $e$ is a heavy chore and the algorithm moves $e$ from $R$ to $E$.
Conversely, if $B_i^*$ is a terminal interval, then every block in $\cP_i$ has cost below $c_i(R)/k$.
Once this holds for all agents, we merge their current cut positions to obtain a common partition $\cP$ of $R$ into at most $n(4k-2)+1$ blocks.

As $R$ shrinks, the heavy-chore thresholds $c_i(R)/(4k)$ may decrease, so repeated removal alone does not bound the number of removals independently of $m$.
We therefore maintain, for each agent $i\in N$, an ordinary array $E_i^*$ of length at most $n-1$ storing her most costly removed chores in $E$, together with a minimum-cost chore $e_i^{\min}\in E_i^*$.
Whenever a chore enters $E$, updating $E_i^*$ and $e_i^{\min}$ uses $O(n)$ comparisons per agent, or $O(n^2)$ in total.
Whenever $|R|>n$, Algorithm~\ref{alg:heavy-items} performs the following checks for each agent $i\in N$ before checking the phase condition or inspecting the trees:
\begin{enumerate}[label=(\arabic*)]
    \item If $c_i(R)=0$, as checked by $\Comp_i(R,\emptyset)=R$, we assign $R$ to agent $i$ and $\emptyset$ to every other agent, and return this allocation together with $E$.
    \item Otherwise, if $|E_i^*|=n-1$ and $\Comp_i(R,\{e_i^{\min}\})=R$, then $c_i(R)\le c_i(e_i^{\min})\le c_i(e)$ for every $e\in E_i^*$.
    We assign $R$ to agent $i$ and one singleton from $E_i^*$ to every other agent, remove $E_i^*$ from $E$, and return the resulting allocation of $M\setminus E$ together with the updated set $E$.
\end{enumerate}
Note that both cases return an EF1 partial allocation of $M\setminus E$.
If neither check terminates the algorithm, then $c_i(R)>0$ for every agent $i\in N$.
If the loop ends with $|R|\le n$, we assign at most one chore from $R$ to each agent and return this allocation together with $E$.
Whenever the algorithm continues after an agent has triggered $n-1$ removals, the check on $E_i^*$ forces her remaining cost to exceed the least cost of those chores, limiting the total number of removals to $O(nk\log k)$ and the number of phase resets to $O(n(n+\log k))$.

\begin{algorithm}[!htb]
\caption{Collection of Heavy Chores}
\label{alg:heavy-items}
\KwIn{Instance $\mathcal{I}=(N,M,\bc)$}
\KwOut{Heavy chores $E$, and either an EF1 allocation $\bX$ or a partition $\cP$ of $M\setminus E$}
$k\gets$ smallest power of two at least $16n^5$\;
Initialize $R\gets M$, $S\gets M$, $E\gets\emptyset$, and $X_i\gets\emptyset, E_i^*\gets\emptyset, \cT_i\gets\emptyset$ for all $i\in N$\;
\While{$|R|>n$}{
    \For{$i\in N$}{
        \lIf{$\Comp_i(R,\emptyset)=R$}{$X_i\gets R$; \Return $(E,\bX)$}
        \If{$|E_i^*|=n-1$ \textbf{and} $\Comp_i(R,\{e_i^{\min}\})=R$}{
            $X_i\gets R$; assign one chore from $E_i^*$ to each agent in $N\setminus\{i\}$\;
            \Return $(E\setminus E_i^*,\bX)$\;
        }
    }
    \lIf{$\Comp_i(R,S\setminus R)=R$ for some $i\in N$}{$S\gets R$; $\cT_j\gets\emptyset$ for all $j\in N$}
    $f\gets\mathrm{false}$\;
    \For{$i\in N$}{
        \If{$\cT_i=\emptyset$}{
            Build $\cT_i$ on $R$ to depth $\log_2(2k)$ via Algorithm~\ref{alg:detector}, initializing $\cP_i$ and subtree records\;
        }
        Let $B_i^*$ be the maximum-cost block in $\cP_i$ at the root of $\cT_i$\;
        \If{$B_i^*=\{e\}$}{
            $R\gets R\setminus\{e\}$, $E\gets E\cup\{e\}$, $f\gets\mathrm{true}$\;
            Delete $e$ from each built $\cT_j$ and $\cP_j$, and update records along the path to the root\;
            For each $j\in N$, update $E_j^*$ and $e_j^{\min}$ to maintain her $\le n-1$ costliest chores in $E$\;
            \textbf{break}\;
        }
    }
    \If{\textbf{not} $f$}{
        $\cP\gets$ partition of $R$ formed by merging the cut positions of $\cP_1,\dots,\cP_n$\;
        \Return $(E,\cP)$\;
    }
}
Assign at most one chore from $R$ to each agent in $N$\;
\Return $(E,\bX)$\;
\end{algorithm}

\begin{samepage}
\begin{lemma}\label{lem:heavy-items}
Algorithm~\ref{alg:heavy-items} uses $O(n^8 \log m)$
comparison queries and returns a heavy item set $E\subseteq M$ of $O(nk\log k)$ chores together with one of the following for $M'=M\setminus E$:
\begin{itemize}
    \item an EF1 allocation $\bX$ of $M'$, or
    \item a partition $\cP$ of $M'$ into $O(nk)$ blocks such that $c_i(B)\le c_i(M')/k$ for all $i\in N$ and $B\in\cP$.
\end{itemize}
\end{lemma}
\end{samepage}
\begin{proof}
We first bound the number of chores added to $E$ before termination.
In each iteration of the \textbf{while} loop, Algorithm~\ref{alg:heavy-items} first performs the early-termination checks.
If neither check terminates the algorithm, it checks the phase condition and inspects the splitting trees.
If no maximum-cost block is a singleton, the collection procedure returns a partition; otherwise, it moves a singleton $e$ maximizing some agent $i$'s block cost from $R$ into $E$.
For each agent $i\in N$, we denote by $E_i$ the set of chores whose removal is triggered by agent $i$.
Suppose $|E_i|\ge n-1$, and let $e_1,\dots,e_{n-1}$ be the first $n-1$ chores added to $E_i$.
Let $e^*\in\{e_1,\dots,e_{n-1}\}$ be a minimum-cost chore among them according to agent $i$, and let $R^*$ denote the remaining pool immediately before $e^*$ is removed.
By the singleton bound above, we have $c_i(e^*)>c_i(R^*)/(4k)$, and thus $c_i(R^*)<4k\cdot c_i(e^*)$.
After $e_1,\dots,e_{n-1}$ have been removed, the remaining pool is a subset of $R^*$.
Before every subsequent removal triggered by agent $i$, these $n-1$ chores are already in $E$, and the early-termination check on $E_i^*$ has not halted the algorithm.
Thus, $c_i(R)>c_i(e_i^{\min})\ge c_i(e^*)$, since $E_i^*$ contains her $n-1$ most costly chores in $E$.
By the same singleton bound, each such removal leaves cost less than $(1-1/(4k))c_i(R)$, while removals triggered by other agents can only further decrease the cost.
Consequently, immediately before the $q$-th subsequent removal triggered by agent $i$, where $q\ge1$, the remaining pool satisfies
$$
    c_i(e^*)<c_i(R)<4k\cdot c_i(e^*)\left(1-\frac{1}{4k}\right)^{q-1}\le 4k\cdot c_i(e^*)\,e^{-(q-1)/(4k)}.
$$
It follows that $q-1<4k\ln(4k)$, so there are $O(k\log k)$ such removals.
Hence, $|E_i|=O(n+k\log k)$ for every agent $i$.
Since $k\ge n$, the total number of removals is $\sum_{i\in N}|E_i|=O(nk\log k)$.
In particular, $|E|=O(nk\log k)$ throughout the algorithm.

We next verify the two possible outputs of Algorithm~\ref{alg:heavy-items}.
\begin{itemize}
    \item \textbf{Case 1: The algorithm returns an allocation $(E,\bX)$.} 
    If $c_i(R)=0$ for some $i\in N$ (assigning $R$ to agent $i$ and $\emptyset$ to all others) or $|R|\le n$ (assigning at most one chore to each agent), $\bX$ is trivially EF1 on $M'$. 
    Otherwise, the maintained set $E_i^*\subseteq E$ contains $n-1$ chores satisfying $c_i(R)\le c_i(e)$ for every $e\in E_i^*$, for some agent $i$.
    The algorithm then removes $E_i^*$ from $E$, assigns $R$ to agent $i$, and gives one chore from $E_i^*$ to each other agent.
    With $E$ denoting the updated set, the allocated chores are exactly $R\cup E_i^*=M\setminus E=M'$.
    Since agent $i$ does not envy any other agent and every other agent receives a single chore, $\bX$ is an EF1 allocation of $M'$.

    \item \textbf{Case 2: The algorithm returns a partition $(E,\cP)$.}
    In this case, the early-termination and phase checks ensure $c_i(R)>c_i(S)/2>0$ for every $i\in N$.
    Each tree $\cT_i$ was built on a subset of $S$ to depth $\log_2(2k)$, so Lemma~\ref{lem:halving} implies that every terminal interval $B$ satisfies $c_i(B\cap R)\le c_i(S)/(2k)<c_i(R)/k$.
    Since no agent's maximum-cost block $B_i^*$ is a singleton, $B_i^*$ is a terminal interval for every $i\in N$, and hence every block in $\cP_i$ has cost at most $c_i(R)/k$.
    Deleting chores preserves the contiguity of each block in the remaining item order, so merging the cut positions across $\cP_1,\dots,\cP_n$ as in Lemma~\ref{lem:detector} produces a common partition $\cP$ of $M'=R$ into at most $n(4k-2)+1$ blocks with $c_i(B)\le c_i(M')/k$ for all $i\in N$ and $B\in\cP$.
    Moreover, since $\sum_{B\in\cP}c_i(B)=c_i(M')>0$, this upper bound implies $|\cP|\ge k$.
\end{itemize}

We now bound the number of phases.
For the remaining complexity analysis, $E$ denotes the collected set before any chores are taken from it for the final partial allocation.
If $E=\emptyset$, no phase reset occurs; otherwise, $|E|\ge1$.
We charge each phase reset to one agent satisfying the halving condition.
For a fixed agent $i\in N$, the remaining pools at the resets charged to her are denoted by $R_1,\dots,R_t$, with $R_0=M$.
Other agents may trigger resets between two of her resets, but
$$
    c_i(R_\ell)\le\frac{1}{2}c_i(R_{\ell-1})
    \qquad\text{for all }\ell\in[t].
$$
Indeed, her remaining cost is at most half its value on the current phase snapshot $S$, which is contained in $R_{\ell-1}$.
We write $h=\ceil{\log_2 |E|}$.
For every $\ell\le t-h$, we have
$$
    c_i(R_{\ell-1}\setminus R_\ell)
    =c_i(R_{\ell-1})-c_i(R_\ell)
    \ge c_i(R_\ell)
    \ge 2^{t-\ell}c_i(R_t)
    \ge |E| \cdot c_i(R_t).
$$
Each batch $R_{\ell-1}\setminus R_\ell$ contains at most $|E|$ chores, so it contains a chore costing agent $i$ at least $c_i(R_t)$.
The batches are disjoint, so these chores are distinct.
If $t-h\ge n-1$, then by her $t$-th reset, $n-1$ distinct chores costing her at least $c_i(R_t)$ have already been collected, so the early-termination check on $E_i^*$ would terminate the algorithm first, a contradiction.
Therefore $t\le n-2+\ceil{\log_2 |E|}$, and across all $n$ agents there are $O(n(n+\log |E|))=O(n(n+\log k))$ phase resets.

Finally, we bound the number of comparison queries.
When a new chore enters $E$, updating $E_i^*$ and $e_i^{\min}$ takes $O(n)$ comparisons per agent, so maintaining these records across all insertions requires $O(n^2 |E|)$ queries in total.
Furthermore, each round uses $O(n)$ additional comparisons to decide whether to stop early or start a new phase.
After each deletion, updating the maximum-block records along the root-to-leaf path in each constructed splitting tree takes $O(\log k)$ comparisons per agent, totaling $O(n|E|\log k)$ queries.
Moreover, each phase constructs at most one tree of depth $\log_2(2k)$ per agent, which requires $O(nk\log m)$ queries by Lemma~\ref{lem:detector} (including the initialization of subtree records).
Thus, the total number of queries is
$$
    O\bigl(n^2k(n+\log k)\log m+n^2|E|+n|E|\log k\bigr)=O(n^8\log m),
$$
where the equality follows from $|E|=O(nk\log k)$, $k=\Theta(n^5)$, and $m>n\ge3$.
\end{proof}

\FloatBarrier
\subsection{The Construction of Balanced Partitions}\label{subsec:balanced-construction}
In this subsection, we combine the small blocks from Section~\ref{subsec:compression} into $n-1$ ordered bundles using $\mathrm{poly}(n)$ comparison queries.
Consider the case in which Algorithm~\ref{alg:heavy-items} returns a partition $\cP$ of the remaining chores $M'=M\setminus E$.
By Lemma~\ref{lem:heavy-items}, we have $k\le|\cP|\le n(4k-2)+1$, $c_i(M')>0$ for every agent $i\in N$, and $c_i(B)\le c_i(M')/k$ for every agent $i\in N$ and block $B\in\cP$.

Our goal is to partition the blocks of $\cP$ into $n-1$ ordered bundles $(P_1,\dots,P_{n-1})$ such that, for every agent $i\in N$, each bundle has cost greater than her proportional share $c_i(M')/n$ even after any one of its chores is removed.
Because every chore in $M'$ belongs to a block of cost at most $c_i(M')/k$, it suffices to ensure that $c_i(P_j)>(1/n+1/k)\cdot c_i(M')$ and $|P_j|\ge2$ for all $i\in N$ and $j\in[n-1]$.
We refer to such a partition $(P_1,\dots,P_{n-1})$ as a \emph{balanced partition} of $M'$.
This guarantee will be used to establish the preference transition in Section~\ref{subsec:balance}.

\paragraph{From Existence to Computation.}
If each block is assigned independently and uniformly at random to one of the $n-1$ bundles, the expected cost of each bundle for agent $i$ is $c_i(M')/(n-1)$, which exceeds her proportional share $c_i(M')/n$.
Because every block has small cost for every agent, a variance bound shows that there exists a partition in which all bundle costs stay sufficiently close to their expectations to satisfy the required lower bounds simultaneously (Lemma~\ref{lem:balanced-partition}).
However, turning this existence guarantee into an efficient comparison-based algorithm faces two challenges: an algorithm cannot directly test numerical cost thresholds using comparison queries, and enumerating all possible partitions of $\cP$ is too expensive.
To address the first challenge, Algorithm~\ref{alg:balanced-partition} constructs a reference bundle for each agent by greedily assigning each block to a minimum-cost bundle among $2n-1$ initially empty bundles and taking the union of the two most costly bundles.
The small-block bound ensures that the cost of this reference bundle lies strictly between the target lower bound and the stronger lower bound guaranteed by the variance argument.
We accept a candidate partition if every agent finds each candidate bundle at least as costly as her reference bundle.
The two bounds ensure that every accepted candidate is balanced and that any partition satisfying the stronger lower bound is accepted.
To address the second challenge, we rely on the fact that the variance bound requires only pairwise independence across blocks rather than full independence.
This allows us to construct an explicit family of polynomially many candidate partitions that contains at least one partition passing all reference-bundle comparisons.
Algorithm~\ref{alg:balanced-partition} selects candidates from this family using auxiliary costs before making reference-bundle comparisons.
Each failed comparison updates these costs and decreases their squared distance to the normalized true costs, limiting the total to $O(n^8)$ comparison queries (Lemma~\ref{lem:construct-balanced-partition}).
\medskip

We now give the existence proof and the deterministic construction.

\begin{lemma}\label{lem:balanced-partition}
If every block $B\in\cP$ satisfies $c_i(B)\le c_i(M')/k$ for all $i\in N$, then the blocks in $\cP$ can be partitioned into $n-1$ bundles $(P_1,\dots,P_{n-1})$, with $|P_j|\ge 2$ for every $j\in[n-1]$, such that
$$
    c_i(P_j)>\left(\frac{1}{n}+\frac{1}{k}\right) \cdot c_i(M')
    \qquad\text{for all } i\in N \text{ and } j\in[n-1].
$$
\end{lemma}
\begin{proof}
To prove the existence of such a partition, we assign each block $B\in\cP$ independently and uniformly at random to one of the $n-1$ bundles $P_1,\dots,P_{n-1}$.
For a fixed agent $i\in N$ and bundle $j\in[n-1]$, the indicator $I_{B,j}\in\{0,1\}$ records whether block $B$ is assigned to $P_j$.
Thus, $\Pr[I_{B,j}=1]=1/(n-1)$ and $c_i(P_j)=\sum_{B\in\cP} c_i(B) I_{B,j}$.

By linearity of expectation, the expected cost is
$$
    \mathbb{E}[c_i(P_j)]=\frac{1}{n-1}\sum_{B\in\cP}c_i(B)=\frac{c_i(M')}{n-1}.
$$
The indicators are independent across blocks, so the variance satisfies
$$
    \operatorname{Var}(c_i(P_j))
    =\sum_{B\in\cP}c_i(B)^2\operatorname{Var}(I_{B,j})
    \le\frac{1}{n-1}\sum_{B\in\cP}c_i(B)^2
    \le\frac{c_i(M')}{(n-1)k}\sum_{B\in\cP}c_i(B)
    =\frac{c_i(M')^2}{(n-1)k},
$$
where the second inequality uses $c_i(B)\le c_i(M')/k$ for all $B\in\cP$.

By linearity of expectation, the total normalized squared deviation satisfies
$$
    \mathbb{E}\!\left[
        \sum_{i=1}^n\sum_{j=1}^{n-1}
        \left(\frac{c_i(P_j)}{c_i(M')}-\frac{1}{n-1}\right)^2
    \right]
    =\sum_{i=1}^n\sum_{j=1}^{n-1}\frac{\operatorname{Var}(c_i(P_j))}{c_i(M')^2}
    \le\frac{n}{k}\le\frac{1}{16n^4},
$$
where the last inequality follows from $k\ge 16n^5$.
Consequently, there exists a partition for which the sum of normalized squared deviations is at most $1/(16n^4)$.
Since every term in the sum is nonnegative, each squared deviation is at most $1/(16n^4)$, which gives
$$
    \left|\frac{c_i(P_j)}{c_i(M')}-\frac{1}{n-1}\right|\le\frac{1}{4n^2}
    \qquad\text{for all } i\in N \text{ and } j\in[n-1].
$$
Consequently, for every agent $i \in N$, the cost of bundle $P_j$ can be bounded by
$$
    c_i(P_j)\ge
    \left(\frac{1}{n-1}-\frac{1}{4n^2}\right)c_i(M')
    >\left(\frac{1}{n}+\frac{1}{k}\right)c_i(M'),
$$
where the last inequality holds because $n\ge 3$ and $k\ge 16n^5$.
Finally, since every single chore $e\in P_j$ belongs to a block of cost at most $c_i(M')/k$, whereas $c_i(P_j)>c_i(M')/k$, every bundle $P_j$ must contain at least two chores, i.e., $|P_j|\ge 2$.
\end{proof}

We now turn the existence guarantee of Lemma~\ref{lem:balanced-partition} into the deterministic polynomial-query procedure in Algorithm~\ref{alg:balanced-partition}.
We index the blocks as $\cP=(B_1,\dots,B_{|\cP|})$.
For each agent $i\in N$, the algorithm starts with $2n-1$ empty bundles and processes the blocks in order, adding each block to a minimum-cost bundle for agent $i$ via comparisons.
The reference bundle $H_i$ is the union of the two highest-cost bundles.
During the greedy construction for agent $i\in N$, the difference between the maximum and minimum bundle costs among the $2n-1$ bundles is initially $0$, and adding a block $B\in\cP$ of cost at most $c_i(M')/k$ to a minimum-cost bundle preserves the invariant that all $2n-1$ bundle costs differ by at most $c_i(M')/k$.
At the end of the process, the average cost of the $2n-1$ bundles is $c_i(M')/(2n-1)$.

Thus, the two most costly bundles have an average cost of at least $c_i(M')/(2n-1)$, while each has cost at most $\bigl(\frac{1}{2n-1}+\frac{1}{k}\bigr) \cdot c_i(M')$.
Consequently, their union $H_i$ satisfies
$$
    \frac{2}{2n-1} \cdot c_i(M')\le c_i(H_i)\le\left(\frac{2}{2n-1}+\frac{2}{k}\right) \cdot c_i(M').
$$
Since $n\ge 3$ and $k\ge 16n^5$, we have 
$$\frac{2}{2n-1}-\frac{1}{n}=\frac{1}{n(2n-1)}>\frac{1}{k} \text{,} \qquad
\frac{1}{n-1}-\frac{2}{2n-1}-\frac{1}{4n^2}>\frac{1}{4n^2}>\frac{2}{k}.
$$
Consequently, for every agent $i\in N$, the reference bundle $H_i$ satisfies
$$
    \left(\frac{1}{n}+\frac{1}{k}\right) \cdot c_i(M')
    <c_i(H_i)
    <\left(\frac{1}{n-1}-\frac{1}{4n^2}\right) \cdot c_i(M').
$$
In particular, any candidate partition $(P_1,\dots,P_{n-1})$ with $c_i(P_j)\ge c_i(H_i)$ for all $i\in N$ and $j\in[n-1]$ satisfies $c_i(P_j)>(1/n+1/k)c_i(M')$ and hence $|P_j|\ge 2$, as required by Lemma~\ref{lem:balanced-partition}.

We choose a prime $p$ satisfying $|\cP|\le p\le 2|\cP|$, which exists by Bertrand's postulate~\cite{books/daglib/0019107}.
For each pair $(a,b)\in\{0,\dots,p-1\}^2$, block $B_t\in\cP$ is assigned to bundle $P_j$, where
$$
    j=1+\bigl(((at+b)\bmod p)\bmod(n-1)\bigr).
$$
This family yields $O(|\cP|^2)$ candidate partitions, and the bundle assignments of distinct blocks are pairwise independent when $(a,b)$ is chosen uniformly at random~\cite{journals/fttcs/Vadhan12}.
Before performing any comparison queries, the algorithm selects a candidate partition using auxiliary cost functions $\hat c_i$.
For each agent $i\in N$, the function $\hat c_i$ is additive over the blocks in $\cP$ and satisfies
$$
    0\le\hat c_i(B)\le\frac{1}{k}\quad(\forall B\in\cP),
    \qquad \hat c_i(M')=1,
    \qquad \hat c_i(H_i)\le\frac{2}{2n-1}+\frac{2}{k}.
$$
Since the normalized true costs $c_i(B)/c_i(M')$ satisfy the auxiliary-cost constraints, a feasible choice of $\hat c_i$ is guaranteed to exist.
Moreover, the algorithm maintains these auxiliary costs without direct access to the true costs.
We set $\gamma=1/(8n^2)$.
Each round selects a candidate satisfying
$$
    \hat c_i(P_j)\ge\frac1{n-1}-\frac1{4n^2}
    \qquad\text{for all }i\in N\text{ and }j\in[n-1].
$$
The pairwise-independence argument below guarantees that the family contains such a candidate.
Together with the auxiliary-cost constraints, this gives
$$
    \hat c_i(P_j)-\hat c_i(H_i)
    \ge\frac1{(n-1)(2n-1)}-\frac1{4n^2}-\frac2k
    >\gamma.
$$
If a comparison reveals $c_i(P_j)<c_i(H_i)$, the auxiliary costs therefore disagree with that comparison by at least $\gamma$.
For this pair $(i,j)$, we first adjust each block's auxiliary cost in one of three ways:
\begin{itemize}
    \item If $B\subseteq P_j\setminus H_i$, its auxiliary cost decreases by $\gamma/|\cP|$.
    \item If $B\subseteq H_i\setminus P_j$, its auxiliary cost increases by $\gamma/|\cP|$.
    \item Otherwise, its auxiliary cost remains unchanged.
\end{itemize}
We then replace these tentative costs by the nearest feasible costs in squared Euclidean distance, subject to the auxiliary-cost constraints.
All other agents' auxiliary costs remain unchanged.

\begin{lemma}\label{lem:auxiliary-update}
For each agent, the update above occurs at most $O(n^4|\cP|/k)$ times.
Each update can be computed exactly in polynomial time without comparison queries.
\end{lemma}
\begin{proof}
We fix an agent $i$ and write $\hat{\bc}_i=(\hat c_i(B))_{B\in\cP}$ and $\bc_i=(c_i(B)/c_i(M'))_{B\in\cP}$ for the auxiliary and normalized true cost vectors.
For a rejected comparison, we have
$$
    \hat c_i(P_j)-\hat c_i(H_i)
    -\frac{c_i(P_j)-c_i(H_i)}{c_i(M')}\ge\gamma.
$$
We write $\by$ for the tentative cost vector after these changes.
In the expansion of $\|\by-\bc_i\|^2-\|\hat{\bc}_i-\bc_i\|^2$, the linear term is $-2\gamma/|\cP|$ times the left-hand side above, since blocks shared by $P_j$ and $H_i$ have unchanged costs.
The quadratic term is at most $\gamma^2/|\cP|$, because at most $|\cP|$ coordinates change, each by $\gamma/|\cP|$.
The feasible set defined by the auxiliary-cost constraints is closed and convex and contains $\bc_i$.
Moving a point to its nearest feasible point cannot increase its distance to $\bc_i$.
Indeed, if $\bx$ is the nearest feasible point to $\by$, minimizing squared distance along the segment from $\bx$ to $\bc_i$ gives $\langle\by-\bx,\bc_i-\bx\rangle\le0$.
Expanding squared distances then gives $\|\bx-\bc_i\|^2\le\|\by-\bc_i\|^2$.
Consequently, each update satisfies
$$
\begin{aligned}
    \|\hat{\bc}_i^{\,\mathrm{new}}-\bc_i\|^2
    &\le\|\by-\bc_i\|^2 \le\|\hat{\bc}_i-\bc_i\|^2-\frac{\gamma^2}{|\cP|}.
\end{aligned}
$$
Both cost vectors are nonnegative, sum to $1$, and have entries at most $1/k$.
Their initial squared distance is therefore at most $2/k$.
Since squared distance is nonnegative, there are at most $2|\cP|/(k\gamma^2)=O(n^4|\cP|/k)$ updates for agent $i$.

For completeness, the nearest feasible costs can be computed using only sorting and rational arithmetic.
Given target values $y_B$, the nearest vector with entries in $[0,1/k]$ summing to $1$ has entries $\min\{1/k,\max\{0,y_B-\lambda\}\}$, where $\lambda$ makes their sum $1$.
Sorting the breakpoints $y_B$ and $y_B-1/k$ determines $\lambda$ by a linear equation on the appropriate interval.
If this vector satisfies the bound on $H_i$, it is already feasible.
Otherwise, that bound is tight at the nearest feasible vector, and the same computation is performed separately on the blocks in $H_i$ and outside $H_i$, with prescribed sums $2/(2n-1)+2/k$ and $1-2/(2n-1)-2/k$.
A feasible initialization is obtained in the same way from the uniform vector.
Each computation uses rational additions and divisions by integers at most $|\cP|$.
Together with the rational constants in the auxiliary-cost constraints and the update step, this increases the common-denominator bit length by $O(\log(nk\cdot |\cP|))$ per update.
\end{proof}

\begin{algorithm}[!htb]
\caption{Balanced Partition}
\label{alg:balanced-partition}
\KwIn{Block partition $\cP=(B_1,\dots,B_{|\cP|})$ of $M'$ returned by Algorithm~\ref{alg:heavy-items}}
\KwOut{Balanced ordered bundles $(P_1,\dots,P_{n-1})$}
\For{$i\in N$}{
    Initialize $2n-1$ empty bundles and process $B_1,\dots,B_{|\cP|}$ in order, adding each block to a minimum-cost bundle for agent $i$\;
    $H_i\gets$ the union of the two highest-cost bundles for agent $i$\;
    Initialize feasible auxiliary costs $\hat c_i$ satisfying the auxiliary-cost constraints\;
}
$p\gets$ a prime satisfying $|\cP|\le p\le2|\cP|$\;
\While{true}{
    \For{$(a,b)\in\{0,\dots,p-1\}^2$}{
        \For{$j\in[n-1]$}{
            $P_j\gets\bigcup\{B_t:t\in[|\cP|],\ 1+(((at+b)\bmod p)\bmod(n-1))=j\}$\;
        }
        \If{$\hat c_i(P_j)\ge1/(n-1)-1/(4n^2)$ for all $i\in N$, $j\in[n-1]$}{
            Stop enumerating and keep this candidate\;
        }
    }
    Compare $H_i$ with $P_j$ for all $i\in N$ and $j\in[n-1]$, stopping at the first failed comparison\;
    \eIf{all comparisons return $H_i$}{
        Order each $P_j$ according to the fixed ordering of $M$ and \Return $(P_1,\dots,P_{n-1})$\;
    }{
        For the failed pair $(i,j)$, decrease $\hat c_i(B)$ by $\gamma/|\cP|$ for $B\subseteq P_j\setminus H_i$, increase it by $\gamma/|\cP|$ for $B\subseteq H_i\setminus P_j$, and leave all other block costs unchanged\;
        Replace these tentative costs by the nearest costs satisfying the auxiliary-cost constraints in squared Euclidean distance\;
    }
}
\end{algorithm}

\begin{lemma}\label{lem:construct-balanced-partition}
Algorithm~\ref{alg:balanced-partition} uses $O(n^8)$ comparison queries to partition the blocks in $\cP$ into $n-1$ bundles $(P_1,\dots,P_{n-1})$, with $|P_j|\ge2$ for every $j\in[n-1]$, such that
$$
    c_i(P_j)>\left(\frac1n+\frac1k\right) \cdot c_i(M')
    \qquad\text{for all }i\in N\text{ and }j\in[n-1].
$$
\end{lemma}
\begin{proof}
By the reference-bundle bound established above, any returned partition $(P_1,\dots,P_{n-1})$ satisfies $c_i(P_j)\ge c_i(H_i)>(1/n+1/k) \cdot c_i(M')$ for all $i\in N$ and $j\in[n-1]$.
Furthermore, since each individual chore has cost at most $c_i(M')/k$, every bundle $P_j$ must contain at least two chores, which implies $|P_j|\ge 2$.

We next show that each round identifies a candidate satisfying the auxiliary-cost threshold.
For the analysis, suppose that $(a,b)$ is chosen uniformly at random from $\{0,\dots,p-1\}^2$.
For any distinct $t,t'\in[|\cP|]$ and any $u,v\in\{0,\dots,p-1\}$, the system of congruences $at+b\equiv u\pmod p$ and $at'+b\equiv v\pmod p$ admits a unique solution $(a,b)$, since $p$ is prime and $0<|t-t'|<p$.
Thus, each residue $(at+b)\bmod p$ is uniformly distributed over $\{0,\dots,p-1\}$, and the residues for distinct blocks are pairwise independent.
Following Lemma~\ref{lem:balanced-partition}, let $I_{B,j}\in\{0,1\}$ be the indicator random variable for the event that block $B\in\cP$ is assigned to $P_j$.
Consequently, for each bundle $j\in[n-1]$, the indicators $\{I_{B,j}\}_{B\in\cP}$ are pairwise independent across blocks.
Moreover, since each bundle index $j$ has either $\lfloor p/(n-1)\rfloor$ or $\lceil p/(n-1)\rceil$ preimages among the $p$ residues, the assignment probabilities $p_j:=\Pr[I_{B,j}=1]$ satisfy
$$
    \sum_{j=1}^{n-1}p_j=1
    \qquad\text{and}\qquad
    \Bigl|p_j-\frac{1}{n-1}\Bigr|\le\frac{1}{p}\le\frac{1}{k},
$$
where the last inequality follows from $p\ge |\cP|\ge k$.
By the auxiliary-cost constraints, for every agent $i\in N$ and bundle $j\in[n-1]$, we have $\mathbb{E}[\hat c_i(P_j)]=p_j\hat c_i(M')=p_j$ and
\[
    \operatorname{Var}\bigl(\hat c_i(P_j)\bigr)
    =\sum_{B\in\cP}\hat c_i(B)^2\operatorname{Var}(I_{B,j})
    = p_j(1-p_j)\sum_{B\in\cP}\hat c_i(B)^2
    \le\frac{p_j(1-p_j)}{k}.
\]
Since $\sum_jp_j^2\ge1/(n-1)$, the expected total squared deviation satisfies
$$
\begin{aligned}
    \mathbb E\!\left[\sum_{i=1}^n\sum_{j=1}^{n-1}
        \left(\hat c_i(P_j)-\frac1{n-1}\right)^2\right]
    &\le\frac nk\left(1-\frac1{n-1}\right)+\frac{n(n-1)}{k^2} \le\frac nk\le\frac1{16n^4},
\end{aligned}
$$
where the final two inequalities use $k\ge16n^5\ge(n-1)^2$.
Hence, some candidate satisfies $|\hat c_i(P_j)-1/(n-1)|\le1/(4n^2)$ for all $i,j$, and the enumeration finds a candidate meeting every auxiliary-cost threshold.
This argument applies to the updated auxiliary costs in every round.

Finally, we bound the total query complexity.
Whenever a candidate is rejected, the algorithm updates the auxiliary costs of some agent.
By Lemma~\ref{lem:auxiliary-update}, the total number of rejected candidates across all agents is at most $O(n^5|\cP|/k)$.
Since verifying a candidate requires at most $n(n-1)$ comparisons, all candidate checks take $O(n^7|\cP|/k)$ queries in total.
Combined with the $O(n^2|\cP|)$ queries needed to construct the reference bundles, as well as $|\cP|=O(nk)$ and $k=\Theta(n^5)$, the total number of queries is $O(n^2|\cP|+ n^7|\cP|/k)=O(n^8)$.
\end{proof}

\section{Proof of Root Monotonicity}\label{app:root-monotonicity}

\begin{proof}[Proof of Lemma~\ref{lem:root-monotonicity-fractional}]
Let $G=G(\mathbf{x})$ and $G'=G(\mathbf{x}')$.
For each agent $i$, we let $$
\mu_i=\min_{j\in[n-1]}\Rank_i(P_j,x_j) \text{,} \qquad \mu_i'=\min_{j\in[n-1]}\Rank_i(P_j,x_j'),$$ 
and denote her unique inactive neighbors in $G$ and $G'$ by $p_i$ and $p_i'$, respectively.
These neighbors exist by Lemma~\ref{lem:robust-x-properties}.

We first make two observations about the interpolation.
If $\mu_i=\mu_i'$, then the inactive minima at the two points are represented by the same prefix, since distinct prefixes have distinct ranks for agent $i$.
In particular, $p_i=p_i'$.
Moreover, if $(i,j)$ is an active edge of $G$, then the fractional part of $x_j$ lies strictly inside $I_i$.
Otherwise, $\Rank_i(P_j,x_j)$ would equal the rank of a prefix of that bundle, contradicting its equality with the inactive minimum, which is represented by a prefix of a different bundle.
Thus, decreasing $x_j$ from an active minimum strictly decreases agent $i$'s rank for $P_j$, since the rank is strictly increasing inside the active interval and nondecreasing everywhere.

Suppose, for a contradiction, that $A=\{j\in[n-1]:x_j>x_j'\}$ is nonempty.
Since $x_r\le x_r'$, we have $r\notin A$.
Starting with $S=A$, repeatedly apply the following rule.
If an agent $i$ satisfies $\mu_i=\mu_i'$ and $p_i\in S$, add every coordinate $j$ such that $x_j=x_j'$ and $(i,j)$ is an active edge of $G$.
Stop when no further coordinate can be added.
Every newly added coordinate is unchanged between the two points.
Furthermore, the root cannot be added, because it is inactive for every agent.
Hence, $\varnothing\ne S\subseteq\{j:x_j\ge x_j'\}$ and $r\notin S$.

Let $\mathcal I=N_G(S)$ be the set of agents adjacent to $S$ in $G$.
We claim that every neighbor in $G'$ of every agent in $\mathcal I$ also belongs to $S$.
Fix $i\in\mathcal I$, and choose a neighbor $j\in S$ in $G$.
Since $x_j'\le x_j$,
\[
    \mu_i'
    \le \Rank_i(P_j,x_j')
    \le \Rank_i(P_j,x_j)
    =\mu_i.
\]
If $\mu_i'<\mu_i$, then for every $k\notin A$ we have $\Rank_i(P_k,x_k')\ge\Rank_i(P_k,x_k)\ge\mu_i>\mu_i'$.
Thus, all neighbors of $i$ in $G'$ lie in $A\subseteq S$.

It remains to consider $\mu_i'=\mu_i$.
We first show that $p_i\in S$.
If $(i,j)$ is inactive in $G$, then $j=p_i$.
Otherwise, $(i,j)$ is active.
In this case $j\notin A$, since decreasing its coordinate from $x_j$ to $x_j'$ would strictly decrease its rank and imply $\mu_i'<\mu_i$.
Therefore, $j\in S\setminus A$ and was added by the rule above.
The agent through whom it was added must be $i$, because the disjoint active intervals allow at most one active edge incident to bundle $j$ in $G$.
That rule required $p_i\in S$, as desired.

Now consider any neighbor $k$ of $i$ in $G'$.
If $(i,k)$ is inactive in $G'$, then $k=p_i'=p_i\in S$.
If $(i,k)$ is active in $G'$, we cannot have $x_k'>x_k$, since decreasing its coordinate from $x_k'$ to $x_k$ would give $\Rank_i(P_k,x_k)<\Rank_i(P_k,x_k')=\mu_i'=\mu_i$, contradicting the definition of $\mu_i$.
If $x_k>x_k'$, then $k\in A\subseteq S$.
Finally, if $x_k=x_k'$, then $(i,k)$ is also an edge of $G$, since its rank at $\bx$ equals $\mu_i'=\mu_i$.
The edge remains active because the coordinate $x_k$ is unchanged.
Since $\mu_i=\mu_i'$ and $p_i\in S$, the closure rule gives $k\in S$.
This proves the claim.

Robustness implies that every nonempty set of bundles has at least one more neighboring agent than its cardinality.
Indeed, after deleting a neighboring agent, the remaining agents must still be able to match all bundles in the set.
Applying this to $S$ in $G$ gives $|\mathcal I|=|N_G(S)|\ge |S|+1$.
On the other hand, $[n-1]\setminus S$ is nonempty because it contains the root.
By the claim, none of its neighbors in $G'$ belongs to $\mathcal I$.
Robustness of $\bx'$ therefore gives
\[
    n-|S|
    =|[n-1]\setminus S|+1
    \le |N_{G'}([n-1]\setminus S)|
    \le n-|\mathcal I|.
\]
Thus, $|\mathcal I|\le |S|$, a contradiction.
Consequently, $A=\varnothing$, and hence $x_j\le x_j'$ for every $j\in[n-1]$.
\end{proof}

\section{Proof of the Extension for a Fixed Root}\label{app:fixed-root-extension}

\begin{proof}[Proof of Lemma~\ref{lem:fixed-root-extension}]
Fix $r$ and $T$, and let $\alpha=T-\frac12$.
We keep the root coordinate fixed at $\alpha$.
Since every active interval lies in $(1/2,1)$, we have $\Rank_i(P_r^+,\alpha)=\Rank_i(P_r^+,T-1)$ for every agent $i\in N$.
For each nonroot bundle $j\ne r$, let $L_j=|P_j^+|$ and define $f_{ij}(z)=\Rank_i(P_j^+,z)-\Rank_i(P_r^+,\alpha)$ for $z\in[0,L_j]$.
Subtracting the fixed root rank preserves each agent's minimum-rank bundles and makes the root's rank zero.

We first show that
\begin{equation}\label{eq:fixed-root-endpoint-signs}
    f_{ij}(0)<0<f_{ij}(L_j)
    \qquad\text{for all }i\in N,\ j\ne r.
\end{equation}
Every labeled empty prefix precedes every nonempty prefix in the prefix order.
Since $T-1\ge1$, this gives $f_{ij}(0)<0$.
For the other inequality, Theorem~\ref{thm:(n-1)-partition} gives $c_i(P_j)>(1/n+1/k)c_i(M')>0$.
Thus, $c_i(P_j^+)=c_i(P_j)+c_i(M)>c_i(M)\ge c_i(\Pre(P_r,T-1))$, so $f_{ij}(L_j)>0$.

We next apply Lemma~\ref{lem:continuous-matching}, this time keeping the root coordinate fixed.
To construct the required selections on the simplex, we first extend each $f_{ij}$ from $[0,L_j]$ to $\bR$ by
\[
    \hat f_{ij}(z)=
    \begin{cases}
        f_{ij}(0)+z, & z<0,\\
        f_{ij}(z), & 0\le z\le L_j,\\
        f_{ij}(L_j)+z-L_j, & z>L_j.
    \end{cases}
\]
Each $\hat f_{ij}$ is continuous and nondecreasing.
Moreover, $\hat f_{ij}(z)-z$ is bounded, since it is constant on each of $(-\infty,0]$ and $[L_j,\infty)$ and continuous on $[0,L_j]$.
Let $\Delta=\{\by\in\bR_{\ge0}^{n-1}:\sum_{j=1}^{n-1}y_j=1\}$.
For an interior point $\by\in\Delta$, we associate each nonroot bundle $j$ with the coordinate $x_j(\by)=\ln(y_r/y_j)$.
To define selections on the entire simplex, set $W_{ir}(\by)=y_r$ and, for $j\ne r$, define
\[
    W_{ij}(\by)=
    \begin{cases}
        y_r\exp\!\left(-\hat f_{ij}\!\left(\ln\frac{y_r}{y_j}\right)\right),
        & y_r>0,\ y_j>0,\\[2mm]
        e^{-f_{ij}(0)}y_j,
        & y_r=0,\\[1mm]
        0,
        & y_r>0,\ y_j=0.
    \end{cases}
\]

We now verify that every $W_{ij}$ is continuous on $\Delta$.
Continuity is immediate when $y_r,y_j>0$.
Writing $z=\ln(y_r/y_j)$ gives $W_{ij}(\by)=y_j\exp\bigl(-(\hat f_{ij}(z)-z)\bigr)$.
Since $\hat f_{ij}(z)-z$ is bounded, this expression tends to zero as $y_j\to0$, including when $y_r$ also tends to zero.
If $y_r\to0$ while $y_j$ tends to a positive value, then $z\to-\infty$ and $\hat f_{ij}(z)-z=f_{ij}(0)$ for all sufficiently small $y_r$.
The expression therefore agrees in the limit with the boundary value $e^{-f_{ij}(0)}y_j$.
Hence, every $W_{ij}$ is continuous on $\Delta$.
The definitions also give $W_{ij}(\by)>0$ if and only if $y_j>0$.

At each $\by\in\Delta$, let agent $i$ select all indices maximizing $W_{ij}(\by)$.
These selections are nonempty and have closed graphs because the functions $W_{ij}$ are continuous.
An index with $y_j=0$ cannot be selected: its score is zero, whereas some positive coordinate has a strictly positive score.
Lemma~\ref{lem:continuous-matching} therefore gives a point $\by^*\in\Delta$ at which removing any single agent leaves a perfect matching between the remaining agents and the bundle indices.
Every bundle index must be selected by some agent at $\by^*$, so $y_j^*>0$ for every $j\in[n-1]$.
Set $x_r=\alpha$ and $x_j=\ln(y_r^*/y_j^*)$ for each $j\ne r$.
Since $y_r^*>0$, maximizing the scores at $\by^*$ is equivalent to minimizing the values $0$ for the root and $\hat f_{ij}(x_j)$ for the nonroot bundles.
Indeed, all scores have the common positive factor $y_r^*$, and $\exp(-z)$ is strictly decreasing.
Thus, the bipartite graph defined by these minima admits a perfect matching after any single agent is removed.

Finally, we show that $0<x_j<L_j$ for every $j\ne r$.
If $x_j\le0$, then $\hat f_{ij}(x_j)\le f_{ij}(0)<0$ for every agent $i$.
Since the root has value zero, it cannot attain any agent's minimum and would have no neighbor, contradicting the matching guarantee.
Similarly, if $x_j\ge L_j$, then $\hat f_{ij}(x_j)\ge f_{ij}(L_j)>0$ for every agent $i$.
Bundle $j$ would then have no neighbor, since the root has a strictly smaller value.
Therefore, all nonroot coordinates lie in $(0,L_j)$, where $\hat f_{ij}=f_{ij}$.
The graph obtained above is consequently the minimum-rank graph $G(\bx)$ in the extended system.
It admits a perfect matching after any single agent is removed, and $x_r=T-\frac12$ is a half-integer.
Hence, $\bx$ is robust.
\end{proof}

\section{Proof of Consistency in Two Systems}\label{sec:agreeing-in-two-systems}

\begin{proof}[Proof of Lemma~\ref{lem:agreeing-in-two-systems}]
By symmetry, it suffices to prove one direction.
Suppose that $(\bt,r)$ is consistent in the first system, and let $\bx$ be a robust vector witnessing this consistency, expressed in block coordinates.
The root coordinate $x_r$ is a half-integer, and Lemma~\ref{lem:robust-x-properties}(3) implies that every nonroot coordinate has fractional part in $(1/2,1)$.
Thus, the coordinatewise ceiling of $\bx$ corresponds to $\bt$, and each coordinate lies in the interior of the interpolation interval ending at $\Pre(P_j^+,t_j)$.
In particular, no $t_j$ is the first cut-point of its bundle.
By hypothesis, $t_j$ therefore has the same immediate predecessor $a_j$ in both systems.

By Lemma~\ref{lem:robust-x-properties}(2), each agent $i$ has a unique inactive edge in $G(\bx)$.
Let $Q_i$ denote the endpoint prefix corresponding to this edge.
It is one of the prefixes $\Pre(P_\ell^+,a_\ell)$ or $\Pre(P_\ell^+,t_\ell)$, and its rank is agent $i$'s minimum at $\bx$.
Now consider an active edge $(i,j)$.
The interpolated rank of bundle $j$ equals the rank of $Q_i$.
Since $Q_i$ belongs to a different bundle and prefixes of distinct bundles have distinct ranks, $Q_i$ lies strictly between $\Pre(P_j^+,a_j)$ and $\Pre(P_j^+,t_j)$ in agent $i$'s prefix order.
Both systems contain these endpoint prefixes and preserve their relative order, so the same strict ordering holds in the second system.

We construct $\bx'$ in the block coordinates of the second system.
Set $x'_r$ to the midpoint of the interpolation interval corresponding to $a_r$ and $t_r$.
For each nonroot bundle $j$, let $(i,j)$ be its unique active edge in $G(\bx)$, given by Lemma~\ref{lem:robust-x-properties}(3).
Choose $x'_j$ within agent $i$'s active interval in the interpolation interval corresponding to $a_j$ and $t_j$, so that her interpolated rank equals her rank of $Q_i$ in the second system.
Such a point exists uniquely in the interior of the active interval, because the interpolation is strictly increasing there and $Q_i$ lies strictly between the two endpoint prefixes.

We next show that the minimum-rank graph at $\bx'$ in the second system equals $G(\bx)$.
Fix an agent $i$ and a bundle $j$ such that $(i,j)$ is not an active edge of $G(\bx)$.
If $j=r$, both $x_j$ and $x'_j$ have fractional part $1/2$.
Otherwise, their fractional parts lie in the active interval of the same agent $i'\neq i$, namely the unique active neighbor of $j$ in $G(\bx)$.
Since the intervals $I_1,\dots,I_n$ are pairwise disjoint and lie in $(1/2,1)$, these fractional parts lie on the same side of $I_i$.
Thus, in each system, agent $i$'s interpolated rank at the corresponding point equals her rank of the same endpoint prefix.
The relative order of these endpoint prefixes is unchanged, so $Q_i$ remains the unique minimum-rank prefix among them.
At every coordinate $j$ where $(i,j)$ is an active edge of $G(\bx)$, our construction makes agent $i$'s interpolated rank equal to her rank of $Q_i$.
Thus, every agent has exactly the same minimum-rank bundles at $\bx'$ as at $\bx$, and the two minimum-rank graphs coincide.

The graph at $\bx'$ therefore admits a perfect matching after any single agent is removed.
Moreover, $x'_r$ is a half-integer, while every nonroot coordinate has fractional part in $(1/2,1)$.
Hence, $\bx'$ is robust, and its two nearest-integer roundings correspond to $\bt$ and the vector obtained by replacing $t_r$ with $a_r$, the cut-point immediately preceding $t_r$ in the second system.
Thus, $(\bt,r)$ is consistent in the second system.
\end{proof}

\section{Extension to Goods}\label{sec:goods}

This appendix presents the algorithm and proof of Theorem~\ref{thm:goods} for indivisible goods with nonnegative additive valuations.

\subsection{Definitions and Basic Lemmas}\label{subsec:goods-model}

We use the same notation for agents, items, allocations, and ordered bundles as in Section~\ref{sec:model}, where each agent $i\in N$ has an additive valuation function $v_i:2^M\to\bR_{\ge0}$ and prefers bundles of larger value.
We write $\bv=(v_1,\dots,v_n)$ for the valuation profile.
An allocation $\bX=(X_1,\dots,X_n)$ is envy-free for goods if $v_i(X_i)\ge v_i(X_j)$ for all $i,j\in N$.

\begin{definition}[EF1 for Goods]\label{def:goods-ef1}
An allocation $\bX=(X_1,\dots,X_n)$ is \emph{envy-free up to one item} (EF1) for goods if for all agents $i,j\in N$, either $X_j=\emptyset$, or there exists an item $e\in X_j$ such that $v_i(X_i)\ge v_i(X_j\setminus\{e\})$.
Any such $e$ is called a \emph{witness item} for agent $i$ with respect to agent $j$.
\end{definition}
The same definitions apply to partial allocations.

\begin{definition}[Comparison Model for Goods]\label{def:goods-comparison}
Algorithms access the valuation functions only through comparison queries.
For any agent $i\in N$ and bundles $X,Y\subseteq M$, the query $\Comp_i^{\mathrm{g}}(X,Y)$ returns the bundle of larger value, breaking ties in favor of the first argument:
$$
    \Comp_i^{\mathrm{g}}(X,Y)=
    \begin{cases}
        X, & \text{if }v_i(X)\ge v_i(Y),\\
        Y, & \text{if }v_i(X)<v_i(Y).
    \end{cases}
$$
\end{definition}
Note that each invocation of $\Comp_i^{\mathrm{g}}$ counts as one comparison query.
For a partial allocation $\bX=(X_1,\dots,X_n)$, each nonempty bundle $X_j$ has a designated witness item $e_j\in X_j$.

\begin{definition}[Feasible Allocation for Goods]\label{def:goods-feasible}
A partial allocation $\bX$ with these designated witness items is \emph{feasible} for goods if
$$
    v_i(X_i)\ge v_i(X_j\setminus\{e_j\})
    \qquad\text{for all }i,j\in N\text{ with }X_j\ne\emptyset.
$$
\end{definition}
This condition implies EF1 and holds for the empty allocation.
For each nonempty bundle $X_j$, the designated item $e_j$ serves as a witness item for all agents comparing their own bundles with $X_j$.

\begin{lemma}[\cite{conf/wine/BuLLST24}]\label{lem:goods-small}
Under nonnegative additive valuations, EF1 allocations for goods can be computed using $O(\log m)$ comparison queries when $n\leq2$.
\end{lemma}
When $m\le n$, assigning at most one good to each agent gives a feasible allocation, with that good as its witness item.
For $n=2$, we apply the binary-search splitting rule of
Lemma~\ref{lem:halving} to agent~1's values to find a good $e$
such that the bundles on either side of $e$ each have value
at most $v_1(M)/2$.
The less valuable side is denoted by $A$ and the other side by $B$.
Then $v_1(A)\le v_1(B)\le v_1(A\cup\{e\})$.
Agent~2 chooses her preferred bundle from $A\cup\{e\}$ and $B$,
and agent~1 receives the other.
Designating $e$ as the witness for $A\cup\{e\}$ and any good
as the witness for $B$, when nonempty, gives a feasible allocation.
This requires $O(\log m)$ comparison queries.

The next lemma uses the standard idea of adding a good to an unenvied bundle~\cite{LiptonEtAl2004}.
We obtain such a bundle through a top-trading cycles reassignment~\cite{ShapleyScarf1974}.

\begin{lemma}[\cite{LiptonEtAl2004,ShapleyScarf1974}]\label{lem:goods-insertion}
Any feasible partial allocation $\bX$ of goods can be extended to include one unallocated good $e$ while preserving feasibility, using $O(n^3)$ comparison queries.
\end{lemma}
Top-trading cycles weakly improve every agent's bundle, so keeping each witness item with its bundle preserves feasibility.
Adding $e$ to an unenvied bundle and designating $e$ as its new witness item also preserves feasibility.
The procedure uses at most $n$ rounds, each requiring $O(n^2)$ comparison queries.
When inserting a set $E$ of goods, we store each agent's pairwise comparisons among the current bundles, using $O(n^3)$ initial queries.
Reassigning bundles only permutes these records, and adding a good changes one bundle, so updating its comparisons with the other bundles uses $O(n^2)$ queries.
Thus, all insertions together use $O(n^3+n^2|E|)$ queries.

\subsection{The Construction of Small Blocks and Balanced Bundles}
\label{subsec:(n-1)-partition-goods}
As in Theorem~\ref{thm:(n-1)-partition}, we set aside a bounded
number of goods and either find an EF1 allocation of the remaining
items or construct $n-1$ ordered bundles for the subsequent search.
We choose $k$ as the smallest power of two that is at least $16n^5$.

For goods, each bundle $P_j$ consists of a nonempty initial part $A_j$ followed
by a final part $B_j$, and $e_j$ denotes the last good of $A_j$.
We require
\begin{equation}\label{eq:goods-preprocessing-shape}
    v_i(A_j)<v_i(P_\ell),
    \qquad
    v_i\!\left(\bigcup_{h=1}^{n-1}B_h\right)
    <v_i(A_j\setminus\{e_j\})
    \qquad(i\in N,\ j,\ell\in[n-1]).
\end{equation}
Thus, every agent values each initial part less than every full bundle,
but still more than the combined final parts after its last good is removed.
These comparisons replace the balanced-bundle condition for chores.

\begin{theorem}\label{thm:goods-preprocessing}
    There exists a deterministic algorithm that makes $O(n^8\log m)$
    comparison queries and outputs a set $E\subseteq M$ with
    $|E|=O(nk\log k)$ together with one of the following outcomes for $M'=M\setminus E$:
    \begin{itemize}
        \item an EF1 allocation of $M'$; or
        \item an ordered partition $(P_1,\ldots,P_{n-1})$ of $M'$,
        with initial and final parts satisfying~\eqref{eq:goods-preprocessing-shape},
        such that $v_i(M')>0$ and $v_i(g)\le v_i(M')/k$
        for every $i\in N$ and $g\in M'$.
    \end{itemize}
\end{theorem}

The first inequality in~\eqref{eq:goods-preprocessing-shape}, with $\ell=j$,
implies $B_j\ne\emptyset$. The second implies $|A_j|\ge2$.
The construction follows the same two steps as for chores.
We first obtain small blocks using the splitting trees and phase resets
of Appendix~\ref{subsec:compression}, with a different rule for early allocations.
We then partition these blocks using the variance argument of
Lemma~\ref{lem:balanced-partition} and the auxiliary-value updates of
Lemma~\ref{lem:auxiliary-update}.
Below, we give the changes needed to prove Theorem~\ref{thm:goods-preprocessing}.
The elementary cases follow from Lemma~\ref{lem:goods-small}, so
assume $n\ge3$ and $m>n$. Throughout, $n$ denotes the initial
number of agents and $k$ keeps its initial value.

\paragraph{The Reduction to Small Blocks.}
The reduction starts with active agents $N'=N$, remaining pool $R=M$, and
set-aside goods $E=\emptyset$. It reuses the splitting trees, dynamic
updates, and phase resets of Algorithm~\ref{alg:heavy-items}, with
values in place of costs. The required comparisons are available
because $\Comp_i^{\mathrm g}(Y,X)=Y$ if and only if $v_i(X)\le v_i(Y)$.
Before inspecting the trees or resetting a phase, an active agent
with $v_i(R)=0$ receives the empty bundle and leaves $N'$; an active
agent with $v_i(e)\ge v_i(R)$ for some $e\in E$ receives $\{e\}$,
with witness $e$, and both she and that good leave $N'$ and $E$,
respectively. We maintain a maximum-value good in $E$ for each active
agent to check the latter condition. Departures do not change $R$,
so the other agents' trees remain valid.
The reduction stops before further preprocessing if $|N'|\le2$ or $|R|\le|N'|$;
the residual instance is then elementary.

The proof of Lemma~\ref{lem:heavy-items} applies to the surviving
trees: each removal satisfies $v_i(e)>v_i(R)/(4k)$ for its triggering
agent, and a partition outcome consists of $|\cP|=O(nk)$ blocks with
\begin{equation}\label{eq:goods-residual-smallness}
    v_i(R)>0,\qquad v_i(D)\le\frac{v_i(R)}k
    \quad(i\in N',\ D\in\cP).
\end{equation}
In particular, $|\cP|\ge k$. Only the removal and phase counts need
adjustment, because goods can leave $E$ through early allocations.
In the removal-count argument, we use the first $n$ goods removed
by an agent instead of the first $n-1$.
While she remains active, at most $n-1$ of them can have been
allocated, so one remains in $E$. If $e^*$ has minimum value among
these $n$ goods, all later removals triggered by her occur with
$v_i(R)>v_i(e^*)$, whereas the pool has already fallen below
$4k\,v_i(e^*)$. The same multiplicative-decrease argument bounds
her subsequent removals by $O(k\log k)$.
Thus, there are $O(nk\log k)$ removals in total, and $|E|=O(nk\log k)$.

At the end of the reduction, at most $n$ removed goods have been
allocated early, so the total number of removals is at most $|E|+n$.
For the phase count, we use this upper bound in the proof of
Lemma~\ref{lem:heavy-items}.
Its disjoint-batch argument shows that an active agent charged at
least $n+\ceil{\log_2(|E|+n)}$ resets would yield $n$ distinct removed
goods each worth at least her current pool.
At most $n-1$ of these goods can have been allocated, so one remains
in $E$ and would trigger her departure before that reset.
Without any removals, there are no resets.
Thus, there are $O(n(n+\log k))$ phases. Maximum-value records in $E$
are updated at each insertion and rescanned after at most $n$
early allocations. Including the tree costs from
Appendix~\ref{subsec:compression}, the total is
$$
    O\!\left(n^2k(n+\log k)\log m+n(|E|+n)\log k+n^2(|E|+n)\right)
    =O(n^8\log m).
$$

\paragraph{The Construction of Balanced Bundles.}
Suppose that $s=|N'|\ge3$ and the reduction returns $\cP$ satisfying
\eqref{eq:goods-residual-smallness}. As in
Appendix~\ref{subsec:balanced-construction}, we maintain additive auxiliary values with
$$
    0\le\hat v_i(D)\le1/k\quad(D\in\cP),\qquad
    \hat v_i(R)=1\quad(i\in N').
$$
Initially, every block value is $1/|\cP|$, and we set $\gamma=1/(8s^2)$.
For any current auxiliary values, we independently assign each block
to one of the parts with probabilities
$$
    \Pr[D\in A_j]=\frac1{s-1}-\frac1{s^2},\qquad
    \Pr[D\in B_j]=\frac1{s^2}\quad(j\in[s-1]).
$$
We write $P_j=A_j\cup B_j$ and $B=\bigcup_jB_j$.
The target values of $P_j,A_j,B$ are their expectations,
$1/(s-1)$, $1/(s-1)-1/s^2$, and $(s-1)/s^2$, respectively.
We denote the total squared deviation by $F$:
$$
\begin{aligned}
F={}&\sum_{i\in N'}\sum_{j=1}^{s-1}
    \left[
      \left(\hat v_i(P_j)-\frac1{s-1}\right)^2
      +\left(\hat v_i(A_j)-\frac1{s-1}+\frac1{s^2}\right)^2
    \right]+\sum_{i\in N'}
      \left(\hat v_i(B)-\frac{s-1}{s^2}\right)^2.
\end{aligned}
$$
The variance calculation from Lemma~\ref{lem:balanced-partition}
gives a bound $p/k$ for a set whose membership probability is $p$,
since $\sum_D\hat v_i(D)^2\le1/k$.
For each agent, the membership probabilities sum to $2$ over all
$P_j$, all $A_j$, and $B$. Therefore, by $k\ge16s^5$, we have
$$
    \mathbb E[F]\le\frac{2s}{k}
    \le\frac1{8s^4}<\frac9{64s^4}.
$$

We fix the block assignments successively, each time choosing a part
that minimizes the conditional expectation of $F$.
At least one choice does not increase it, since the current
conditional expectation is the weighted average over the choices.
Each conditional expectation is computable from the known auxiliary
values as a sum of conditional variances and squared deviations of
conditional means. Thus, this deterministic procedure requires no
comparison queries and produces $F\le2s/k$.
Every deviation is consequently less than $3/(8s^2)$, so
$$
\begin{aligned}
    \hat v_i(P_\ell)-\hat v_i(A_j)
       &>\frac1{4s^2}>\gamma,\\
    \hat v_i(A_j)-\hat v_i(B)
       &>\frac1{s(s-1)}-\frac3{4s^2}>\frac1k+\gamma
\end{aligned}
$$
for all active agents and bundle indices.
We preserve the original order of blocks within each part and goods
within each block, and place $A_j$ before $B_j$. Since $\hat v_i(A_j)>1/k$, each $A_j$ contains at least
two blocks, hence at least two goods. Its last good is denoted by $e_j$.

\paragraph{Verification and Auxiliary-Value Updates.}
We check the inequalities in~\eqref{eq:goods-preprocessing-shape}
for the active agents and the $s-1$ bundles.
For each agent, we compare her minimum-value $P_\ell$ with her
maximum-value $A_j$, and her minimum-value $A_j\setminus\{e_j\}$
with $B$. Finding these extrema and checking the strict inequalities
uses $O(s^2)$ queries in total. If all inequalities hold, we return the
partition. In particular, $v_i(P_j)>v_i(A_j)$ implies $B_j\ne\emptyset$.

Otherwise, a failed first inequality gives $X=P_\ell,Y=A_j$ with
$v_i(X)\le v_i(Y)$. For a failed second inequality, we use
$X=A_j,Y=B$. Additivity and~\eqref{eq:goods-residual-smallness} give
$$
    v_i(A_j)-v_i(B)\le v_i(e_j)\le v_i(R)/k.
$$
In either case, the auxiliary-value bounds above yield
$$
    \hat v_i(X)-\hat v_i(Y)
      -\frac{v_i(X)-v_i(Y)}{v_i(R)}\ge\gamma.
$$
Both $X$ and $Y$ are unions of whole blocks, even in the second
case, so no auxiliary value for $e_j$ is needed.
We apply the update from Lemma~\ref{lem:auxiliary-update}: we decrease each block value in $X\setminus Y$ by $\gamma/|\cP|$, increase each block value in $Y\setminus X$ by the same amount, and project onto the constraints
$0\le\hat v_i(D)\le1/k$, $\hat v_i(R)=1$.
Only agent $i$'s auxiliary values change.

The proof of that lemma uses only this margin and the fact that the
projection set contains the normalized true values.
It therefore applies here: the initial squared-distance potential
is at most $2/k$, and each update decreases it by at least
$\gamma^2/|\cP|$. There are $O(s^4|\cP|/k)$ updates per agent and
$O(s^5|\cP|/k)$ rejections in total. Including the final successful
verification, the construction uses $O(s^7|\cP|/k)=O(n^8)$ queries.
Conditional-expectation selection and projection use only known
auxiliary values and require no preference queries.

\paragraph{The Two Outcomes.}
If no agent has left and the reduction returns small blocks, the
construction above gives the second outcome with $R=M'=M\setminus E$.
Otherwise, we complete the residual instance using
Lemma~\ref{lem:goods-small} or the conditional search and recovery
procedures in Sections~\ref{subsec:goods-search}
and~\ref{subsec:goods-recovery}.
These procedures use only the bundle properties proved above.
Their output combines with the early allocations to give the first outcome,
as verified in Section~\ref{subsec:goods-completion}.
This completion is performed at most once and costs
$O(n^5\log n\log m+n^2)$ further queries.
Since $|E|=O(nk\log k)$, the reduction, construction, and any
residual completion together satisfy the claimed bounds.

\subsection{Existence of a Consistent Pair at the Boundary}
\label{subsec:goods-transition}

Throughout this subsection and the next two, we fix $n\ge3$ agents and
an ordered partition $(P_1,\ldots,P_{n-1})$ of $M'$ satisfying the
second outcome of Theorem~\ref{thm:goods-preprocessing}.
We use only the stated conditions on these bundles, not the
algorithm that constructs them.
In particular, the arguments also apply to a residual instance
with fewer agents, provided that its parameter satisfies $k\ge16n^5$.

For an integer prefix-length vector $\bt$, we write
$$
    R(\bt)=M'\setminus\bigcup_{j=1}^{n-1}\Pre(P_j,t_j).
$$
For each agent, we order the labeled prefixes by value, prefix length, and bundle index, with smaller values receiving smaller ranks.
The labeled empty prefixes precede all nonempty
prefixes and are ordered by bundle index.
We extend the ranks continuously over $[0,|P_j|]$ using the same
pairwise disjoint intervals $I_1<\cdots<I_n\subset(1/2,1)$ as in
the chore construction.
For goods, we define $G(\bx)$ by
$$
    (i,j)\in E(G(\bx))
    \quad\Longleftrightarrow\quad
    j\in\argmax_{\ell\in[n-1]}\{\Rank_i(P_\ell,x_\ell)\}.
$$
Robustness and consistency have the same definitions as in the
main text, with this maximum-rank graph in place of the
minimum-rank graph.
As in the main text, the original system consists of the nonempty prefixes of $P_1,\dots,P_{n-1}$.
Including the labeled empty prefixes shifts every agent's interpolated ranks on $[1,|P_j|]$ by the same constant $n-1$, so it does not change the maximum-rank graph on $\prod_j[1,|P_j|]$.
The edge-counting proof of Lemma~\ref{lem:robust-x-properties}
applies unchanged: at a robust vector, every agent has exactly
one inactive edge and every nonroot bundle has exactly one active
edge, and the graph is a tree.
Thus, the half-integer root is unique, all nonroot coordinates
round upward, and a consistent pair satisfies $\bt=\ceil{\bx}$.

For $\bt\in\prod_j\{1,\ldots,|P_j|\}$, we define
$$
    \Phi(\bt)=1
    \quad\Longleftrightarrow\quad
    v_i(R(\bt))<
    \max_{j\in[n-1]}\{v_i(\Pre(P_j,t_j-1))\}
    \quad\text{for every }i\in N.
$$
This predicate is nondecreasing coordinatewise and can be evaluated
using $O(n^2)$ comparison queries.
Unlike the chore predicate, it requires an inequality for every
agent, and the prefix on its right-hand side has one good removed.
These choices will ensure that every agent assigned a prefix
satisfies EF1 toward the remaining bundle.
As before, a consistent pair $(\bt,r)$ is at the boundary when
$\Phi(\bt^{-r})=0$ and $\Phi(\bt)=1$.

\begin{samepage}
\begin{lemma}\label{lem:goods-boundaries}
    For $\bx\in\prod_{j=1}^{n-1}[1,|P_j|]$, the following properties hold.
    \begin{enumerate}
        \item If $x_j=1$ and $(i,j)\in E(G(\bx))$ for some agent
        $i$, then $\Phi(\ceil{\bx})=0$.
        \item If some coordinate is full and every bundle has a
        neighbor in $G(\bx)$, then $\Phi(\floor{\bx})=1$.
    \end{enumerate}
\end{lemma}
\end{samepage}

\begin{proof}
For the first assertion, we use the indicated agent $i$.
The maximum-rank condition implies
$$
    v_i(\Pre(P_\ell,\floor{x_\ell}))
    \le v_i(\Pre(P_j,1))\le\frac{v_i(M')}{k}
    \qquad(\ell\in[n-1]).
$$
Rounding upward adds at most one good, so every upper-rounded
prefix has value at most $2v_i(M')/k$.
Consequently, since $k>2n$,
$$
    v_i(R(\ceil{\bx}))
    \ge\left(1-\frac{2(n-1)}k\right)v_i(M')
    >\frac{2v_i(M')}{k}
    \ge\max_{\ell\in[n-1]}\{v_i(\Pre(P_\ell,\ceil{x_\ell}-1))\}.
$$
Thus, agent $i$ witnesses $\Phi(\ceil{\bx})=0$.

For the second assertion, suppose that $x_\ell=|P_\ell|$.
For every nonfull bundle $j$, we choose an incident agent $i_j$.
By the maximum-rank condition and
\eqref{eq:goods-preprocessing-shape},
$$
    v_{i_j}(\Pre(P_j,\ceil{x_j}))
    \ge v_{i_j}(P_\ell)>v_{i_j}(A_j).
$$
Since $A_j$ is an initial part of $P_j$, this implies
$\ceil{x_j}>|A_j|$ and hence $\floor{x_j}\ge|A_j|$.
The latter inequality also holds for full coordinates.
Therefore,
$$
    R(\floor{\bx})\subseteq\bigcup_j B_j,
    \qquad
    A_j\setminus\{e_j\}
    \subseteq\Pre(P_j,\floor{x_j}-1).
$$
For every agent $i$ and bundle $j$, it follows that
$$
    v_i(R(\floor{\bx}))
    \le v_i\!\left(\bigcup_h B_h\right)
    <v_i(A_j\setminus\{e_j\})
    \le v_i(\Pre(P_j,\floor{x_j}-1)).
$$
This proves $\Phi(\floor{\bx})=1$.
\end{proof}

\paragraph{Reusing the Chore Transition Argument.}
We record the general form of the argument in
Lemma~\ref{lem:continuous-transition}.
For each agent and bundle $j$, we consider a continuous function on
$[1,U_j]$, where $U_j\ge2$ is an integer. Agents select the bundle indices at which
their functions attain the minimum.
The Boolean predicate $\Psi$ is coordinatewise nondecreasing on
$\prod_j\{1,\ldots,U_j\}$.
The same argument gives a point $\mathbf z\in\prod_j[3/2,U_j)$ whose
nearest-integer roundings attain both values of $\Psi$ and whose
minimum graph is robust, provided that
\begin{enumerate}
    \item $\Psi(\mathbf u)=0$ whenever some $u_j=1$; and
    \item if $z_j=U_j$ and $j$ is a minimum choice for some agent,
    then $\Psi(\floor{\mathbf z})=1$.
\end{enumerate}
Here, ``robust'' for the graph means that deleting any agent leaves
a perfect matching.
The proof of Lemma~\ref{lem:continuous-transition} applies with these
two conditions: the first keeps the transition boundary away from
the lower endpoints, and the second supplies the boundary condition
for Lemma~\ref{lem:continuous-matching} and excludes the upper endpoints
at the resulting matching point.
We verify these conditions for goods by reflecting coordinates and negating ranks.

\begin{lemma}\label{lem:goods-crossing}
    There exists $\bx\in\prod_{j=1}^{n-1}(1,|P_j|)$ whose
    nearest-integer roundings attain both values of $\Phi$ and
    whose maximum-rank graph admits a perfect matching after any
    agent is removed.
    Consequently, there exists a consistent pair $(\bt,r)$ at
    the boundary with $t_j\ge2$ for every $j$.
\end{lemma}

\begin{proof}
For this existence argument, we append one artificial good to each
$P_j$, giving it value $v_i(M')$ for every agent $i$, and write
$P_j^+$ for the extended bundle.
Each full extended bundle $P_j^+$ follows every prefix of the original bundles because $v_i(P_j)>0$.
Thus, the extended rank functions agree with the rank functions defined above on $[0,|P_j|]$.
We extend $\Phi$ to $\bt\in\prod_j\{1,\ldots,|P_j|+1\}$ by
$$
    \overline\Phi(\bt)=
    \begin{cases}
        \Phi(\bt),& t_j\le |P_j|\text{ for every }j,\\
        1,& t_j=|P_j|+1\text{ for some }j.
    \end{cases}
$$
This monotone auxiliary predicate agrees with $\Phi$ whenever
no artificial good is included.

We reflect each coordinate by $z_j=|P_j|+2-x_j$ and use the
transformed rank function $-\Rank_i(P_j^+,|P_j|+2-z)$.
For integer $\mathbf u\in\prod_j[1,|P_j|+1]$, we define
$$
    \Psi(\mathbf u)
    =1-\overline\Phi\bigl((|P_j|+2-u_j)_{j=1}^{n-1}\bigr).
$$
These transformed rank functions are continuous and nondecreasing.
Their minimum-rank graph at $\mathbf z$ is the maximum-rank graph of the extended bundles at $\bx$, and $\Psi$ is coordinatewise nondecreasing.
If some $u_j=1$, its reflected coordinate is $|P_j|+1$, so
$\Psi(\mathbf u)=0$.

To verify the other boundary condition, suppose $z_j=|P_j|+1$
and $j$ is a minimum choice for agent $i$.
In the goods coordinates, $x_j=1$ and $j$ is a maximum choice
for $i$.
No coordinate $x_\ell$ can satisfy $x_\ell\ge |P_\ell|$, since
$$
    v_i(P_\ell)>v_i(A_j)\ge v_i(\Pre(P_j,1))
$$
would contradict that maximum choice.
Thus, $1\le x_\ell<|P_\ell|$ for every $\ell$, and Lemma~\ref{lem:goods-boundaries}(1) gives $\Phi(\ceil{\bx})=0$.
Since $\floor{z_j}=|P_j|+2-\ceil{x_j}$,
we obtain $\Psi(\floor{\mathbf z})=1$.
The general chore transition argument above therefore applies.

We reflect the resulting point back to $\bx$.
The maximum-rank graph has the required matching property, the
nearest-integer roundings attain both values of
$\overline\Phi$, and
$$
    1<x_j\le |P_j|+\frac12\qquad(j\in[n-1]).
$$
We show that in fact $x_j<|P_j|$ for every $j$.
Suppose otherwise, and define $y_j=\min\{x_j,|P_j|\}$.
The extended rank functions are constant on $[|P_j|,|P_j|+1/2]$
because every active interval lies in $(1/2,1)$.
Capping the coordinates therefore preserves the maximum-rank
graph.
The point $\by$ has a full coordinate and every bundle has a
neighbor, so Lemma~\ref{lem:goods-boundaries}(2) gives
$\Phi(\floor{\by})=1$.
Every nearest-integer rounding of $\bx$ either uses an artificial item, in which case $\overline\Phi=1$ by definition, or belongs to $\prod_j\{1,\ldots,|P_j|\}$ and is coordinatewise at least $\floor{\by}$, in which case $\overline\Phi=\Phi=1$ by monotonicity.
This contradicts the mixed roundings.

Hence, $1<x_j<|P_j|$ for all $j$.
All nearest-integer roundings now use original goods, and
$\overline\Phi$ agrees with $\Phi$ on them.
There is at least one half-integer coordinate, so $\bx$ is robust.
The structural lemma gives a unique half-integer root $r$ and
forces every other coordinate to round upward.
Its two nearest-integer roundings are therefore
$\bt^{-r}$ and $\bt$, where $\bt=\ceil{\bx}$.
Monotonicity and the mixed roundings imply
$\Phi(\bt^{-r})=0$ and $\Phi(\bt)=1$.
Since $x_j>1$, we have $t_j\ge2$ for every $j$.
\end{proof}

\subsection{Root Monotonicity and Search by Item Grouping}
\label{subsec:goods-search}

The search for goods reuses the root monotonicity and
candidate-weighted grouping arguments from the main text.
We verify the required properties for maximum-rank choices.

\begin{lemma}\label{lem:goods-root-monotonicity}
    For a root $r$ and robust goods vectors $\bx,\bx'$ with half-integer $r$-th coordinates,
    $$
        x_r\le x_r'\quad\Longrightarrow\quad\bx\le\bx'.
    $$
    Consequently, for consistent goods pairs $(\bt,r)$ and
    $(\bt',r)$,
    $$
        t_r\le t_r'\quad\Longrightarrow\quad\bt\le\bt'.
    $$
    In particular, a fixed root coordinate determines at most
    one robust vector, and a fixed root prefix length determines
    at most one consistent pair.
\end{lemma}

\begin{proof}
As in the proof of Lemma~\ref{lem:goods-crossing}, we use the
reflected coordinates $z_j=|P_j|+2-x_j$ and negate the ranks.
The transformed rank functions are nondecreasing and have distinct endpoint ranks.
Their minimum-rank graph at $\mathbf z$ is the maximum-rank graph at $\bx$.
Their active intervals are the disjoint reflections $1-I_i$.
Although these lie in $(0,1/2)$, the proof of
Lemma~\ref{lem:root-monotonicity-fractional} requires only
pairwise disjointness, strict increase inside the active
intervals, distinct endpoint ranks, and inactivity at the common
half-integer root.
All these properties are preserved.
If $x_r\le x_r'$, then $z_r'\le z_r$.
Applying that proof to $(\mathbf z',\mathbf z)$ yields $\mathbf z'\le\mathbf z$,
and reflecting back gives $\bx\le\bx'$.
For consistent pairs, $x_r=t_r-1/2$ and $\bt=\ceil{\bx}$ in the goods coordinates, so taking coordinatewise ceilings gives the claimed implication.
Applying either implication in both directions proves uniqueness.
\end{proof}

\paragraph{Extended Prefix Ranges and Common Systems.}
We append an artificial good to the end of each bundle $P_j$, with value $v_i(M')$ for every agent $i$, and include the labeled empty prefixes, as above.
For an integer root cut $T\in\{2,\dots,|P_r|\}$, the fixed root rank at $T-1/2$ lies strictly between the ranks of the empty prefix $\Pre(P_j^+,0)$ and the full extended bundle $P_j^+$ for every nonroot bundle $j$.
After reflecting coordinates and negating ranks, the rank
differences from the fixed root are continuous and nondecreasing,
with strictly negative lower endpoints and strictly positive
upper endpoints.
The fixed-root construction in the proof of
Lemma~\ref{lem:fixed-root-extension} therefore applies and gives
a robust completion with that root coordinate.
That construction uses only continuity, monotonicity, and these
endpoint inequalities.
The same reasoning applies to a grouped system, using block indices for fractional coordinates and keeping $T-1$ and $T$ as adjacent cut-points of the root bundle.

The proof of Lemma~\ref{lem:agreeing-in-two-systems} also applies,
with each agent's unique inactive maximum in place of her
inactive minimum.
For a proposed active edge $(i,j)$, agent $i$'s rank at her inactive edge must lie strictly between her ranks of the two endpoint prefixes of the interpolation interval for bundle $j$.
Preserving the endpoint order preserves these inequalities and
allows the crossing to be reconstructed inside the same active
interval.
Thus, preserving the endpoint prefixes of every selected interpolation interval preserves consistency and the maximum-rank graph, even though the numerical ranks and fractional crossing positions may change.
We use the system conventions of Section~\ref{subsec:items-grouping}:
ties are broken by original prefix length and bundle index, ranks
are assigned within each system, and consistency uses block coordinates.

\begin{lemma}\label{lem:goods-search}
    Applied to the ordered bundles above, with the goods predicate $\Phi$ and maximum-rank graphs,
    Algorithm~\ref{alg:coarse-prefix-search} finds a consistent
    boundary pair $(\bt,r)$ with $t_j\ge2$ for every $j$, using
    $O(n^5\log n\log m)$ comparison queries.
\end{lemma}

\begin{proof}
The coarse-pair computation uses each agent's maximum-rank endpoint
prefix in place of her minimum-rank endpoint prefix and checks the
resulting maximum-rank graph.
The preceding fixed-root argument ensures that a consistent coarse
pair exists at every prescribed root cut.
The learned prefix orders determine these maxima and the comparisons
between endpoint prefixes, so finding this pair requires no further queries.
The full extended bundles follow all prefixes of the original bundles and have the same relative order as the full original bundles.
Since all full original bundles are included in the coarse system, no additional queries are needed to place the full extended bundles in the rankings.

The candidate-weighted grouping of
Lemma~\ref{lem:candidate-weighted-grouping} is independent of
preferences and bounds the exceptional candidates by less than
one quarter of the current set.
Lemmas~\ref{lem:agreeing-in-two-systems} and~\ref{lem:goods-root-monotonicity}
give the same implications
for every non-exceptional consistent candidate as in
Lemma~\ref{lem:exceptional}.
Since the goods predicate is monotone, the same filtering cases preserve every candidate $\bt$ for which $(\bt,r)$ is consistent and at the boundary in the original system.
In evaluating this predicate, $t_j-1$ always removes one original
good, even if it is not a retained cut-point in the coarse system.
If a coarse pair $(\bw,r)$ satisfies $w_j=|P_j|+1$ for some $j$, we apply the corresponding filtering rule in Algorithm~\ref{alg:coarse-prefix-search} before evaluating $\Phi$.

When the algorithm returns, no artificial item is used, every
selected interval contains one original good, and
$\Phi(\bw^{-r})=0$, $\Phi(\bw)=1$.
We check that no returned coordinate equals $1$.
Suppose $w_j=1$, and choose an agent $i$ adjacent to $j$ in the
robust coarse graph.
The upper endpoint of this interval is the first good of $P_j$.
Thus, the maximum-rank condition implies
$$
    v_i(\Pre(P_\ell,w_\ell-1))\le\frac{v_i(M')}{k}
    \qquad(\ell\in[n-1]).
$$
Since the selected intervals are singletons, each full returned prefix
has value at most $2v_i(M')/k$.
As in Lemma~\ref{lem:goods-boundaries}(1),
$$
    v_i(R(\bw))
    \ge\left(1-\frac{2(n-1)}k\right)v_i(M')
    >\frac{2v_i(M')}{k},
$$
contradicting $\Phi(\bw)=1$.
Consequently, every $w_j\ge2$.
The selected interpolation intervals have the same endpoint prefixes as the corresponding intervals in the original system, so Lemma~\ref{lem:agreeing-in-two-systems} transfers consistency to that system.

Lemma~\ref{lem:goods-crossing} guarantees a consistent pair at the
boundary in the original system for some root. Its vector belongs
to the initial candidate set and survives every filtering step.
The candidate counts, query cost per round, and fraction eliminated
in each nonterminal round are the same as in Theorem~\ref{thm:coarse-prefix-search}.
Thus, the search succeeds with the same $O(n^5\log n\log m)$ query bound.
\end{proof}

As in the main text, this is a comparison-query bound, not a
polynomial bound on total running time.
The enumeration of coarse configurations and the maintenance of
candidate vectors use the learned order tables without further
preference queries.

\subsection{Recovering an EF1 Allocation}
\label{subsec:goods-recovery}

For a consistent pair $(\bt,r)$ at the boundary, the partition
induced by $\bt^{-r}$ admits an EF1 allocation.
The last good of $\Pre(P_r,t_r)$ serves as the witness item for
the remaining bundle, and each prefix uses its own last good.
We recover the allocation from the pair by comparing these bundles.

\begin{lemma}\label{lem:goods-crossing-alloc}
    Given a consistent boundary pair $(\bt,r)$ with $t_j\ge2$
    for every $j$, a feasible EF1 allocation of $M'$ and its
    designated witness items can be computed using $O(n^2)$
    comparison queries.
\end{lemma}

\begin{proof}
We consider the partition induced by $\bt^{-r}$, whose bundles are
$$
    Q_j=\Pre(P_j,t_j^{-r})\quad(j\in[n-1]),
    \qquad Q_n=R(\bt^{-r}).
$$
For every $j<n$, we designate the last good of $Q_j$ as its witness
item and write $\hat Q_j$ for the bundle after its removal.
For $Q_n$, the witness item is the last good $e$ of $\Pre(P_r,t_r)$, so
$$
    \hat Q_n=Q_n\setminus\{e\}=R(\bt).
$$
We construct a bipartite graph $H$ by
$$
    (i,j)\in E(H)
    \quad\Longleftrightarrow\quad
    v_i(Q_j)\ge\max_{\ell\in[n]}\{v_i(\hat Q_\ell)\}.
$$
Every perfect matching in $H$ gives a feasible allocation with
these designated witnesses.

To prove that such a matching exists, we use a robust vector $\bx$
witnessing the consistency of $(\bt,r)$.
For every edge $(i,j)$ of $G(\bx)$ and every $\ell<n$,
$$
    \Rank_i(P_j,t_j^{-r})
    \ge\Rank_i(P_j,x_j)
    \ge\Rank_i(P_\ell,x_\ell)
    \ge\Rank_i(P_\ell,t_\ell-1).
$$
The first inequality uses equality of the root rank at the
half-integer and its lower prefix, and upward rounding at all
other coordinates.
Thus, every such edge satisfies
$$
    v_i(Q_j)\ge
    \max_{\ell\in[n-1]}\{v_i(\Pre(P_\ell,t_\ell-1))\}.
$$
Since $\Phi(\bt^{-r})=0$, there is an agent $i^*$ with
$$
    v_{i^*}(Q_n)
    \ge\max_{\ell\in[n-1]}
       \{v_{i^*}(\Pre(P_\ell,t_\ell^{-r}-1))\}
    =\max_{\ell\in[n-1]}\{v_{i^*}(\hat Q_\ell)\}.
$$
Monotonicity also gives $v_{i^*}(Q_n)\ge v_{i^*}(\hat Q_n)$,
so $(i^*,n)$ is an edge of $H$.
Robustness gives a perfect matching between the agents other than
$i^*$ and the $n-1$ prefix bundles in $G(\bx)$.
For each matched pair $(i,j)$,
the prefix bound above dominates every trimmed prefix, because $t_\ell^{-r}-1\le t_\ell-1$.
It also dominates $\hat Q_n$, since $\Phi(\bt)=1$ gives
$$
    v_i(Q_j)
    \ge\max_{\ell\in[n-1]}\{v_i(\Pre(P_\ell,t_\ell-1))\}
    >v_i(R(\bt))=v_i(\hat Q_n).
$$
These matching edges thus belong to $H$.
Together with $(i^*,n)$, they form a perfect matching in $H$.

For each agent, finding a maximum-value trimmed bundle and
comparing all $n$ full bundles against it uses $O(n)$ queries.
Consequently, $H$ can be constructed with $O(n^2)$ queries.
A perfect matching in $H$ then gives the allocation and its
witnesses without additional preference queries or knowledge of
$G(\bx)$.\qedhere
\end{proof}

\subsection{The Complete Algorithm and Query Complexity}
\label{subsec:goods-completion}

We now combine preprocessing, search, and allocation recovery to
prove Theorem~\ref{thm:goods}.
The preceding three subsections apply whenever the ordered bundles
satisfy~\eqref{eq:goods-preprocessing-shape} and the small-item bound.
We use them to allocate the final pool, then reinsert the set-aside goods.

The reduction maintains an active set $N'$, a remaining pool $R$,
a set $E$ of set-aside goods, and empty or singleton bundles
assigned to agents who have left $N'$.
We run this reduction until the final pool is ready to be allocated.
If the residual instance is elementary, we allocate it using
Lemma~\ref{lem:goods-small}.
Otherwise, we apply the ordered-bundle construction from
Theorem~\ref{thm:goods-preprocessing} to $N'$ and $R$, followed
by Lemmas~\ref{lem:goods-search} and~\ref{lem:goods-crossing-alloc}.
In these conditional applications, the number of agents is
$s=|N'|$, but $k$ keeps its initial value and satisfies
$k\ge16s^5$.
Finally, we reinsert the goods still in $E$.

\begin{algorithm}[!htbp]
\caption{Computation of EF1 Allocations for Additive Goods}
\label{alg:goods}
\KwIn{Instance $(N,M,\bv)$ with nonnegative additive valuations}
\KwOut{An EF1 allocation of $M$}
\If{$n\le2$ \textbf{or} $m\le n$}{
    Compute an allocation using Lemma~\ref{lem:goods-small} and
    \Return it\;
}
$k\gets$ the smallest power of two at least $16n^5$\;
Run the reduction in the proof of
Theorem~\ref{thm:goods-preprocessing}, stopping before allocating
the final pool; retain $N'$, $R$, $E$, and all early allocations
with their designated witnesses\;
\If{$R\ne\emptyset$}{
    \eIf{$|N'|\le2$ \textbf{or} $|R|\le|N'|$}{
        Allocate $R$ among $N'$ and assign witnesses using
        Lemma~\ref{lem:goods-small}\;
    }{
        Apply the ordered-bundle construction in the proof of
        Theorem~\ref{thm:goods-preprocessing} to obtain
        $P_1,\ldots,P_{|N'|-1}$ for the residual instance\;
        Find $(\bt,r)$ using the goods version of
        Algorithm~\ref{alg:coarse-prefix-search} from
        Lemma~\ref{lem:goods-search}\;
        Allocate $R$ among $N'$ and assign witnesses using
        Lemma~\ref{lem:goods-crossing-alloc}\;
    }
}
Combine the residual allocation with the early allocations\;
Insert all goods in $E$ using Lemma~\ref{lem:goods-insertion},
maintaining the pairwise comparisons among current bundles\;
\Return the resulting allocation\;
\end{algorithm}

\paragraph{Correctness and the Preprocessing Outcomes.}
Whenever an agent $i$ leaves $N'$, her assigned empty or singleton
bundle $X_i$ satisfies $v_i(X_i)\ge v_i(R)$ at that time.
Every bundle subsequently formed from the residual pool is a
subset of this pool, so she does not envy any such bundle.
Each previously assigned bundle is empty or becomes empty after
its designated witness is removed, so every agent satisfies the
feasibility comparison with it.
By Lemma~\ref{lem:goods-small} or
Lemma~\ref{lem:goods-crossing-alloc}, the final pool has a
feasible allocation among the active agents.
Combining it with the early allocations therefore gives a
feasible allocation of $M\setminus E$.
In particular, the first outcome of
Theorem~\ref{thm:goods-preprocessing} can return this stronger
feasible allocation together with its witnesses.

If no agent has left and the reduction reaches the ordered-bundle
construction, these bundles give the second outcome of
Theorem~\ref{thm:goods-preprocessing}.
Otherwise, the allocation of $M\setminus E$ above gives its first outcome.
For the complete goods algorithm, both branches are followed by
reinsertion of $E$.
Lemma~\ref{lem:goods-insertion} preserves feasibility during
these insertions, so the resulting allocation of all goods is
EF1.

\paragraph{Query Complexity.}
Here, $n$ and $m$ denote the original numbers of agents and goods.
The reduction analysis in the proof of
Theorem~\ref{thm:goods-preprocessing} gives
$O(n^8\log m)$ queries and $|E|=O(nk\log k)$.
If needed, the ordered-bundle construction is performed once.
For $s\le n$ active agents and $|\cP|=O(nk)$ residual small blocks,
its query cost is $O(s^7|\cP|/k)=O(n^8)$.
Lemmas~\ref{lem:goods-search} and~\ref{lem:goods-crossing-alloc}
then allocate the final pool using
$O(n^5\log n\log m+n^2)$ further queries.
The elementary branch costs at most $O(\log m)$ queries.
By Lemma~\ref{lem:goods-insertion}, maintaining the pairwise
comparisons permits all final insertions using
$O(n^3+n^2|E|)=O(n^3+ n^3k\log k)$ queries.
Substituting $k=\Theta(n^5)$ and using $m>n\ge3$, the total is
$O(n^8\log m)$.
This proves Theorem~\ref{thm:goods}.

\newpage
\bibliographystyle{alpha}
\AddToHook{env/thebibliography/begin}{\DeclareUrlCommand\path{\urlstyle{same}}}
\phantomsection
\addcontentsline{toc}{section}{\refname}
\bibliography{ref}

@inproceedings{conf/focs/BuchbinderF24,
  author        = {Niv Buchbinder and
                   Moran Feldman},
  title         = {Deterministic Algorithm and Faster Algorithm for Submodular Maximization subject
                   to a Matroid Constraint},
  booktitle     = {{FOCS}},
  pages         = {700--712},
  publisher     = {{IEEE} Computer Society},
  year          = {2024},
  doi           = {10.1109/FOCS61266.2024.00050},
  url           = {https://arxiv.org/abs/2408.03583}
}

@inproceedings{conf/stoc/BuchbinderF25,
  author        = {Niv Buchbinder and
                   Moran Feldman},
  title         = {Extending the Extension: Deterministic Algorithm for Non-monotone Submodular
                   Maximization},
  booktitle     = {{STOC}},
  pages         = {1130--1141},
  publisher     = {{ACM}},
  year          = {2025},
  doi           = {10.1145/3717823.3718120},
  url           = {https://arxiv.org/abs/2409.14325}
}

@article{FoxWoodruff2026,
  author        = {James Fox and
                   David P. Woodruff},
  title         = {Symmetric Submodular Minimization from Comparisons},
  journal       = {arXiv preprint arXiv:2610.10518},
  year          = {2026},
  doi           = {10.48550/arXiv.2610.10518},
  eprint        = {2610.10518},
  archivePrefix = {arXiv},
  primaryClass  = {cs.DS},
  url           = {https://arxiv.org/abs/2610.10518}
}

@article{FeigeFine2026,
  author        = {Uriel Feige and
                   Assaf Fine},
  title         = {Simultaneous {EF1} and Approximate {MMS} Allocations for Submodular Valuations},
  journal       = {arXiv preprint arXiv:2606.06451},
  year          = {2026},
  doi           = {10.48550/arXiv.2606.06451},
  eprint        = {2606.06451},
  archivePrefix = {arXiv},
  primaryClass  = {cs.GT},
  url           = {https://arxiv.org/abs/2606.06451}
}

@article{AkramiEtAl2026,
  author        = {Hannaneh Akrami and
                   Uriel Feige and
                   Ryoga Mahara and
                   Kurt Mehlhorn and
                   Nidhi Rathi},
  title         = {The Power of Share-Based Notions in Proving Envy-Based Fairness Guarantees},
  journal       = {arXiv preprint arXiv:2602.11732},
  year          = {2026},
  doi           = {10.48550/arXiv.2602.11732},
  eprint        = {2602.11732},
  archivePrefix = {arXiv},
  primaryClass  = {cs.GT},
  url           = {https://arxiv.org/abs/2602.11732}
}

@article{Psomas2026,
  author        = {Alexandros Psomas},
  title         = {Cutting a Cake For Human Consumption: An Elementary {$n$}-agent Cake Cutting
                   Protocol},
  journal       = {arXiv preprint arXiv:2610.30370},
  year          = {2026},
  doi           = {10.48550/arXiv.2610.30370},
  eprint        = {2610.30370},
  archivePrefix = {arXiv},
  primaryClass  = {cs.GT},
  url           = {https://arxiv.org/abs/2610.30370}
}

@article{journals/jacm/ChenLY25,
  author        = {Xi Chen and
                   Yuhao Li and
                   Mihalis Yannakakis},
  title         = {Computing a Fixed Point of Contraction Maps in Polynomial Queries},
  journal       = {J. {ACM}},
  volume        = {72},
  number        = {4},
  pages         = {25:1--25:20},
  year          = {2025},
  doi           = {10.1145/3744738},
  url           = {https://arxiv.org/abs/2403.19911}
}

@inproceedings{conf/focs/ChenLY26,
  author        = {Xi Chen and
                   Yuhao Li and
                   Mihalis Yannakakis},
  title         = {The Mystery Deepens: On the Query Complexity of {Tarski} Fixed Points},
  booktitle     = {{FOCS}},
  year          = {2026},
  note          = {To appear},
  url           = {https://arxiv.org/abs/2604.00268}
}

@article{journals/ai/AmanatidisABFLMVW23,
  author        = {Georgios Amanatidis and
                   Haris Aziz and
                   Georgios Birmpas and
                   Aris Filos{-}Ratsikas and
                   Bo Li and
                   Herv{\'{e}} Moulin and
                   Alexandros A. Voudouris and
                   Xiaowei Wu},
  title         = {Fair division of indivisible goods: Recent progress and open questions},
  journal       = {Artif. Intell.},
  volume        = {322},
  pages         = {103965},
  year          = {2023},
  doi           = {10.1016/j.artint.2023.103965}
}

@inproceedings{BhaskarEtAl2021,
  author        = {Umang Bhaskar and
                   A. R. Sricharan and
                   Rohit Vaish},
  title         = {On Approximate Envy-Freeness for Indivisible Chores and Mixed Resources},
  booktitle     = {{APPROX-RANDOM}},
  series        = {LIPIcs},
  volume        = {207},
  pages         = {1:1--1:23},
  publisher     = {Schloss Dagstuhl - Leibniz-Zentrum f{\"{u}}r Informatik},
  year          = {2021},
  doi           = {10.4230/LIPIcs.APPROX/RANDOM.2021.1}
}

@inproceedings{conf/wine/BuLLST24,
  author        = {Xiaolin Bu and
                   Zihao Li and
                   Shengxin Liu and
                   Jiaxin Song and
                   Biaoshuai Tao},
  title         = {Logarithmic Comparison-Based Query Complexity for Fair Division of Indivisible
                   Goods},
  booktitle     = {{WINE}},
  series        = {Lecture Notes in Computer Science},
  volume        = {15534},
  pages         = {348--365},
  publisher     = {Springer},
  year          = {2024},
  doi           = {10.1007/978-3-032-08560-3_20},
  url           = {https://arxiv.org/abs/2404.18133}
}

@inproceedings{LinEtAl2026,
  author        = {Zehan Lin and
                   Shengxin Liu and
                   Biaoshuai Tao and
                   Shengwei Zhou},
  title         = {Comparison-Based Fair Division of Indivisible Chores},
  booktitle     = {{SODA}},
  year          = {2027},
  note          = {To appear},
  url           = {https://arxiv.org/abs/2609.08687}
}

@inproceedings{LiptonEtAl2004,
  author        = {Richard J. Lipton and
                   Evangelos Markakis and
                   Elchanan Mossel and
                   Amin Saberi},
  title         = {On Approximately Fair Allocations of Indivisible Goods},
  booktitle     = {{EC}},
  pages         = {125--131},
  publisher     = {{ACM}},
  year          = {2004},
  doi           = {10.1145/988772.988792}
}

@article{ShapleyScarf1974,
  author        = {Lloyd Shapley and
                   Herbert Scarf},
  title         = {On Cores and Indivisibility},
  journal       = {J. Math. Econ.},
  volume        = {1},
  number        = {1},
  pages         = {23--37},
  year          = {1974},
  doi           = {10.1016/0304-4068(74)90033-0}
}

@article{journals/siamdm/OhPS21,
  author        = {Hoon Oh and
                   Ariel D. Procaccia and
                   Warut Suksompong},
  title         = {Fairly Allocating Many Goods with Few Queries},
  journal       = {{SIAM} J. Discret. Math.},
  volume        = {35},
  number        = {2},
  pages         = {788--813},
  year          = {2021},
  doi           = {10.1137/20M1313349}
}

@article{journals/corr/YamagataS26,
  author        = {Tatsuhito Yamagata and
                   Hanna Sumita},
  title         = {Witness-Certified Fair Division with Comparison Queries},
  journal       = {arXiv preprint arXiv:2608.16109},
  year          = {2026},
  doi           = {10.48550/arXiv.2608.16109},
  eprint        = {2608.16109},
  archivePrefix = {arXiv},
  primaryClass  = {cs.GT},
  url           = {https://arxiv.org/abs/2608.16109}
}

@article{liu2024mixed,
  author        = {Liu, Shengxin and
                   Lu, Xinhang and
                   Suzuki, Mashbat and
                   Walsh, Toby},
  title         = {Mixed fair division: A survey},
  journal       = {J. Artif. Intell. Res.},
  volume        = {80},
  pages         = {1373--1406},
  year          = {2024}
}

@article{journals/sigecom/AzizLMW22,
  author        = {Haris Aziz and
                   Bo Li and
                   Herv{\'{e}} Moulin and
                   Xiaowei Wu},
  title         = {Algorithmic fair allocation of indivisible items: a survey and new questions},
  journal       = {{SIGecom} Exch.},
  volume        = {20},
  number        = {1},
  pages         = {24--40},
  year          = {2022}
}

@article{journals/teco/CaragiannisKMPS19,
  author        = {Ioannis Caragiannis and
                   David Kurokawa and
                   Herv{\'{e}} Moulin and
                   Ariel D. Procaccia and
                   Nisarg Shah and
                   Junxing Wang},
  title         = {The Unreasonable Fairness of Maximum {Nash} Welfare},
  journal       = {{ACM} Trans. Econ. Comput.},
  volume        = {7},
  number        = {3},
  pages         = {12:1--12:32},
  year          = {2019}
}

@article{journals/siamdm/PlautR20,
  author        = {Benjamin Plaut and
                   Tim Roughgarden},
  title         = {Almost Envy-Freeness with General Valuations},
  journal       = {{SIAM} J. Discret. Math.},
  volume        = {34},
  number        = {2},
  pages         = {1039--1068},
  year          = {2020}
}

@article{steinhaus1948problem,
  author        = {Steinhaus, Hugo},
  title         = {The problem of fair division},
  journal       = {Econometrica},
  volume        = {16},
  pages         = {101--104},
  year          = {1948}
}

@article{journals/siamcomp/PlautR20,
  author        = {Benjamin Plaut and
                   Tim Roughgarden},
  title         = {Communication Complexity of Discrete Fair Division},
  journal       = {{SIAM} J. Comput.},
  volume        = {49},
  number        = {1},
  pages         = {206--243},
  year          = {2020}
}

@inproceedings{conf/sigecom/Feige25,
  author        = {Uriel Feige},
  title         = {Low communication protocols for fair allocation of indivisible goods},
  booktitle     = {{EC}},
  pages         = {358--382},
  publisher     = {{ACM}},
  year          = {2025}
}

@inproceedings{conf/ec/BranzeiN19,
  author        = {Simina Br{\^{a}}nzei and
                   Noam Nisan},
  title         = {Communication Complexity of Cake Cutting},
  booktitle     = {{EC}},
  pages         = {525},
  publisher     = {{ACM}},
  year          = {2019}
}

@article{journals/iandc/LiMSS25,
  author        = {Zihan Li and
                   Pasin Manurangsi and
                   Jonathan Scarlett and
                   Warut Suksompong},
  title         = {Complexity of round-robin allocation with potentially noisy queries},
  journal       = {Inf. Comput.},
  volume        = {306},
  pages         = {105332},
  year          = {2025}
}

@inproceedings{journals/corr/abs-2602-06361,
  author        = {Zihan Li and
                   Yan Hao Ling and
                   Jonathan Scarlett and
                   Warut Suksompong},
  title         = {Envy-Free Allocation of Indivisible Goods via Noisy Queries},
  booktitle     = {{ICML}},
  year          = {2026}
}

@book{books/daglib/0019107,
  author        = {Martin Aigner and
                   G{\"u}nter M. Ziegler},
  title         = {Proofs from {THE BOOK}},
  edition       = {Third},
  publisher     = {Springer},
  year          = {2004},
  isbn          = {978-3-540-40460-6},
  doi           = {10.1007/978-3-662-05412-3},
  url           = {https://doi.org/10.1007/978-3-662-05412-3},
  bibsource     = {dblp computer science bibliography, https://dblp.org}
}

@book{books/daglib/RobertsonW98,
  author        = {Jack Robertson and
                   William A. Webb},
  title         = {Cake-Cutting Algorithms: Be Fair if You Can},
  publisher     = {A K Peters},
  year          = {1998}
}

@article{journals/talg/EdmondsP11,
  author        = {Jeff Edmonds and
                   Kirk Pruhs},
  title         = {Cake Cutting Really is Not a Piece of Cake},
  journal       = {{ACM} Trans. Algorithms},
  volume        = {7},
  number        = {4},
  pages         = {51:1--51:12},
  year          = {2011}
}

@inproceedings{conf/nips/BranzeiN22,
  author        = {Simina Br{\^{a}}nzei and
                   Noam Nisan},
  title         = {The Query Complexity of Cake Cutting},
  booktitle     = {{NeurIPS}},
  volume        = {35},
  year          = {2022}
}

@article{journals/dam/EvenP84,
  author        = {Shimon Even and
                   Azaria Paz},
  title         = {A Note on Cake Cutting},
  journal       = {Discret. Appl. Math.},
  volume        = {7},
  number        = {3},
  pages         = {285--296},
  year          = {1984}
}

@inproceedings{conf/ijcai/Procaccia09,
  author        = {Ariel D. Procaccia},
  title         = {Thou Shalt Covet Thy Neighbor's Cake},
  booktitle     = {{IJCAI}},
  pages         = {239--244},
  year          = {2009}
}

@article{journals/combinatorics/Stromquist08,
  author        = {Walter Stromquist},
  title         = {Envy-Free Cake Divisions Cannot be Found by Finite Protocols},
  journal       = {Electron. J. Comb.},
  volume        = {15},
  number        = {1},
  pages         = {R11},
  year          = {2008}
}

@inproceedings{conf/focs/AzizM16,
  author        = {Haris Aziz and
                   Simon Mackenzie},
  title         = {A Discrete and Bounded Envy-Free Cake Cutting Protocol for Any Number of Agents},
  booktitle     = {{FOCS}},
  pages         = {416--427},
  publisher     = {{IEEE} Computer Society},
  year          = {2016},
  doi           = {10.1109/FOCS.2016.52},
  url           = {https://arxiv.org/abs/1604.03655}
}

@article{MeunierSu2019,
  author        = {Fr{\'e}d{\'e}ric Meunier and
                   Francis Edward Su},
  title         = {Multilabeled Versions of {Sperner}'s and {Fan}'s Lemmas and Applications},
  journal       = {{SIAM} J. Appl. Algebra Geom.},
  volume        = {3},
  number        = {3},
  pages         = {391--411},
  year          = {2019},
  doi           = {10.1137/18M1192548},
  url           = {https://doi.org/10.1137/18M1192548}
}

@inproceedings{conf/soda/DehghaniFHY18,
  author        = {Sina Dehghani and
                   Alireza Farhadi and
                   MohammadTaghi Hajiaghayi and
                   Hadi Yami},
  title         = {Envy-free Chore Division for An Arbitrary Number of Agents},
  booktitle     = {{SODA}},
  pages         = {2564--2583},
  publisher     = {{SIAM}},
  year          = {2018},
  doi           = {10.1137/1.9781611975031.164},
  url           = {https://doi.org/10.1137/1.9781611975031.164}
}

@inproceedings{conf/soda/HaeuplerHIRTT25,
  author        = {Bernhard Haeupler and
                   Richard Hlad{\'{i}}k and
                   John Iacono and
                   V{\'{a}}clav Rozhon and
                   Robert E. Tarjan and
                   Jakub T{\v{e}}tek},
  title         = {Fast and Simple Sorting Using Partial Information},
  booktitle     = {{SODA}},
  pages         = {3953--3973},
  publisher     = {{SIAM}},
  year          = {2025},
  doi           = {10.1137/1.9781611978322.134},
  url           = {https://arxiv.org/abs/2404.04552}
}

@inproceedings{conf/stoc/Cohen-AddaddGGL26,
  author        = {Vincent Cohen-Addad and
                   Tommaso d'Orsi and
                   Anupam Gupta and
                   Guru Guruganesh and
                   Euiwoong Lee and
                   Renato {Paes Leme} and
                   Debmalya Panigrahi and
                   Madhusudhan Reddy Pittu and
                   Jon Schneider and
                   David P. Woodruff},
  title         = {Combinatorial Optimization using Comparison Oracles},
  booktitle     = {{STOC}},
  pages         = {1995--2006},
  publisher     = {{ACM}},
  year          = {2026},
  doi           = {10.1145/3798129.3800904},
  url           = {https://arxiv.org/abs/2511.15142}
}

@inproceedings{conf/stoc/AzizM16,
  author        = {Haris Aziz and
                   Simon Mackenzie},
  title         = {A Discrete and Bounded Envy-Free Cake Cutting Protocol for Four Agents},
  booktitle     = {{STOC}},
  pages         = {454--464},
  publisher     = {{ACM}},
  year          = {2016},
  doi           = {10.1145/2897518.2897522},
  url           = {https://arxiv.org/abs/1508.05143}
}

@inproceedings{conf/soda/ArndtMIP27,
  author        = {Stephen Arndt and
                   Benjamin Moseley and
                   Sungjin Im and
                   Kirk Pruhs},
  title         = {No, Cake Cutting Really is a Piece of Cake},
  booktitle     = {{SODA}},
  year          = {2027},
  note          = {To appear},
  url           = {https://arxiv.org/abs/2606.07238}
}

@inproceedings{conf/focs/HollenderR23,
  author        = {Alexandros Hollender and
                   Aviad Rubinstein},
  title         = {Envy-Free Cake-Cutting for Four Agents},
  booktitle     = {{FOCS}},
  pages         = {113--122},
  publisher     = {{IEEE} Computer Society},
  year          = {2023},
  doi           = {10.1109/FOCS57990.2023.00015},
  url           = {https://arxiv.org/abs/2311.02075}
}

@inproceedings{conf/soda/0001SW25,
  author        = {Aditya Anand and
                   Thatchaphol Saranurak and
                   Yunfan Wang},
  title         = {Deterministic Edge Connectivity and Max Flow using Subquadratic Cut Queries},
  booktitle     = {{SODA}},
  pages         = {124--142},
  publisher     = {{SIAM}},
  year          = {2025},
  doi           = {10.1137/1.9781611978322.4},
  url           = {https://arxiv.org/abs/2410.18704}
}

@inproceedings{conf/focs/Rutschmann26,
  author        = {Daniel Rutschmann},
  title         = {Sorting under Partial Information with Optimal Preprocessing Time via Unified
                   Bound Heaps},
  booktitle     = {{FOCS}},
  year          = {2026},
  note          = {To appear},
  url           = {https://arxiv.org/abs/2604.12653}
}

@inproceedings{conf/soda/GeHKV27,
  author        = {Luise Ge and
                   Daniel Halpern and
                   Gregory Kehne and
                   Yevgeniy Vorobeychik},
  title         = {Linear Social Choice with Few Queries: A Moment-Based Approach},
  booktitle     = {{SODA}},
  year          = {2027},
  note          = {To appear},
  url           = {https://arxiv.org/abs/2603.19510}
}

@book{Moulin2003,
  author        = {Herv{\'e} Moulin},
  title         = {Fair Division and Collective Welfare},
  publisher     = {{MIT} Press},
  year          = {2003}
}

@article{Foley1967,
  author        = {Duncan K. Foley},
  title         = {Resource Allocation and the Public Sector},
  journal       = {Yale Econ. Essays},
  volume        = {7},
  number        = {1},
  pages         = {45--98},
  year          = {1967}
}

@inproceedings{BalcanVitercikWhite2016,
  author        = {Maria-Florina Balcan and
                   Ellen Vitercik and
                   Colin White},
  title         = {Learning Combinatorial Functions from Pairwise Comparisons},
  booktitle     = {{COLT}},
  series        = {Proceedings of Machine Learning Research},
  volume        = {49},
  pages         = {310--335},
  publisher     = {PMLR},
  year          = {2016},
  url           = {https://proceedings.mlr.press/v49/balcan16b.html}
}

@article{KaneLovettMoran2019,
  author        = {Daniel M. Kane and
                   Shachar Lovett and
                   Shay Moran},
  title         = {Near-optimal Linear Decision Trees for {$k$-SUM} and Related Problems},
  journal       = {J. {ACM}},
  volume        = {66},
  number        = {3},
  pages         = {16:1--16:18},
  year          = {2019},
  doi           = {10.1145/3285953},
  url           = {https://arxiv.org/abs/1705.01720}
}

@article{YeBai2026,
  author        = {Qilin Ye and
                   Yannan Bai},
  title         = {Cutting Down the Tower: Single-Exponential Envy-Free Cake Cutting},
  journal       = {arXiv preprint arXiv:2609.05191},
  year          = {2026},
  doi           = {10.48550/arXiv.2609.05191},
  eprint        = {2609.05191},
  archivePrefix = {arXiv},
  primaryClass  = {cs.GT},
  url           = {https://arxiv.org/abs/2609.05191}
}

@article{MackenzieSuzuki2026,
  author        = {Simon Mackenzie and
                   Mashbat Suzuki},
  title         = {Exact Envy-Free Cake Cutting with Polynomial Expected Queries},
  journal       = {arXiv preprint arXiv:2610.28535},
  year          = {2026},
  doi           = {10.48550/arXiv.2610.28535},
  eprint        = {2610.28535},
  archivePrefix = {arXiv},
  primaryClass  = {cs.GT},
  url           = {https://arxiv.org/abs/2610.28535}
}

@article{ArndtPruhsTran2026,
  author        = {Stephen Arndt and
                   Kirk Pruhs and
                   Trung Tran},
  title         = {An {$\Omega(n \log n)$} Randomized Lower Bound for Cutting a Cake into
                   Proportionally Fair Pieces},
  journal       = {arXiv preprint arXiv:2605.21829},
  year          = {2026},
  doi           = {10.48550/arXiv.2605.21829},
  eprint        = {2605.21829},
  archivePrefix = {arXiv},
  primaryClass  = {cs.DS},
  url           = {https://arxiv.org/abs/2605.21829}
}

@article{BranzeiPaparasRecker2020,
  author        = {Simina Br{\^a}nzei and
                   Dimitris Paparas and
                   Nicholas Recker},
  title         = {Searching, Sorting, and Cake Cutting in Rounds},
  journal       = {arXiv preprint arXiv:2012.00738},
  year          = {2020},
  doi           = {10.48550/arXiv.2012.00738},
  eprint        = {2012.00738},
  archivePrefix = {arXiv},
  primaryClass  = {cs.DS},
  url           = {https://arxiv.org/abs/2012.00738}
}

@inproceedings{conf/aaai/BalkanskiBKP14,
  author        = {Eric Balkanski and
                   Simina Br{\^a}nzei and
                   David Kurokawa and
                   Ariel D. Procaccia},
  title         = {Simultaneous Cake Cutting},
  booktitle     = {{AAAI}},
  pages         = {566--572},
  publisher     = {{AAAI} Press},
  year          = {2014},
  doi           = {10.1609/aaai.v28i1.8802},
  url           = {https://ojs.aaai.org/index.php/AAAI/article/view/8802}
}

@article{journals/fttcs/Vadhan12,
  author        = {Salil P. Vadhan},
  title         = {Pseudorandomness},
  journal       = {Found. Trends Theor. Comput. Sci.},
  volume        = {7},
  number        = {1--3},
  pages         = {1--336},
  year          = {2012},
  doi           = {10.1561/0400000010}
}

\end{document}